\documentclass[preprint,12pt]{elsarticle}

\usepackage{caption}
\usepackage{amsmath}
\usepackage{amssymb}
\usepackage{amsthm}
\usepackage{mathrsfs}

\usepackage{epstopdf}

\usepackage{subfig}

\usepackage{booktabs} 
\usepackage{multirow} 

\usepackage{url}
\usepackage{hyperref}
\usepackage{xcolor}
\usepackage{cleveref} 

\usepackage{lineno}

\usepackage{threeparttable}
\usepackage{color}
\usepackage{epstopdf}
\usepackage{float}
\usepackage[top=2.5cm, bottom=2.5cm, left=2cm, right=2cm]{geometry}
\renewcommand{\vec}[1]{\boldsymbol{#1}}

\biboptions{numbers,sort&compress}

\journal{Journal of Computational Physics}

\begin{document}


\begin{frontmatter}



\title{Adaptive unified gas-kinetic scheme for diatomic gases with rotational and vibrational nonequilibrium}


\author[a]{Yufeng Wei}
\author[a]{Wenpei Long}
\author[a,b,c]{Kun Xu\corref{cor1}}

\cortext[cor1] {Corresponding author.}
\ead{makxu@ust.hk}

\address[a]{Department of Mathematics, Hong Kong University of Science and Technology, Hong Kong, China}
\address[b]{Department of Mechanical and Aerospace Engineering, Hong Kong University of Science and Technology, Hong Kong, China}
\address[c]{HKUST Shenzhen Research Institute, Shenzhen, 518057, China}
\begin{abstract}
Multiscale non-equilibrium physics at large variations of local Knudsen number are encountered in applications of aerospace engineering and micro-electro-mechanical systems, such as high-speed flying vehicles and low pressure of the encapsulation. An accurate description of flow physics in all flow regimes within a single computation requires a genuinely multiscale method. The adaptive unified gas-kinetic scheme (AUGKS) is developed for such multiscale flow simulation. The AUGKS applies discretized velocity space to accurately capture the non-equilibrium physics in the multiscale UGKS, and adaptively employs continuous distribution functions following Chapman--Enskog expansion to efficiently recover near-equilibrium flow region in GKS. The UGKS and GKS are dynamically connected at the cell interface through the fluxes from the discretized and continuous gas distribution functions, which avoids any buffer zone between them. In this study, the AUGKS method with rotation and vibration non-equilibrium is developed based on a multiple temperature relaxation model. The real gas effect in different flow regimes has been properly captured. To capture aerodynamic heating accurately, the heat flux modifications from the rotation and vibration modes are also included in the current scheme. Unstructured discrete particle velocity space is adopted to further improve the computational performance of the AUGKS. Numerical tests, including Sod tube, normal shock structure, high-speed flow around the two-dimensional cylinder and three-dimensional sphere and space vehicles, and an unsteady nozzle plume flow from the continuum flow to the background vacuum, have been conducted to validate the current scheme. In comparison with the original UGKS, the current scheme speeds up the computation, reduces the memory requirement, and maintains the equivalent accuracy for multiscale flow simulation, which provides an effective tool for non-equilibrium flow simulations, especially for the flows at low and medium speed.
\end{abstract}

\begin{keyword}
Unified gas-kinetic scheme \sep
Adaptive velocity space \sep
Multiscale transport \sep
Diatomic gas\sep
Vibrational non-equilibrium
\end{keyword}

\end{frontmatter}



\section{Introduction}\label{sec:intro}
Multiscale flows are commonly encountered in applications of aerospace engineering and micro-electro-mechanical systems (MEMS). For high-speed flying vehicles, the highly compressed gas at the leading edge and the strong expansion wave in the trailing edge can cover the whole flow regimes \cite{bird1994molecular,xu2021unified}. In MEMS, the small size of the structure and the low pressure of the encapsulation result in significant rarefaction effects of gas \cite{senturia1997simulating,alexeenko2003numerical,wang2004simulations,wang2022investigation}. Additionally, in real flow physics when a diatomic molecule reaches its characteristic temperature, its rotational and vibrational modes are activated with significant impact on heating and force of flow field \cite{boyd2017nonequilibrium}. Therefore, the development of efficient and accurate multiscale simulation methods with the inclusion of real gas effects is of great importance.

For multiscale flows across all Knudsen regimes, the description of particle collisions and free streaming are equally important. It presents challenges for Navier--Stokes equations and necessitates the gas kinetic theory. The Boltzmann equation is the fundamental governing equation in rarefied gas dynamics. Theoretically, it can capture multiscale flow physics in all Knudsen regimes, with the enforcement of resolving the flow physics in the particle mean free path and mean collision time scale. For highly non-equilibrium flow, there are mainly two kinds of numerical methods to solve the Boltzmann equation, i.e. the stochastic particle method and the deterministic method. The stochastic methods employ discrete particles to simulate the statistical behavior of molecular gas dynamics \cite{bird1994molecular,fan2001statistical,shen2006rarefied,sun2002direct,baker2005variance,homolle2007low,degond2011moment,pareschi2000asymptotic,ren2014asymptotic,dimarco2011exponential}. This kind of Lagrangian-type scheme achieves high computational efficiency and robustness in rarefied flow simulation, especially for hypersonic flow. However, it suffers from statistical noise in the low-speed simulation due to its intrinsic stochastic nature. Meanwhile, in the near continuum flow regime, the treatment of intensive particle collisions makes the computational cost very high. The deterministic approaches apply a discrete distribution function to solve the kinetic equations and naturally obtain accurate solutions without statistical noise \cite{chu1965kinetic,JCHuang1995,Mieussens2000,tcheremissine2005direct,Kolobov2007,LiZhiHui2009,ugks2010,wu2015fast,aristov2012direct,
li2004study,li2019gas,ugks2010,guo2013discrete,chen2017unified,chen2015comparative}. At the same time, the deterministic method can achieve high efficiency by using numerical acceleration techniques, such as implicit algorithms \cite{yang1995rarefied,Mieussens2000,zhu2016implicit,zhu2017implicit,zhu2018implicit,jiang2019implicit}, memory reduction techniques \cite{chen2017unified}, and adaptive refinement method \cite{chen2012unified}, fast evaluation of the Boltzmann collision term \cite{mouhot2006fast,wu2013deterministic}. Asymptotic preserving (AP) schemes \cite{filbet2010class,dimarco2013asymptotic} can be developed to release the stiffness of the collision term at the small Knudsen number case. However, for most AP schemes only the Euler solution in the hydrodynamic limit is recovered. Additionally, for hypersonic and rarefied flow, the deterministic methods have to discretize the particle velocity space with a high resolution to capture nonequilibrium distribution, which brings huge memory consumption and computational cost, especially for the three-dimensional calculation. Moreover, for both stochastic and deterministic methods, once the gas evolution process is split into collisionless free transport and instant collision, a numerical dissipation proportional to the time step is usually unavoidable. Therefore, the mesh size and the time step in these schemes have to be less than the mean free path and the particle mean collision time, respectively, to avoid the physical dissipation being overwhelmingly taken over by the numerical one in the continuum regime, such as the laminar boundary layer computation at high Reynolds number. In order to remove the constraints on the mesh size and time step in the continuum flow regime, the DVM-based unified gas-kinetic scheme (UGKS) \cite{ugks2010}, particle-based unified wave-particle (UGKWP) method \cite{liu2019unified,zhu2019unified}, discrete UGKS \cite{guo2021progress} and discrete UGKWP method \cite{yang2023discrete} with the coupled particle transport and collision in the flux evaluation have been constructed successfully. At the same time, the multiscale particle methods have been constructed as well \cite{fei2020unified,fei2021efficient}. All these multiscale methods have the unified preserving (UP) property in capturing the Navier--Stokes  solution in the continuum regime \cite{guo2023unified}.

Due to the large discretized particle velocity space used in the UGKS for three-dimensional computation, its memory requirements and computational cost limit its efficient applications. In recent years, an adaptive unified gas-kinetic scheme (AUGKS) has been proposed \cite{xiao2020velocity}, which incorporates dynamically coupled continuous and discrete particle velocity spaces within a unified framework. In the near-equilibrium region, the scheme adopts continuous gas distribution functions following the Chapman--Enskog expansion. As a result, only macroscopic variables need to be stored and updated in these regions. In non-equilibrium flow regimes, the evolution of the gas distribution function is directly represented in the discrete velocity space. This adaptive scheme reduces memory requirements and speeds up computations in near-equilibrium flows compared to the original UGKS while providing the same physical solution. Moreover, by using distribution functions throughout the computational domain with deterministic approaches, the AUGKS eliminates the need for domain decomposition in the physical space to distinguish fluid and kinetic solvers. In other words, no buffer zone is needed in AUGKS.

In this paper, we present the AUGKS with the inclusion of vibrational mode for diatomic gas. The vibrational kinetic model \cite{zhang2015vib,wei2023unified} is used to describe the relaxation process from non-equilibrium to the equilibrium state \cite{bird1994molecular}, where three equilibrium states are employed to take into account the elastic and inelastic collisions and the detailed energy exchange between the translational, rotational and vibrational degrees of freedom. To obtain accurate aerodynamic heating simulation results, additional heat flux modification constructed by Hermite and Laguerre polynomial corrections is also included in the current scheme. To further improve the computational performance of AUGKS, unstructured discrete particle velocity space is adopted. In this paper, to clearly present the algorithm development, the scheme with the vibrational relaxation model will be constructed and validated in cases from one-dimensional to three-dimensional flow simulations.

The paper is organized as follows. Section \ref{sec:vib} presents the kinetic model of diatomic gas with molecular vibration. Then the AUGKS method with molecular vibration will be presented in Section \ref{sec:method}. Numerical validation of the current method will be carried out in Section \ref{sec:case}
and a conclusion will be drawn in Section \ref{sec:conclusion}.

\section{Kinetic model equation for diatomic gas}\label{sec:vib}

\subsection{Kinetic model with molecular transition, rotation, and vibration}

Considering molecular rotation and vibration, the kinetic model equation for diatomic gases can be written as
\begin{equation}\label{eq:BGK-vib}
\frac{\partial f}{\partial t}+\vec{u} \cdot \frac{\partial f}{\partial \vec{r}}
=\frac{g_{t}-f}{\tau}
+\frac{g_{t r}-g_{t}}{Z_{r} \tau}
+\frac{g_{M  }-g_{tr}}{Z_{v} \tau},
\end{equation}
where $f = f\left( {{\vec{r}},{\vec{u}}, {\vec{\xi}}, {\varepsilon_v}, t} \right)$ is the distribution function for gas molecules at physical space location $\vec{r}$ with microscopic translational velocity $\vec{u}$, rotational motion ${\vec{\xi}}$, and vibrational energy $\varepsilon_{v}$ at time $t$.
$\tau$ is the mean collision time or relaxation time to represent the mean time interval of two successive collisions.
The rotational and vibrational relaxation times are defined as
\begin{equation*}
	{\tau _{rot}} = {Z_r}\tau,
\end{equation*}
\begin{equation*}
	{\tau _{vib}} = {Z_v}\tau,
\end{equation*}
where $Z_r$ and  $Z_v$ are the rotational and vibration collision numbers, respectively.

The elastic collision process of molecules' translational motions and the inelastic collision process of internal energy exchange are described by the right-hand side of Eq.~\eqref{eq:BGK-vib} with three equilibrium states. The equilibrium state $g_t$ with three different temperatures for molecular translation, rotation, and vibration gives
\begin{equation*}
	g_{t}= \rho \left(\frac{\lambda_{t}}{\pi}\right)^{\frac{3}{2}} e^{-\lambda_{t} {\vec{c}}^2}
	\left(\frac{\lambda_{r}}{\pi}\right) e^{-\lambda_{r} {\vec{\xi}}^{2}} \frac{4 \lambda_{v}}{K_v(\lambda_v)} e^{-\frac{4 \lambda_{v}}{K_v(\lambda_v)} \varepsilon_{v}},
\end{equation*}
where $\vec{c} = \vec{u} - \vec{U}$ denotes the peculiar velocity, and ${\vec{c}}^2 = (u - U)^2 + (v-V)^2 + (w-W)^2$ and ${\vec{\xi}}^2 = \xi_{1}^2 + \xi_{2}^2$. The intermediate equilibrium state $g_{tr}$ has the same temperature $\lambda_{tr}$ of molecular translation and rotation, but a different temperature $\lambda_v$ for vibration, which indicates complete energy exchange between translational and rotational degrees of freedom, and a frozen process of vibrational energy
\begin{equation*}
	g_{tr}=\rho\left(\frac{\lambda_{tr}}{\pi}\right)^{\frac{3}{2}} e^{-\lambda_{t r} {\vec{c}}^2}
	\left(\frac{\lambda_{t r}}{\pi}\right) e^{-\lambda_{t r} {\vec{\xi}}^2} \frac{4 \lambda_{v}}{K_v(\lambda_v)} e^{-\frac{4 \lambda_{v}}{K_v(\lambda_v)} \varepsilon_{v}}.
\end{equation*}
After sufficient collisions, the equilibrium state with equal-partitioned energy for each degree of freedom
\begin{equation*}\label{eq:g-M}
	g_{M}=\rho\left(\frac{\lambda_M}{\pi}\right)^{\frac{3}{2}} e^{-\lambda_M {\vec{c}}^2}
	\left(\frac{\lambda_M}{\pi}\right) e^{-\lambda_M {\vec{\xi}}^2} \frac{4 \lambda_M}{K_v(\lambda_M)} e^{-\frac{4 \lambda_M}{K_v(\lambda_M)} \varepsilon_{v}},
\end{equation*}
will be reached.

In these equilibrium states, $\lambda$ is computed from the corresponding internal energy.
Specifically, we have
\begin{eqnarray*}
&\lambda_{t}=&\frac{3 \rho}{4} / (\rho E_t), \\
&\lambda_{r}=&\frac{K_r \rho}{4} / (\rho E_r), \\
&\lambda_{v}=&\frac{K_v(\lambda_v)\rho}{4} / (\rho E_v), \\
&\lambda_{t r}=&\frac{(3+K_r) \rho}{4} / (\rho E_{tr}), \\
&\lambda_M=&\frac{[3+K_r+K_v(\lambda_M)] \rho}{4} / (\rho E_M),
\end{eqnarray*}
and
\begin{eqnarray*}
&\rho E_{t}&= \frac{1}{2} \int {{\vec{c}}^2 f {\rm d}{\vec{\Xi}}}, \\
&\rho E_{r}&= \frac{1}{2} \int{{\vec{\xi}}^{2} f {\rm d} \vec{\Xi}}, \\
&\rho E_{v}&= \int {\varepsilon_{v} f {\rm d}\vec{\Xi}}, \\
&\rho E_{t r} &=\frac{1}{2} \int{({\vec{c}}^2 + {\vec{\xi}}^{2}) f	{\rm d}{\vec{\Xi}}}, \\
&\rho E_M     &= \int {\left[\frac{1}{2}({\vec{c}}^{2} + \vec{\xi}^{2}) + \varepsilon_{v}\right] f {\rm d}\vec{\Xi}},
\end{eqnarray*}
where
\begin{equation*}
\int{ (\cdot) {\rm d}\vec{\Xi}} = \int_{-\infty}^{\infty} {\rm d}\vec{u} \int_{-\infty}^{\infty} {\rm d}{\vec{\xi}} \int_{0}^{\infty} {(\cdot) {\rm d}{\varepsilon_v}},
\end{equation*}
and $K_r$ and $K_v(\lambda)$ denote the number of rotational and vibrational degrees of freedom, respectively. $\lambda_t$, $\lambda_r$, $\lambda_v$, $\lambda_{tr}$, $\lambda_M$ are associated with the
translational temperature $T_t$, rotational temperature $T_r$, vibrational temperature $T_v$, the translation-rotation average temperature $T_{tr}$ and the fully relaxed temperature $T_M$, respectively by $\lambda = m / (2 k_B T)$, where $m$ is molecular mass, $k_B$ is the Boltzmann constant. It should be noted that the number of vibrational degrees of freedom $K_v(\lambda)$ is determined by the vibrational temperature in each equilibrium state, i.e.,
\begin{equation}\label{eq:Kv}
	K_{v}(\lambda)=\frac{4\Theta_{v}k_B\lambda/m}{e^{2\Theta_{v}k_B\lambda/m}-1},
\end{equation}
where $\Theta_v$ is the characteristic temperature of vibration for diatomic gases, e.g.,  $3371$ K for nitrogen and $2256$ K for oxygen \cite{shen2006rarefied}.

With the above three equilibrium states, the energy exchange between molecular translation, rotation, and vibration can be well described by adjusting the collision numbers $Z_r$ and $Z_v$. Experimental observation shows that the rotational relaxation is faster than the vibrational one, i.e., $1 < Z_r < Z_v$. From the relaxation terms on the right-hand side of Eq.~\eqref{eq:BGK-vib}, the relaxation process can be divided into three stages. Firstly, the non-equilibrium distribution function $f$ has different translational, rotational, and vibrational temperatures. After time $\tau$, the elastic collisions drive the distribution function $f$ approaching the translational equilibrium state $g_t$. In the second stage, the inelastic collisions happen within time $Z_r \tau$ to exchange the translational and rotational energy, which drives the distribution function approaching the rotational equilibrium state $g_{tr}$ with the same translational and rotational temperature $T_{tr}$. In the last stage, gas molecules encounter sufficient elastic and inelastic collisions within time $Z_v \tau$, and the internal energy is fully exchanged between each degree of freedom. At this time, the full equilibrium state $g_M$ with the same temperature $T_M$ for translation, rotation, and vibration is achieved.

\subsection{Heat flux modification for kinetic model}

The current kinetic model Eq.~\eqref{eq:BGK-vib} inherits the advantages of the Bhatnagar--Gross--Krook (BGK) \cite{BGK1954} model, such as its simplicity and the guarantee of entropy increase. However, it also shares the disadvantage of the BGK model in that the Prandtl number is always equal to 1. To ensure that the model captures correct thermal conductivity and viscosity coefficient simultaneously, the Hermite and Laguerre expansions are introduced to adjust the relaxation rates of heat flux \cite{zhang2015vib}. The kinetic model with the inclusion of modified equilibrium distribution is given by
\begin{equation} \label{eq:BGK-vib-Pr}
	\frac{\partial f}{\partial t}+\vec{u} \cdot \frac{\partial f}{\partial \vec{r}}
	=\frac{g_{t}^{+}-f}{\tau}
	+\frac{g_{tr}^{+}-g_{t}^{+}}{Z_{r} \tau}
	+\frac{g_{M}^{+}-g_{tr}^{+}}{Z_{v} \tau}.
\end{equation}
For ease of solution, it is rewritten in a BGK-type form
\begin{equation}\label{eq:BGK-type}
	\frac{\partial f}{ \partial t} + \vec{u} \cdot \frac{\partial f}{\partial \vec{r}} = \frac{ g - f}{\tau },
\end{equation}
where $g$ is the effective equilibrium state, defined as the convex combination of three modified equilibrium distribution function
\begin{equation}\label{eq:g}
	g= \left( 1 - \frac{1}{Z_r} \right){g_t^+} + \left( \frac{1}{Z_r} - \frac{1}{Z_v} \right){g_{tr}^+} + \frac{1}{Z_v} {g_M^+}
\end{equation}
with
\begin{eqnarray*}
	g_t^{+}&=&g_t\left[1+Q_t\left(\lambda_t\right)+Q_r\left(\lambda_t, \lambda_r\right)+Q_v\left(\lambda_t, \lambda_v\right)\right], \\
	g_{t r}^{+}&=&g_{t r}\left[1+\omega_0 Q_t\left(\lambda_{t r}\right)+\omega_1 Q_r\left(\lambda_{t r}, \lambda_{t r}\right)+\omega_2 Q_v\left(\lambda_{t r}, \lambda_v\right)\right], \\
	g_M^{+} &=& g_M\left[1+\omega_3 Q_t\left(\lambda_M\right)+\omega_4 Q_r\left(\lambda_M, \lambda_M\right)+\omega_5 Q_v\left(\lambda_M, \lambda_M\right)\right],
\end{eqnarray*}
where $\omega_0$ to $\omega_5$ are coefficients to obtain the right relaxation rate of heat flux.$Q_t(\lambda)$, $Q_r(\lambda_1, \lambda_2)$, and $Q_v(\lambda_1, \lambda_2)$ are terms of orthogonal polynomial constructed by the Hermite and Laguerre expansions around Maxwellian equilibrium states
\begin{eqnarray*}
	Q_t(\lambda)&=&\frac{4 \vec{q}_t \cdot \vec{c} \lambda^2(1-{\rm Pr})}{5 \rho}\left(2 \lambda \vec{c}^2-5\right) ,\\
	Q_r(\lambda_1, \lambda_2)&=&\frac{4 \vec{q}_r \cdot \vec{c} \lambda_1 \lambda_2( 1 - \sigma)}{\rho}\left(\lambda_2 \vec{\xi}^2 -1 \right), \\
	Q_v(\lambda_1, \lambda_2)&=&\frac{8 \vec{q}_v \cdot \vec{c} \lambda_1 \lambda_2(1 - \sigma)}{K_v\left(\lambda_2\right) \rho}\left(\frac{4 \lambda_2 \varepsilon_v}{K_v\left(\lambda_2\right)} - 1\right),
\end{eqnarray*}
where $\sigma$ depends on the inter-molecular potential, and the translational heat flux $\vec{q}_t$, rotational heat flux $\vec{q}_r$, and vibrational heat flux $\vec{q}_v$ are
\begin{eqnarray*}
	\vec{q}_t&=& \frac{1}{2} \int \vec{c}^2 f {\rm d}{\vec{\Xi}}, \\
	\vec{q}_r&=& \frac{1}{2} \int \vec{c} \vec{\xi}^2 f {\rm d}{\vec{\Xi}}, \\
	\vec{q}_v&=&  \int \vec{c} \varepsilon_v f {\rm d}{\vec{\Xi}}.
\end{eqnarray*}
For a uniform flow, the thermal flux relaxation rate corresponding to the model equation Eq.~\eqref{eq:BGK-vib-Pr} is given by
\begin{eqnarray*}
		\frac{\partial \vec{q}_t}{\partial t}&=&-\left(\operatorname{Pr}+\frac{1-\omega_0}{3 Z_r}+\frac{\omega_0-\omega_3}{3 Z_v}\right) \frac{\vec{q}_t}{\tau}, \\
		\frac{\partial \vec{q}_r}{\partial t}&=&-\left[\sigma+\frac{(1-\sigma)\left(1-\omega_1\right)}{Z_r}+\frac{(1-\sigma)\left(\omega_1-\omega_4\right)}{Z_v}\right] \frac{\vec{q}_r}{\tau}, \\
		\frac{\partial \vec{q}_v}{\partial t}&=&-\left[\sigma+\frac{(1-\sigma)\left(1-\omega_2\right)}{Z_r}+\frac{(1-\sigma)\left(\omega_2-\omega_5\right)}{Z_v}\right] \frac{\vec{q}_v}{\tau} .
\end{eqnarray*}

\subsection{Reduced distribution function}

In gas kinetic models, the computational cost grows exponentially as the number of degrees of freedom increases. For monatomic gases, which only possess translational degrees of freedom, the computational cost for $N$ particles is typically proportional to $N^3$. However, diatomic molecules have three additional degrees of freedom associated with molecular rotational velocities $\xi_1$ and $\xi_2$ and vibrational energy $\varepsilon_v$, resulting in a computational cost of $N^6$, which is three orders of magnitude higher than that of monatomic gases. Nevertheless, since our focus is solely on the macroscopic rotational and vibrational energies, the specific motions and energy exchange processes of these additional degrees of freedom can be disregarded. By introducing simplified distribution functions, the computational cost can be reduced to the order of $3N^3$
\begin{eqnarray*}
	&G =& \int_0^\infty \int_{-\infty}^\infty  f  {\rm d}\vec{\xi} {\rm d}\varepsilon_v,\\
	&R =& \int_0^\infty  \int_{-\infty}^\infty \vec{\xi}^2f{\rm d} \vec {\xi} {\rm d}\varepsilon_v,\\
	&V =& \int_0^\infty  \int_{-\infty}^\infty  \varepsilon_v f {\rm d} \vec {\xi} {\rm d} \varepsilon_v,
\end{eqnarray*}
where $G$, $R$, and $V$ are reduced functions of $(\vec{r}, \vec{u}, t)$ denote the mass, rotational energy, and vibrational energy distribution functions within translational velocity space $\vec{u}$, respectively. Therefore, macroscopic variables $\vec{W}$, i.e. the densities of mass, momentum, total energy, rotational energy, and vibrational energy can be rewritten as
\begin{equation*}
	{\vec{W}} = \left( \begin{array}{c}
		\rho \\
		\rho \vec{U}\\
		\rho E\\
		\rho E_r\\
		\rho E_v
		\end{array} \right) = \int_{-\infty}^\infty  {\left( \begin{array}{c}
		G\\
		\vec{u} G\\
		\frac{\vec{u}^2}{2}G + \frac{1}{2}R + V\\
		\frac{1}{2}R\\
		V
		\end{array} \right){\rm d}\vec{u} }.
\end{equation*}
Integrating the Eq.~\eqref{eq:BGK-vib-Pr} by rotational and vibrational degrees of freedom, the relaxation model can be presented by reduced distribution functions
\begin{eqnarray*}
		\frac{\partial G}{\partial t}+\vec{u} \cdot \frac{\partial G}{\partial \vec{r}}&=&\frac{G_t^+-G}{\tau}+\frac{G_{tr}^+-G_{t}^+}{Z_r \tau}+\frac{G_M^+-G_{tr}^+}{Z_v \tau}, \\
		\frac{\partial R}{\partial t}+\vec{u} \cdot \frac{\partial R}{\partial \vec{r}}&=&\frac{R_{t}^{+}-R}{\tau}+\frac{R_{tr}^{+}-R_{t}^{+}}{Z_{r} \tau}+\frac{R_{M}^{+}-R_{tr}^{+}}{Z_{v} \tau}, \\
		\frac{\partial V}{\partial t}+\vec{u} \cdot \frac{\partial V}{\partial \vec{r}}&=&\frac{V_{t}^{+}-V}{\tau}+\frac{V_{tr}^{+}-V_{t}^{+}}{Z_{r} \tau}+\frac{V_{M}^{+}-V_{tr}^{+}}{Z_{v} \tau},
\end{eqnarray*}
where the reduced modified equilibrium distribution functions are
\begin{eqnarray*}
		G_t^+ & =& G_t        \left[   1 + Q_t( \lambda_t ) \right], \\
		R_t^+  &=& R_t        \left[  1 + Q_t( \lambda_t ) + Q_r^\prime( \lambda_t,\lambda_r ) \right], \\
		V_t^+  &=& V_t        \left[  1 + Q_t( \lambda_t ) + Q_v^\prime( \lambda_t,\lambda_v) \right], \\
		G_{tr}^+ & = & G_{tr} \left[  1 + \omega_0 Q_t( \lambda_{tr} ) \right], \\
		R_{tr}^+ & = & R_{tr} \left[  1 + \omega_0 Q_t( \lambda_{tr} ) + \omega_1 Q_r^\prime( \lambda_{tr},\lambda_r ) \right], \\
		V_{tr}^+ & = & V_{tr} \left[  1 + \omega_0 Q_t( \lambda_{tr} ) + \omega_2 Q_v^\prime( \lambda_{tr},\lambda _v ) \right], \\
		G_M^+ & = & G_M       \left[  1 + \omega_3 Q_t( \lambda_M ) \right],\\
		R_M^+& =& R_M 		  \left[  1 + \omega_3 Q_t( \lambda_M ) + \omega_4 Q_r^\prime ( \lambda_M,\lambda_M ) \right], \\
		V_M^+ & =& V_M        \left[  1 + \omega_3 Q_t( \lambda_M ) + \omega_5 Q_v^\prime ( \lambda_M,\lambda_M ) \right],
\end{eqnarray*}
with
\begin{eqnarray*}
	Q_r^\prime(\lambda_1, \lambda_2)&=&\frac{4 \vec{q}_r \cdot \vec{c} \lambda_1 \lambda_2( 1 - \sigma )}{\rho}, \\
	Q_v^\prime(\lambda_1, \lambda_2)&=&\frac{8 \vec{q}_v \cdot \vec{c} \lambda_1 \lambda_2( 1 - \sigma )}{K_v\left(\lambda_2\right) \rho},
\end{eqnarray*}
and the reduced Maxwellian distribution functions
\begin{align*}
	G_t =& \rho \left( \frac{\lambda_t}{\pi} \right)^{\frac{3}{2}} e^{ - \lambda_t \vec{c}^2},
	&R_t &= \frac{1}{\lambda_r} G_t,
	&V_t &= \frac{K_v (\lambda_v)}{4 \lambda_v} G_t, \\
	G_{tr} =& \rho
			  \left( \frac{\lambda_{tr}}{\pi} \right)^{\frac{3}{2}}
		      e^{ -\lambda_{tr} \vec{c}^2 },
	&R_{tr} &= \frac{1}{\lambda_{tr}} G_{tr},
	&V_{tr} &= \frac{K_v(\lambda_v)}{4 \lambda_v} G_{tr}, \\
	{G_M} = & \rho\left(\frac{\lambda_M}{\pi}\right)^{\frac{3}{2}} e^{-\lambda_M \vec{c}^2},
	&R_M &= \frac{1}{\lambda_M} G_M,
	&V_M &=  \frac{K_v(\lambda_M)}{4 \lambda_M} G_M.
\end{align*}

\section{Numerical method}\label{sec:method}
The governing equations for the finite volume method in the context of rotational and vibrational modes can be described by choosing $\vec{W} = (\rho, \rho {\vec U}, \rho E, \rho E_r, \rho E_v)^T$, i.e., densities of mass, momentum, energy, rotational energy, and vibrational energy as the five independent variables. Within a discrete finite volume cell $i$ and a discrete time scale $\Delta t = t^{n+1} - t^n$, the governing equations can be expressed as a set of conservation laws
\begin{equation}\label{eq:update-macro}
\vec{W}_i^{n + 1}
=
\vec{W}_i^n
- \frac{\Delta t}{\Omega_i}
\sum\limits_{j \in N(i)} {\vec{F}_{ij}{\mathcal A}_{ij}}
+ {\vec{S}}_i,
\end{equation}
where $\Omega_i$ denotes the volume of cell $i$, $N(i)$ is the set of all interface-adjacent neighboring cells of cell $i$, and $j$ is one of the neighboring cells of $i$. The interface between them is labeled as $ij$, having an area of $\mathcal{A}_{ij}$. The source term $\vec{S}_i$ represents the contributions from translational, rotational, and vibrational energy exchange. $\vec{F}_{ij}$ is the macroscopic flux  crossing the interface $ij$.

It is important to note that Eq.~\eqref{eq:update-macro} represents the conservation law of macroscopic quantities at the discrete scale, which is a fundamental physical law valid at all spatial and temporal scales. Therefore, the key point in developing numerical methods within the finite volume framework is to accurately describe the physical evolution of the flow at the current discrete scale, which critically depends on the construction of flux functions across interfaces. According to the gas kinetic theory, the macroscopic flux can be determined by taking moments of the microscopic distribution function flux at the interface
\begin{equation*}
	\vec{F}_{ij} =  \int \mathcal{F}_{ij} {\vec{\psi}} {\rm d}\vec{\Xi},
\end{equation*}
where $\vec{\psi} = \left( 1, \vec{u}, \frac{1}{2} {\vec{u}}^2 + \frac{1}{2} {\vec{\xi}}^2 + {\varepsilon_v}, \frac{1}{2} {\vec{\xi}}^2, {\varepsilon_v} \right)^{T}$. The time-averaged microscopic distribution function flux $\mathcal{F}_{ij}$ can be expressed as
\begin{equation*}
	\mathcal{F}_{ij} = \frac{1}{\Delta t} \int_0^{\Delta t} \vec{u} \cdot \vec{n}_{ij} f_{ij}(t) {\rm d}t,
\end{equation*}
where ${\vec{n}_{ij}}$ is the normal vector of the cell interface $ij$, and $f_{ij}(t)$ is the time-dependent gas distribution function on the interface. The evolution of the microscopic distribution function can be described by the kinetic mode Eq.~\eqref{eq:BGK-type}. Along the characteristic line, the integral solution of the kinetic model equation gives
\begin{equation}\label{eq:BGK-int}
f(\vec{r},t) =
\frac{1}{\tau}\int_{0}^t e^{-(t-t')/\tau}
g(\vec{r}^\prime,t^\prime)
{\rm d} t^\prime
+ e^{-t/\tau}f_0(\vec{r}-\vec{u}t),
\end{equation}
where $ f_0(\vec{r}) $ is the initial distribution function at the beginning of each step $t_n$, and $g(\vec{r}, t)$ is the effective equilibrium state distributed in space and time around $\vec{r}$ and $t$. The integral solution describes an evolution process from non-equilibrium to equilibrium state through particle collision.

There are two numerical schemes, i.e., unified gas-kinetic scheme (UGKS) and gas-kinetic scheme (GKS), applying this integral solution to construct the flux function, but their distinct methodologies yield unique characteristics and scopes of application. The UGKS uses discretized velocity space for distribution functions to capture non-equilibrium physics, rendering it applicable to multiscale simulations in all flow regimes. The construction of the UGKS be discussed in \ref{subsec:ugks}. However, the large number of discrete velocity points required makes the UGKS computationally demanding, with significant computational and memory costs. Thus, its efficiency requires further improvement. The GKS employs continuous distribution function following the Chapman--Enskog expansion to represent the initial distribution function $f_0(\vec{r})$ at time $t_n$, making the macroscopic fluxes only dependent on macroscopic variables and their gradients. In this way, the computational cost is equivalent to a Navier--Stokes solver. However, the Chapman--Enskog expansion's constraint means that GKS can only address viscous and thermal issues within the near-continuum flow regime. Details are elaborated in Section \ref{subsec:gks}.

The AUGKS takes advantages of the multiscale property of the UGKS and the high computational efficiency of the GKS. By introducing a non-equilibrium criterion to distinguish the continuum and rarefied flow regions, the AUGKS adopts continuous velocity space with GKS in the near-equilibrium region and discretized velocity space with UGKS in the non-equilibrium region. The consistent physical description of gas evolution process and the unified numerical framework of these two deterministic schemes used enable the AUGKS to simulate without the need for a buffer zone. This velocity space adaptation accelerates computation and reduces memory consumption. The construction of AUGKS is discussed in Section \ref{subsec:augks}. Additionally, for the source term $\vec{S}_i$ in Eq.~\ref{eq:update-macro} representing energy exchange between translational, rotational, and vibrational degrees of freedom, special treatment is necessary and will be discussed in detail in Section \ref{subsec:source}.

\subsection{Unified gas-kinetic scheme}\label{subsec:ugks}
To capture strong non-equilibrium physics of rarefied flows, UGKS employs a discrete velocity distribution function in flux computation, enabling the depiction of arbitrary form distributions. This makes UGKS a numerical scheme capable of simulating multiscale flows from continuum to free molecular flow regimes. To achieve second-order accuracy, the initial distribution function $f_0(\vec r)$ and equilibrium state $g({\vec r}, t)$ in the integral solution of the kinetic model Eq.~\eqref{eq:BGK-int} are expanded with discretized forms
\begin{equation}\label{eq:ugks-k}
	\begin{aligned}
		g_k(\vec{r}, t) &= g_{0,k} + \vec{r} \cdot \frac{\partial g_{0,k}}{\partial \vec{r}} + \frac{\partial g_{0,k}}{\partial t} t, \\
		f_{0,k}(\vec{r}) &= f_{k}^{l,r} + \vec{r} \cdot \frac{\partial f_{k}^{l,r}}{\partial \vec{r}},
	\end{aligned}	
\end{equation}
where $f_{k}^{l,r}$ is the distribution function of $k$-th discrete particle velocity $\vec{u}_k$ constructed by distribution functions $f_k^l$ and $f_k^r$ interpolated from cell centers to the left and right sides of the interface
\begin{equation*}
	f_{k}^{l,r}= f_{k}^l H\left[\bar{u}_{ij,k}\right]+f_k^r\left(1-H\left[\bar{u}_{ij,k}\right]\right),
\end{equation*}
where $\bar{u}_{ij,k}=\vec{u_k}\cdot\vec{n}_{ij}$ is the particle velocity projected on the normal direction of the cell interface $\vec{n}_{ij}$, and $H[x]$ is the Heaviside function. The equilibrium state $g_{0,k}$ is set to Maxwellian distribution function with macroscopic variables ${\vec W}_{ij}$ coming from the colliding particles from both sides of the cell interface
\begin{equation*}
 \vec{W}_{ij} = \sum g_{0,k} \vec{\psi}_k \mathcal{V}_k = \sum f_{k}^{l,r} \vec{\psi}_k \mathcal{V}_k,
\end{equation*}
where $\vec{\psi}_k = \left( 1, \vec{u}_k, \frac{1}{2} {\vec{u}}^2_k + \frac{1}{2} {\vec{\xi}}^2 + {\varepsilon_{v}}, \frac{1}{2} {\vec{\xi}}^2, {\varepsilon_v} \right)^{T}$, and $\mathcal{V}_k$ represents the volume of velocity space unit $k$ or the integral weight at the discrete velocity point $\vec{u}_k$.

Our previous study\cite{xu2021rot} has shown that the temporal and spatial gradients of the terms $g_t^+/g_t$, $(g_{tr}^+ - g_t^+)/Z_r$, and $(g_{M}^+ - g_{tr}^+)/Z_v$ in Eq.~\eqref{eq:g} are of the order of $\mathcal{O}(\tau^2)$ and $\mathcal{O}(t\tau)$, respectively. In the continuum regime where $\Delta t > \tau$, these terms can be neglected when recovering the Navier-Stokes limit. In other words, only translational equilibrium state $g_t$ from $g_0$ are considered in second-order terms, and the gradients of the equilibrium states are obtained from the gradient of macroscopic flow variables $\partial \vec{W}_{ij}/\partial{\vec r}$, respectively (see \ref{sec:appendix-1}). As to the temporal gradient, the compatibility condition on Eq.~\eqref{eq:BGK-type}
\begin{equation*}
	\int {(g - f) \vec{\psi} {\rm d}\Xi} = 0,
\end{equation*}
is employed to give
\begin{equation*}
	\frac{\partial\vec{W}_{ij}}{\partial t}
	= - \int \vec{u} \cdot \frac{\partial g_{t}}{\partial \vec{r}} \vec{\psi} {\rm d}\vec{\Xi}.
\end{equation*}
Correspondingly, the temporal gradient of equilibrium state $\partial_t g_t$ can be evaluated from the above $\partial \vec{W}_{ij}/\partial{t}$. The spatial gradients of distribution function $\partial_{\vec r} f^{l,r}$ are given from reconstruction. After determining all variables of Eq.~\ref{eq:ugks-k}, the time-dependent flux for microscopic distribution function across cell interface $ij$ can be constructed from the integral solution
\begin{equation} \label{eq:ugks-micro-flux}
	\begin{aligned}
	\mathcal{F}_{ij,k}
	&= \vec{u}_k \cdot \vec{n}_{ij}
	\left(C_1 g_{0,k}
	+ C_2 \vec{u}_k \cdot \frac{\partial g_{t,k}}{\partial \vec{r}}
	+ C_3 \frac{\partial g_{t,k}}{\partial t} \right) +
	\vec{u}_k \cdot \vec{n}_{ij} \left(C_4 f^{l,r}_{k}
	+ C_5 \vec{u}_k \cdot \frac{\partial f^{l,r}_k}{\partial \vec{r}} \right) \\
	&= {\mathcal{F}}_{ij,k}^{eq} + {\mathcal{F}}_{ij,k}^{fr},
	\end{aligned}
\end{equation}
where $\mathcal{F}^{fr}_{ij,k}$ and $\mathcal{F}^{eq}_{ij,k}$ are the microscopic fluxes in the free transport and collision processes, respectively, and $C_1$ to $C_5$ are integrated time coefficients
\begin{equation*}
	\begin{aligned}
	C_1 &= 1 - \frac{\tau}{\Delta t} \left( 1 - e^{-\Delta t / \tau} \right) , \\
	C_2 &= -\tau + \frac{2\tau^2}{\Delta t} - e^{-\Delta t / \tau} \left( \frac{2\tau^2}{\Delta t} + \tau\right) ,\\
	C_3 &=  \frac12 \Delta t - \tau + \frac{\tau^2}{\Delta t} \left( 1 - e^{-\Delta t / \tau} \right) , \\
	C_4 &= \frac{\tau}{\Delta t} \left(1 - e^{-\Delta t / \tau}\right), \\
	C_5 &= \tau  e^{-\Delta t / \tau} - \frac{\tau^2}{\Delta t}(1 -  e^{-\Delta t / \tau}).
	\end{aligned}
	\end{equation*}
For the macroscopic flux, the particle collision flux denoted by $g_{0,k}$ can be obtained by integrating the corresponding Maxwellian distribution through $\vec{W}_{ij}$ analytically, while the moment of free transport flux $\vec{F}_{ij}^{fr}$ is obtained discretely
\begin{equation} \label{eq:ugks-macro-flux}
	\vec{F}_{ij}
	= \int \mathcal{F}_{ij}^{eq} {\vec{\psi}} {\rm d}\vec{\Xi} +
	   \sum \mathcal{F}_{ij,k}^{fr} {\vec{\psi}}_k \mathcal{V}_k.
\end{equation}
To record non-equilibrium physics in simulation, the UGKS requires information from the discrete distribution function when calculating the microscopic distribution function flux. Therefore, the discrete distribution function of the cell needs to be recorded and updated. Within the finite volume framework, the update of the distribution function is obtained from the integral $ \int_0^{\Delta t} \int_{\Omega_i}(\cdot) \mathrm{d} \Omega \mathrm{d} t / \Omega_i$ over the space-time control volume of the kinetic model Eq.~\eqref{eq:BGK-type}, and it depends on the microscopic distribution function flux and particle collisions
\begin{equation*}\label{eq:fvm}
	{f}_i^{n + 1}
	=
	{f}_i^n
	- \frac{\Delta t}{\Omega_i}
	\sum\limits_{j \in N(i)} {{\mathcal{F}}_{ij}{\cal A}_{ij}}
	+ \int_0^{\Delta t} \frac{g-f}{\tau} {\rm d}{t},
\end{equation*}
where the trapezoidal rule is used for the time integration of the collision term, yielding a semi-implicit equation for the update of the distribution function
\begin{equation}\label{eq:update-micro}
	f_{i,k}^{n+1} = \left( 1 + \frac{\Delta t}{2\tau_i^{n+1}} \right)^{-1}
	\left[
		f_{i,k}^n
		- \frac{\Delta t}{\Omega_i}\sum_{j \in N(i)} \mathcal{F}_{ij,k} \mathcal{A}_{ij}
		+ \frac{\Delta t}{2}
		\left(
			\frac{g_{i,k}^{n+1}}{\tau_i^{n+1}}+\frac{g_{i,k}^n-f_{i,k}^n}{\tau_i^n}
		\right)
	\right],
\end{equation}
where the equilibrium state on $t^{n+1}$ step should be set after updateing macroscopic variables $W_i^{n+1}$.

In summary, to describe the strong non-equilibrium flow, the UGKS adopts discrete velocity space to depict distribution functions in any form, which enables the multiscale simulations in all flow regimes. However, for the high-speed rarefied flow, the UGKS requires a sufficiently large range and high resolution of the discrete velocity space to avoid ray effects in the simulation. The huge amount of velocity space discretization results in high computational resource requirements, large computational costs, and significant memory consumption. Therefore, there is still room for improvement in the computational efficiency of UGKS.

\subsection{Gas-kinetic scheme}\label{subsec:gks}
For nearly continuum flows with high particle collision frequencies and short relaxation times, the particle distribution function approaches local thermodynamic equilibrium, and the Navier--Stokes equations can be used to describe macroscopic fluid motion. As a Navier--Stokes solver, GKS directly applies the first-order Chapman--Enskog expansion to construct the initial distribution function $f_0(\vec r)$ to recover macroscopic transport coefficients, and Taylor expansions for the equilibrium state $g({\vec r}, t)$ to achieve second-order accuracy of the integral solution of the kinetic model Eq.~\eqref{eq:BGK-int}
\begin{equation}\label{eq:gks-expansion}
\begin{aligned}
	&g(\vec{r}, t) = g_0 + \vec{r} \cdot \frac{\partial g_0}{\partial \vec{r}} + \frac{\partial g_0}{\partial t} t, \\
	&f_0(\vec{r}) =
	g^{l,r}-\tau\left(\vec{u} \cdot \frac{\partial g^{l,r}}{\partial \vec{r}}+\frac{\partial g^{l,r}}{\partial t}\right),
\end{aligned}	
\end{equation}
with $g^{l, r}$ constructed by the equilibrium distribution functions $g^l$ and $g^r$ which are interpolated from the left and right cell center to the interface
\begin{equation*}
	g^{l, r}=g^l H\left[\bar{u}_{i j}\right]+g^r\left(1-H\left[\bar{u}_{i j}\right]\right).
\end{equation*}
The equilibrium state $g_0$ is also obtained through collisions on both sides of the interface
\begin{equation*}
\vec{W}_{ij} = \int g_0 \vec{\psi} {\rm d} \vec{\Xi}
		     = \int g^{l,r} \vec{\psi} {\rm d} \vec{\Xi}.
\end{equation*}
Similar to the construction of the gradients of the equilibrium states in UGKS, $\partial_{\vec r} g^{l,r}$ are obtained from the gradient of macroscopic flow variables $\partial \vec{W}_{ij}^{l}/\partial{\vec r}$ and $\partial \vec{W}_{ij}^{r}/\partial{\vec r}$. The temporal gradients $\partial_t g^{l,r}$ are also evaluated from $\partial \vec{W}_{ij}^{l}/\partial{\vec t}$ and $\partial \vec{W}_{ij}^{r}/\partial{\vec t}$ by employing the compatibility condition
\begin{equation*}
	\begin{aligned}
	\frac{\partial\vec{W}^{l}_{ij}}{\partial t}
	= - \int \vec{u} \cdot \frac{\partial g^l_t}{\partial \vec{r}} \vec{\psi} {\rm d}\vec{\Xi}, \\
	\frac{\partial\vec{W}^{r}_{ij}}{\partial t}
	= - \int \vec{u} \cdot \frac{\partial g^r_t}{\partial \vec{r}} \vec{\psi} {\rm d}\vec{\Xi}.
	\end{aligned}
\end{equation*}
Thus far, the microscopic flux $\mathcal{F}_{ij}$ for distribution function on the cell interface for GKS flux can be fully determined
\begin{equation} \label{eq:gks-micro-flux}
	\begin{aligned}
	\mathcal{F}_{ij}
	&= \vec{u} \cdot \vec{n}_{ij}
	\left(C_1 g_0
	+ C_2 \vec{u} \cdot \frac{\partial g_t}{\partial \vec{r}}
	+ C_3 \frac{\partial g_t}{\partial t} \right)
	+ \vec{u} \cdot \vec{n}_{ij}
	\left(C_4 g^{l,r}
	+ C_5^\prime \vec{u} \cdot \frac{\partial g_t^{l,r}}{\partial \vec{r}}
	+ C_6 \frac{\partial g_t^{l,r}}{\partial t} \right) \\
	&= {\mathcal{F}}_{ij}^{eq} + {\mathcal{F}}_{ij}^{fr},
	\end{aligned}
\end{equation}
where $\mathcal{F}^{fr}_{ij}$ and $\mathcal{F}^{eq}_{ij}$ are the microscopic fluxes in the free transport and collision processes, respectively. Additional integrated time coefficients are
\begin{equation*}
\begin{aligned}
C_5^\prime &= \tau  e^{-\Delta t / \tau} - \frac{2 \tau^2}{\Delta t}(1 -  e^{-\Delta t / \tau}) , \\
C_6 &= - \frac{\tau^2}{\Delta t}\left( 1- e^{-\Delta t / \tau} \right).
\end{aligned}
\end{equation*}
The macroscopic fluxes can be obtained from the moments of the flux of distribution function
\begin{equation} \label{eq:gks-macro-flux}
	\vec{F}_{ij} = \int \mathcal{F}_{ij}^{eq} {\vec{\psi}} {\rm d}\vec{\Xi} +
	   \int \mathcal{F}_{ij}^{fr} {\vec{\psi}} {\rm d}\vec{\Xi}.
\end{equation}
It can be seen from Eq.~\eqref{eq:gks-micro-flux}, the GKS flux adaptively couples the upwind and central characteristics based on local physics. Additionally, the initial distribution function and equilibrium state described by GKS in Eq.~\eqref{eq:gks-expansion} follow specific distributions. Therefore, the flux for gas evolution only depends on macroscopic physical quantities and their derivatives, and the distribution function can be described directly in continuous velocity space. The moments of the distribution function in GKS are obtained directly from the Maxwellian distribution and its Taylor expansion based on macroscopic quantities, resulting in computational costs comparable to those of traditional Navier--Stokes solvers. However, since the form of the distribution function is specified, the GKS can only solve the viscosity and heat transfer problems for nearly continuous flows.

\subsection{Adaptive unified gas-kinetic scheme}\label{subsec:augks}
To reduce the computational costs in the simulations of multiscale flows, the hybrid of particle and traditional CFD method based on physical space partitioning is proposed. The particle/CFD hybrid methods adopt the particle method in rarefied flow regions and the CFD method in continuous flow. However, the distinction between those stochastic and deterministic methods necessitates the overlap buffer zones to distinguish and connect different solvers. Contrary to conventional hybrid methods, the AUGKS eliminates the need for additional buffer zones. This is primarily attributed to its unified framework in describing gas evolution process by the use of gas distribution function with deterministic approaches. The AUGKS applies discretized velocity space, i.e., UGKS, to accurately capture the non-equilibrium physics, which empowers the scheme with the capability to simulate multiscale flows. In the near-equilibrium flow regime, the validity of the Chapman--Enskog expansion lends credibility to the use of the GKS with a continuous distribution function, resulting in computational load and memory reductions.

Here, the gradient-length local Knudsen number is employed as the criterion for velocity space adaptation
\begin{equation}\label{eq:KnGll}
	\mathrm{Kn}_{Gll}=\frac{l}{\rho /|\nabla \rho|},
\end{equation}
where $l$ is the local mean free path of gas molecules. When ${\rm Kn}_{Gll}$ is less than a switching criterion $C_t$, the flow region is regarded as near-equilibrium flow and the GKS with continuous distribution function is adopted. Otherwise, the UGKS with discretized distribution function is used.

The primary issue lies in calculating the flux at the interfaces of adjacent cells in UGKS-GKS partitions. On the adjacent interface, the update of the GKS side merely demands the macroscopic flux $\vec{F}_{ij}$, while the UGKS side necessitates the update of both the macroscopic variables $\vec{W}_i$ and the microscopic distribution function $f_{i,k}$, thereby requiring the microscopic flux $\mathcal{F}_{ij,k}$ with a discrete form. To accommodate the evolution on either side of the interface, the partition of GKS side must supply the corresponding discrete distribution function for the evolution of the UGKS.
\begin{figure}[H]
	\centering
	\subfloat[]{\includegraphics[width=0.4\textwidth]{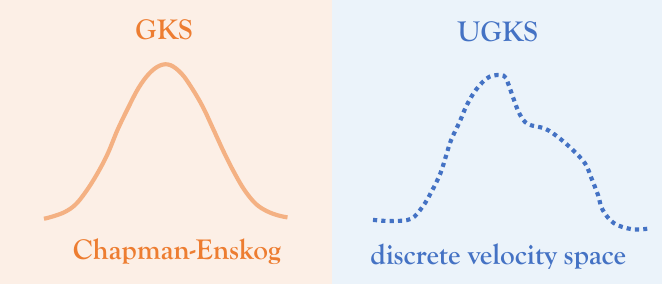}}
	\hspace{5mm}
	\subfloat[]{\includegraphics[width=0.4\textwidth]{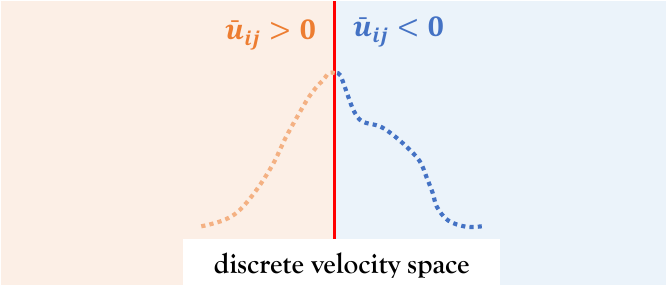}}
	\caption{The the distribution functions on (a) adjacent cells and (b) adjacent interface. }
	\label{fig:augks}
\end{figure}
As shown in Fig.~\ref{fig:augks}, presuming the cell using GKS is on the left side and the UGKS is on the right side of adjacent interface, the integral solution of kinetic model Eq.~\eqref{eq:BGK-int} under the framework of AUGKS is expressed as
\begin{equation*}\label{eq:bgk-augks}
	\begin{aligned}
	&g_k(\vec{r}, t) = g_{0,k} + \vec{r} \cdot \frac{\partial g_{0,k}}{\partial \vec{r}} + \frac{\partial g_{0,k}}{\partial t} t, \\
	&f_{0,k}(\vec{r})= g^G_k H\left[\bar{u}_{ij,k}\right]
				     + f^U_k \left(1-H\left[\bar{u}_{i j}\right]\right).
	\end{aligned}	
\end{equation*}
where the initial distribution function on the GKS side $g^G_k$ follows Chapman--Enskog expansion determined by the macroscopic physical quantities and their gradients. To assist the evaluation of microscopic flux on the UGKS side, the discretized form should be provided
\begin{equation}\label{eq:CE-k}
g^G_k = g^l_k-\tau\left(\vec{u}_k \cdot \frac{\partial g^l_k}{\partial \vec{r}}+\frac{\partial g^l_k}{\partial t}\right).
\end{equation}
The initial distribution function on the UGKS side $f^U_k$ is obtained from the distribution function $f_k^r$ interpolated to the interface
\begin{equation*}
f^U_k= f_k^r + \vec{r} \cdot \frac{\partial f_k^r}{\partial \vec{r}}.
\end{equation*}
The equilibrium state $g_{0,k}$ caused by the colliding particles is constructed from the Chapman--Enskog expansion on the GKS side and the distribution function on the UGKS side
\begin{equation*}
	\vec{W}_{ij}
	= \sum g_{0,k} \vec{\psi}_k \mathcal{V}_k
	= \sum \left[
		g^G_k H\left[\bar{u}_{ij,k}\right]
	+   f^U_k \left( 1 -H\left[\bar{u}_{ij,k}\right]  \right)
	\right]
	\vec{\psi}_k \mathcal{V}_k.
\end{equation*}
Therefore, the microscopic distribution function flux at the interface is
\begin{equation} \label{eq:augks-micro-flux}
	\begin{aligned}
	\mathcal{F}_{ij,k} &= \vec{u}_k \cdot \vec{n}_{ij}
	\left(C_1 g_{0,k}
	+ C_2 \vec{u}_k \cdot \frac{\partial g_{t,k}}{\partial \vec{r}}
	+ C_3 \frac{\partial g_{t,k}}{\partial t} \right) \\
	&+
	\vec{u}_k \cdot \vec{n}_{ij}
	\left[
		\left(
	  C_4 g_k^l
	+ C_5^\prime \vec{u}_k \cdot \frac{\partial g^l_k}{\partial \vec{r}}
	+ C_6 \frac{\partial g^l_k}{\partial t}
	\right) H\left[\bar{u}_{ij,k}\right]
	+
	\left(
		C_4 f_k^r
	  + C_5\vec{u}_k \cdot \frac{\partial f^r_k}{\partial \vec{r}}
	  \right) \left( 1-H\left[\bar{u}_{ij,k}\right] \right)
	\right]	
	 \\
	&= {\mathcal{F}}_{ij,k}^{eq} + {\mathcal{F}}_{ij,k}^{fr}.
	\end{aligned}
\end{equation}
With the calculation of the macroscopic flux using Eq.~\eqref{eq:ugks-macro-flux} as the same as the UGKS, the evaluation of the flux at the adjacent interface for the AUGKS is complete.

\subsection{Treatment of source term}\label{subsec:source}
During the collision process, inelastic collisions will happen, leading to the energy exchange between the degrees of freedom of molecular translation, rotation and vibration. As a result, source terms appear in the macroscopic governing equations, i.e.,
\begin{equation*} \label{eq:source}
\vec{S} = \int_{t^n}^{t^{n+1}} \int \frac{g-f}{\tau} \vec{\psi}
{\rm{d}} \vec{\Xi} {\rm{d}} t
= \int_{t^{n}}^{t^{n+1}} \vec{s} {\rm{d}} t,
\end{equation*}
where $\vec{s}$ can be expressed as
\begin{equation*}
{\vec{s}} = \left(0,\vec{0},0,
		\frac{\rho E_{r}^{tr} - \rho E_r}{Z_r \tau} + \frac{\rho E_r^M - \rho E_{r}^{tr}}{Z_v \tau},
		\frac{\rho E_v^M - \rho E_v}{Z_v \tau} \right)^T.
\end{equation*}
The intermediate equilibrium energy $\rho E_r^{tr}$ is determined under the assumption $\lambda_r = \lambda_t = \lambda_{tr}$, and thus
\begin{equation}\label{eq:rhoErTR-rhoEvTR}
	\rho E_r^{tr} = \frac{K_r\rho}{4 \lambda_{tr}},\quad
	\quad \lambda_{tr} = 	
	\frac{( K_r + 3 )\rho }{ 4 ( \rho E - \frac{1}{2} \rho {\vec{U}}^2 - \rho E_v )}.
\end{equation}
The rotational and vibrational energy at the full equilibrium state $\rho E_r^M$ and $\rho E_v^M$ are determined under the assumption $\lambda_v = \lambda_r = \lambda_t = \lambda_{M}$, and thus
\begin{equation}\label{eq:rhoErM-rhoEvM}
	\rho E_r^M = \frac{K_r \rho}{4 \lambda_M},\quad
	\rho E_v^M = \frac{K_v(\lambda_M) \rho}{4 \lambda_M},\quad
	{\text{ and }}
	\quad {\lambda_{M}} = 	
	\frac{\left[K_v(\lambda_M) + K_r + 3\right]\rho}{4\left(\rho E - \frac{1}{2} \rho {\vec{U}}^2\right)}.
\end{equation}

With consideration of numerical stability, the source term is usually treated in an implicit way, such as the trapezoidal rule for rotational and vibrational energies
\begin{equation*}
\begin{aligned}
S_r &= \frac{\Delta t}{2} \left(s_r^n + s_r^{n+1}\right) \\
 	&= \frac{\Delta t}{2} \left[\frac{(\rho E_r^{tr})^n - (\rho {E_r})^n}{Z_r \tau} + \frac{(\rho E_r^M)^n - (\rho E_r^{tr})^n}{Z_v \tau}\right] \\
    &+\frac{\Delta t}{2} \left[\frac{(\rho E_r^{tr})^{n+1} - (\rho {E_r})^{n+1}}{Z_r \tau} + \frac{(\rho E_r^M)^{n+1} - (\rho E_r^{tr})^{n+1}}{Z_v \tau}\right],\\
S_v &=\frac{\Delta t}{2} \left(s_v^n + s_v^{n+1}\right) = \frac{\Delta t}{2} \left[\frac{(\rho E_v^M)^n - (\rho E_v)^n}{Z_v \tau} + \frac{(\rho E_v^M)^{n+1} - (\rho E_v)^{n+1}}{Z_v \tau}\right].
\end{aligned}	
\end{equation*}
During updating macroscopic variables, based on the fluxes, the conservative flow variables $\rho^{n+1}$, $(\rho \vec{U})^{n+1}$, $(\rho E)^{n+1}$ can be updated directly. Then $\lambda_M^{n+1}$, $(\rho E_r^M)^{n+1}$ and $(\rho E_v^M)^{n+1}$ can be obtained from the updated conservative flow variables by Eq.~\eqref{eq:rhoErM-rhoEvM}, and the vibrational energy $(\rho E_v)^{n+1}$ with implicit source term can be solved in an explicit way without iterations
\begin{equation} \label{eq:update-Ev}
(\rho E_v)^{n+1} = \left(1 + \frac{\Delta t}{2 Z_v \tau}\right)^{-1} \left[(\rho {E_v})^\dagger +
\frac{\Delta t}{2} \left(s_v^n + \frac{(\rho E_v^M)^{n+1}}{Z_v \tau} \right) \right].
\end{equation}
Similarly, $\lambda_{tr}^{n+1}$ and $(\rho E_r^{tr})^{n+1}$ can be obtained by Eq.~\eqref{eq:rhoErTR-rhoEvTR} with the updated $(\rho E_v)^{n+1}$, then the rotational energy $(\rho E_r)^{n+1}$ can be renewed by
\begin{equation} \label{eq:update-Er}
(\rho {E_r})^{n+1} = \left(1 + \frac{\Delta t}{2 Z_r \tau}\right)^{-1}
\left[ (\rho {E_r})^\dagger + \frac{\Delta t}{2} \left(s_r^n  + \frac{(\rho E_r^{tr})^{n+1}}{Z_r \tau} + \frac{(\rho E_r^M)^{n+1} - (\rho E_r^{tr})^{n+1}}{Z_v \tau}\right) \right].
\end{equation}
Here $(\rho E_v)^\dagger$ and $(\rho E_r)^\dagger$ are the updated intermediate vibrational and rotational energies with inclusion of the fluxes only. It would be noticed that in Eq.~\eqref{eq:rhoErM-rhoEvM} the vibrational degrees of freedom $K_v(\lambda_M)$ rely on the full equilibrium temperature $\lambda_M$. The explicit expression of $\lambda_M$ cannot be given due to the complexity of function $K_v(\lambda_M)$. In the current study, $\lambda_M$ is computed by iterations
\begin{equation*}
\lambda_M^{i + 1} = \frac{\left[ K_v(\lambda_M^i) + K_r + 3\right] \rho}{4 \left(\rho E - \frac{1}{2} \rho \vec{U}^2 \right)} \quad \text{with} \quad
\lambda_M^0 = \frac{\left(K_r + 3\right)\rho }{4\left(\rho E - \frac{1}{2} \rho \vec{U}^2\right)}.
\end{equation*}
Numerical tests show that the relative error can approach $\mathcal{O}(10^{-16})$ after $5 \sim 6$ iterations.

\subsection{Summary of algorithm}
The algorithm of AUGKS considering rotation and vibration is specifically summarized as follows:

\begin{description}
    \item[Step 1] Calculate gradients. Use the least squares method to get the gradients of macroscopic variables.
    \item[Step 2] Evaluate velocity adaptation. Apply Eq.~\eqref{eq:KnGll} to derive the partition criteria to decompose the physical domain into continuous (GKS) and discretized (UGKS) distribution functions computational regions.
    \item[Step 3] Allocate memory. Store macroscopic variables and their gradients in the GKS region, and macroscopic variables, distribution functions, and their gradients in the UGKS region. Record the distribution function using Eq.~\eqref{eq:CE-k} for UGKS cells at the initial time $t = 0$ or for cells transitioning from GKS to UGKS between $t^{n-1}$ and $t^n$. On the other hand, for cells transitioning from UGKS to GKS between $t^{n-1}$ and $t^n$, clear the distribution function memory.
    \item[Step 4] Limit gradients. Use the limiter to compute the gradients of the macroscopic quantities and distribution functions.
    \item[Step 5] Evaluate fluxes. Calculate the macroscopic flux Eq.~\eqref{eq:gks-macro-flux} for GKS interfaces, and the microscopic flux Eq.~\eqref{eq:ugks-micro-flux} and macroscopic flux Eq.~\eqref{eq:ugks-macro-flux} for UGKS. On the UGKS-GKS adjacent interface, convert the GKS side to a discrete distribution function form Eq.~\eqref{eq:CE-k}, and then evaluate the microscopic flux Eq.~\eqref{eq:augks-micro-flux} and macroscopic flux Eq.~\eqref{eq:ugks-macro-flux}.
    \item[Step 7] Update flow fields. Apply Eq.~\eqref{eq:update-macro}, Eq.~\eqref{eq:update-Ev} and Eq.~\eqref{eq:update-Er} to obtain the macroscopic quantities of the full field at the $n + 1$ step. Then calculate Eq.~\eqref{eq:update-micro} to update the microscopic distribution function for the UGKS cells.
\end{description}

\section{Numerical Validation}\label{sec:case}
In this section, the AUGKS method with molecular vibration (AUGKS-vib) will be used in the following test cases. Multiscale simulations across all flow regimes will be validated using Sod tube tests at Knudsen numbers ranging from $10^{-5}$ to $10$, and the scheme's capacity to capture significant non-equilibrium effects in the vibrational mode will be demonstrated by simulating shock structures. The AUGKS-vib method's computational accuracy and velocity space adaptation in hypersonic flow at different Mach number will be quantitatively evidenced, especially in terms of aerodynamic heating, through flow simulations around a cylinder. The scheme's accuracy will be further affirmed by examining flow across a sphere. The capability to handle flow simulations involving complex geometries under pyramid and tetrahedral meshes with accuracy and velocity space adaptation will be showcased through a study of flow around a space vehicle. The advantage of AUGKS-vib, i.e., no buffer zone in the scheme, will be reflected in the simulation of an unsteady nozzle plume expanding to an extreme vacuum background within a single computation.

In the test cases, the dynamic viscosity is calculated from the translational temperature by the power law
\begin{equation*}\label{eq:pow-law}
	\mu  = \mu_{ref} \left( \frac{T}{T_{ref}} \right)^{{\omega}},
\end{equation*}
where $\mu_{ref}$ is the reference dynamic viscosity at the temperature $T_{ref}$, and $\omega$ is the viscosity coefficient index. Since most of the cases are external flow, the determination of the initial condition at different free-stream Knudsen number ${\rm Kn}_\infty$ will be provided here first, with the definition
\begin{equation*}
	{\rm Kn}_\infty = \frac{l_\infty}{L_{ref}},
\end{equation*}
where $l_\infty$ is the mean free path of the free-stream flow, and the $L_{ref}$ is the geometric characteristic length. For a specific gas, the density in the free stream corresponding to a given Knudsen number is
\begin{equation*}
	\rho_\infty = \frac{4\alpha(5-2\omega)(7-2\omega)}{5(\alpha+1)(\alpha+2)}\sqrt{\frac{m}{2\pi k_B T}}\frac{\mu_\infty}{L_{ref}{\rm Kn}_\infty},
\end{equation*}
where $m$ is the molecular mass.

Unless otherwise specified, a diatomic nitrogen molecule with molecular mass $m=4.65\times 10^{-26}$ kg, $\alpha=1.0$, $\omega=0.74$, reference dynamic viscosity $\mu_{ref} = 1.65\times 10^{-5}$ ${\rm Nsm^{-2}}$, and reference temperature $T_{ref} = 273$ K is employed in these test cases. For heat flux modification, the Prandtl number is taken as ${\rm Pr} = 2/3$, the self-diffusion coefficient $\sigma$ is 0.64516, the coefficients are $\omega_0 = 0.2354$, $\omega_1 = 0.3049$, $\omega_2=\omega_3=\omega_4=\omega_5=0.3$. For non-dimensional cases, the freestream or upstream values are used to non-dimensionalize the flow variables, i.e.,
\begin{equation*}
	\begin{aligned}
	\rho_0 &= \rho_\infty, \quad U_0 = \sqrt{2 k_B T_{\infty} / m}, \\
	T_0 &= T_\infty, \quad \text{or} \quad p_0 = p_\infty.
	\end{aligned}
\end{equation*}
In addition, according to the reference \cite{tumuklu2016particle}, the vibrational collision number can be evaluated by
\begin{equation}\label{eq:Zv-tumuklu}
	Z_v
	=
	\frac{5}{ 5 + K_v(\lambda_v) }
	\frac{C_1}{ T_t^\omega }
	\exp\left( C_2 T_t^{-1/3} \right), 	
\end{equation}
and the rotational collision number is computed by
\begin{equation} \label{eq:Zr-tumuklu}
	Z_r = \frac{Z_v}{Z_v + Z_r^{DSMC}},
\end{equation}
with
\begin{equation*}
	Z_r^{DSMC}
	= \frac{3}{5}
	\frac{ Z_r^\infty }
	     { 1 + ( \sqrt{\pi} / 2 ) \left( \sqrt{ T^\ast / T_t } \right) + ( T^\ast / T_t ) ( { \pi^2/4 + \pi } )},
\end{equation*}
where $C_1 = 6.5$ and $C_2 = 220.0$ are adopted. If not stated otherwise, the criterion for velocity space adaptation is adopted $C_t = 0.01$.

\subsection{Sod tube}
The Sod shock tube problem is computed at different Knudsen numbers to verify the capability of the AUGKS method for simulating the continuum and rarefied flows. The non-dimensional initial condition is
\begin{equation*}
	(\rho, U, V, W, p)= \begin{cases}(1,0,0,0,1), & 0<x<0.5, \\ (0.125,0,0,0.1), & 0.5<x<1.\end{cases}
\end{equation*}
The spatial discretization is carried out by a three-dimensional structured mesh with $100 \times 5 \times 5$ uniform cells. $101\times 7 \times 7$ velocity points are used to discretize the velocity space with the range of ($-5$, $5$). The inlet and outlet of the tube are treated as far field, and the side walls are set as symmetric planes. The Courant--Friedrichs--Lewy (CFL) number is taken as 0.95. Constant values of $Z_r=3.5$ and $Z_v = 10$ are used for all cases. The results at the time $t = 0.12$ are investigated.

\begin{figure}[H]
	\centering
	\subfloat[]{\includegraphics[width=0.33\textwidth]{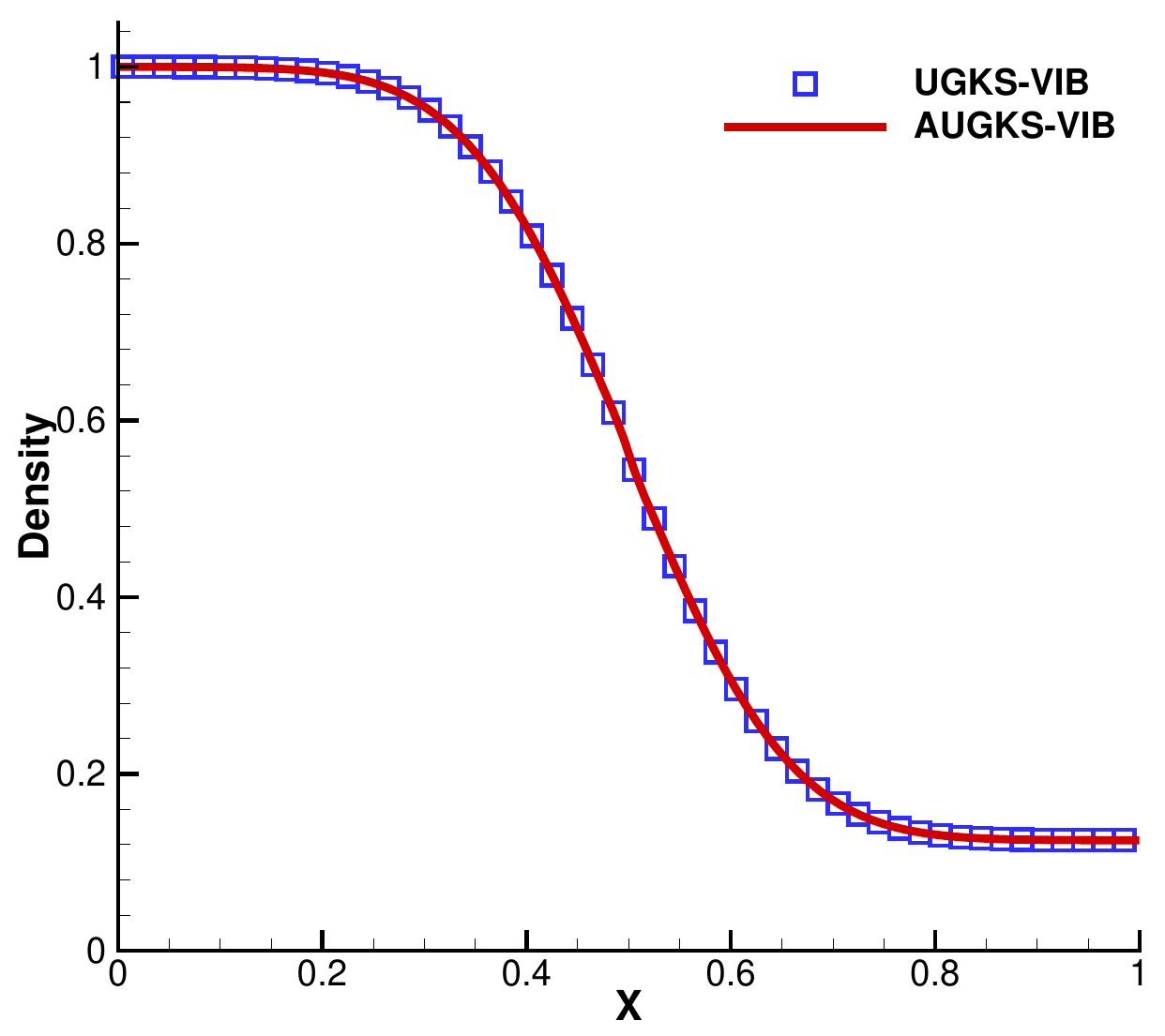}}
	\subfloat[]{\includegraphics[width=0.33\textwidth]{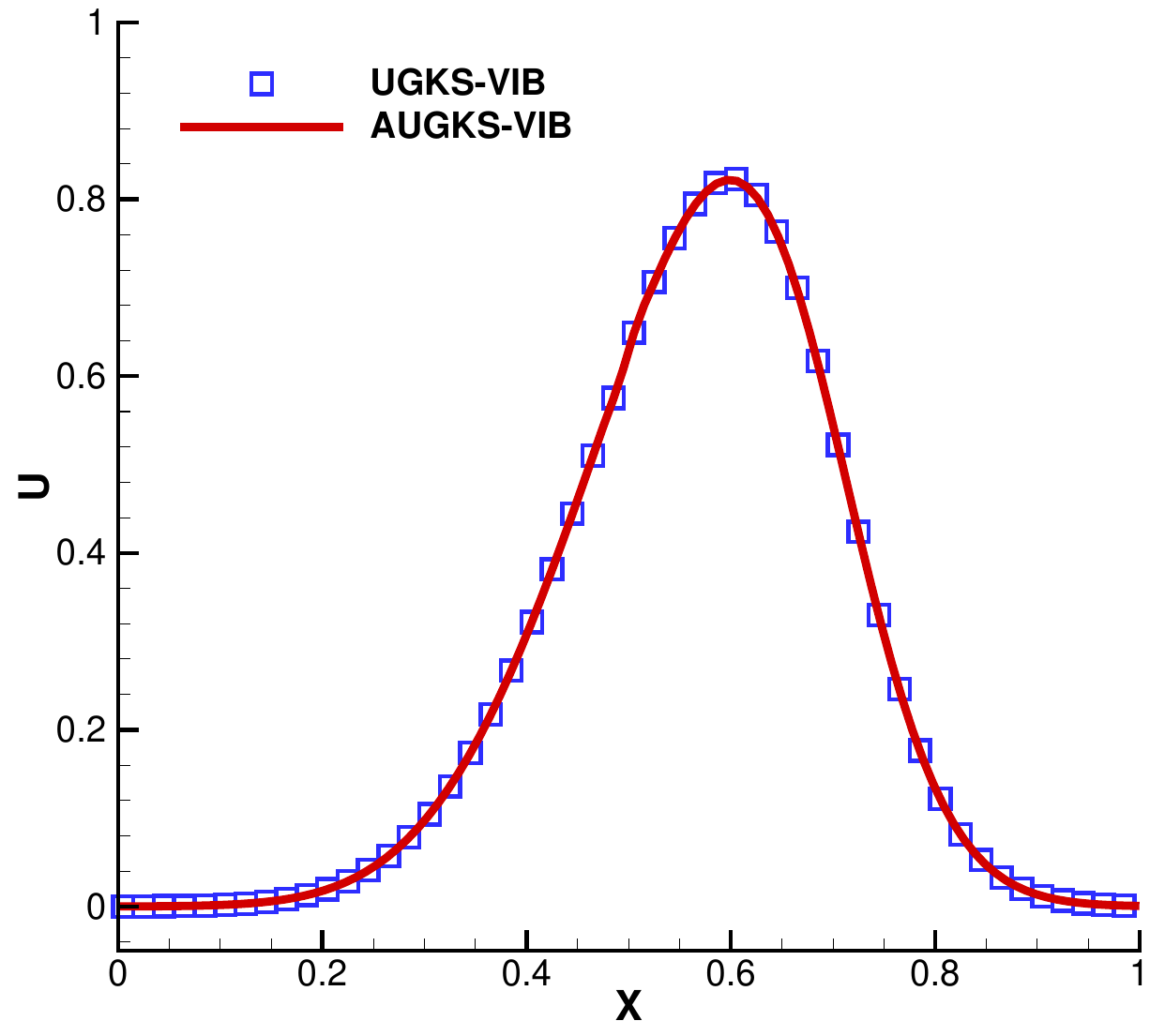}}
	\subfloat[]{\includegraphics[width=0.33\textwidth]{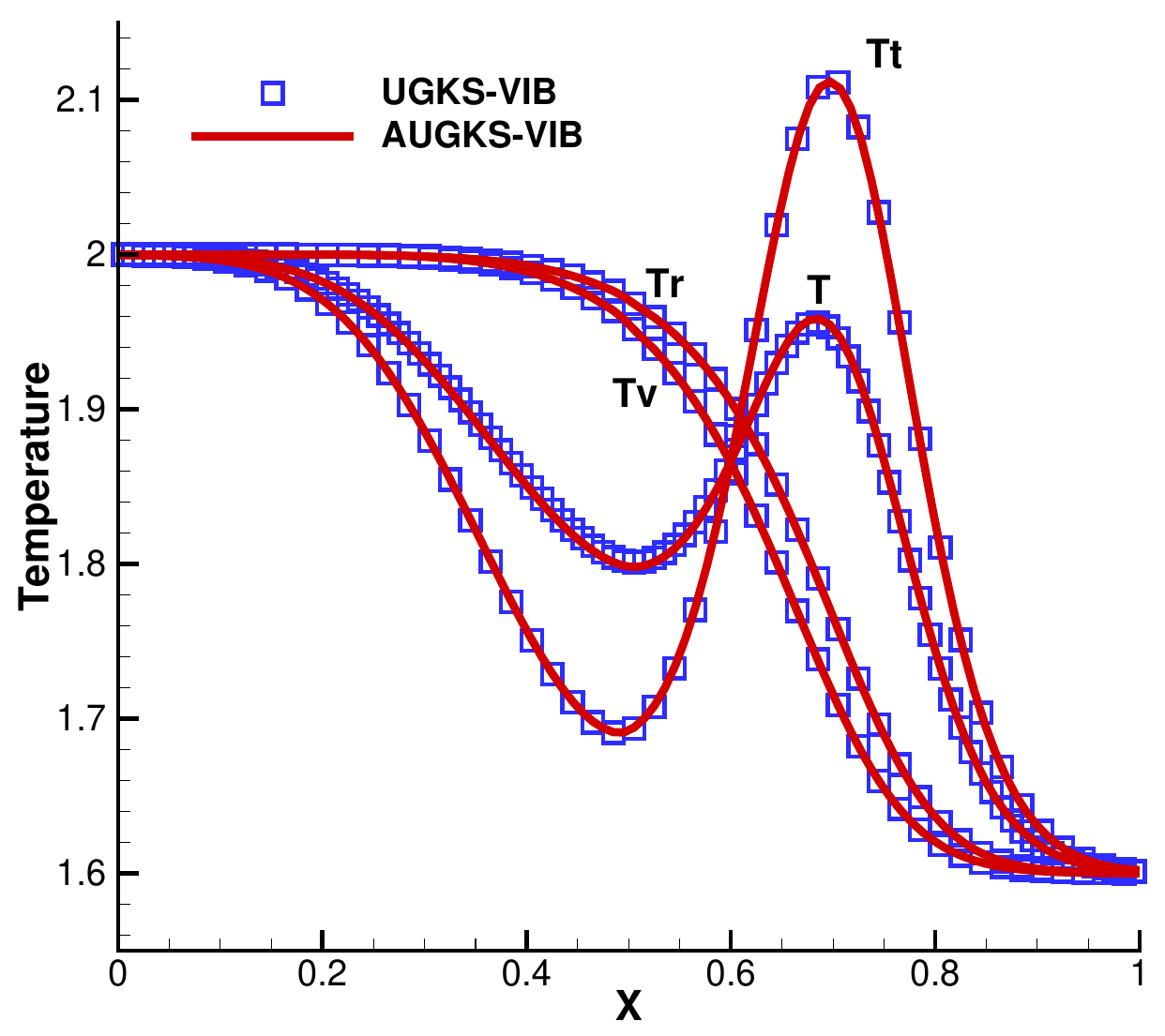}}
	\caption{Sod tube at ${\rm Kn}_\infty = 10$. (a) Density,
	(b) velocity, and (c) temperatures, compared with UGKS results using the same vibrational relaxation model.}
	\label{fig:sod-kn10}
\end{figure}

\begin{figure}[H]
	\centering
	\subfloat[]{\includegraphics[width=0.33\textwidth]{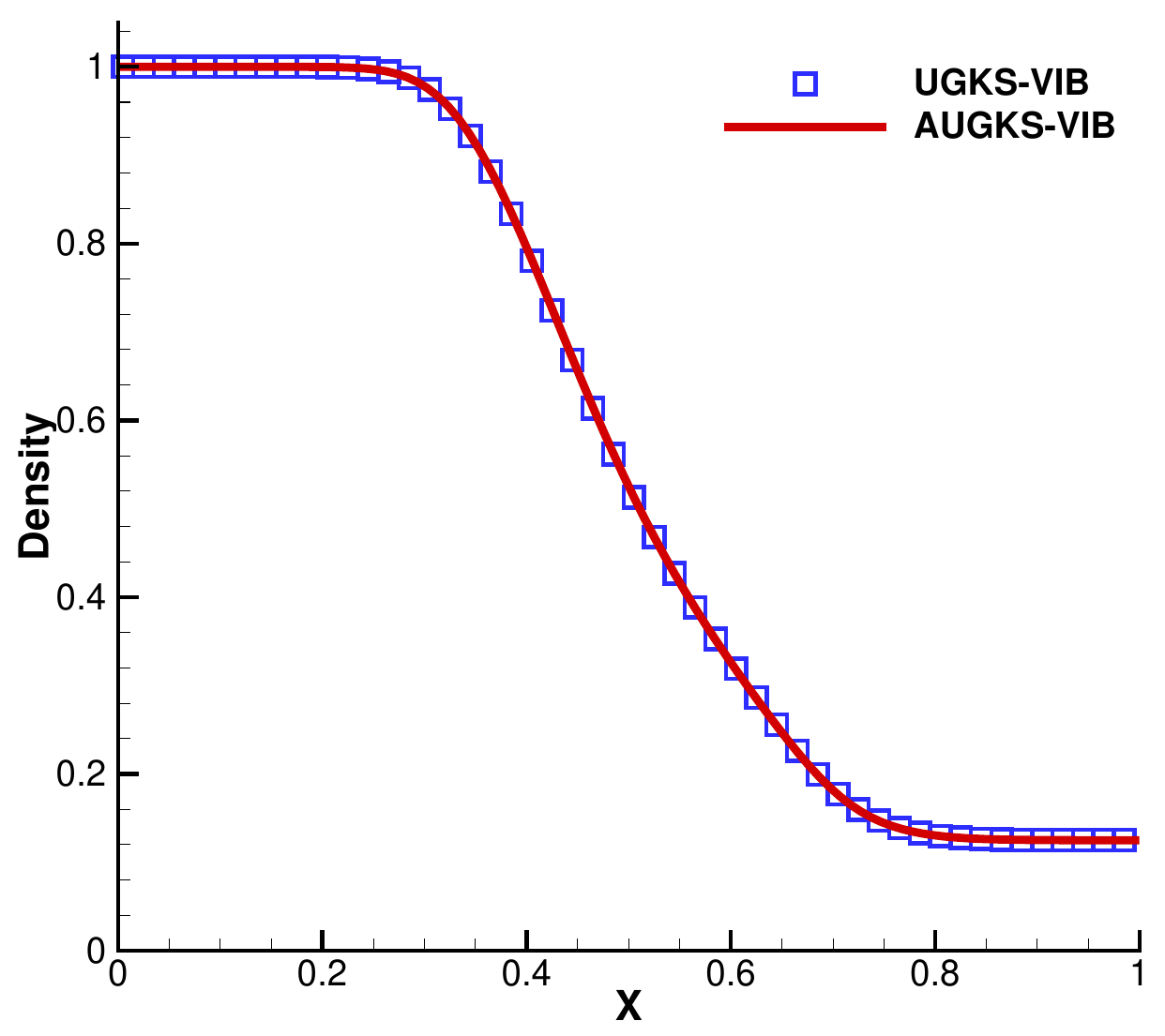}}
	\subfloat[]{\includegraphics[width=0.33\textwidth]{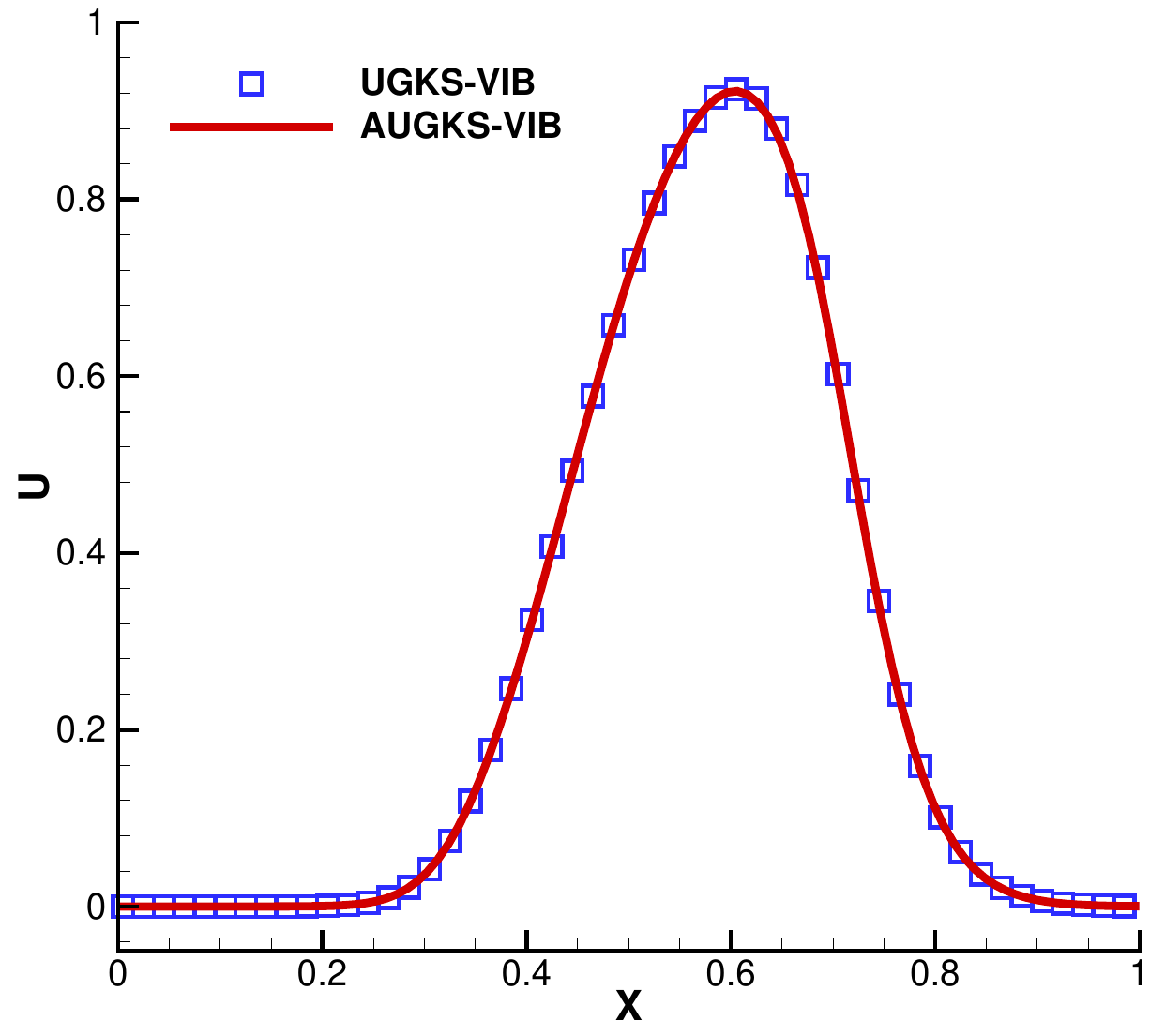}}
	\subfloat[]{\includegraphics[width=0.33\textwidth]{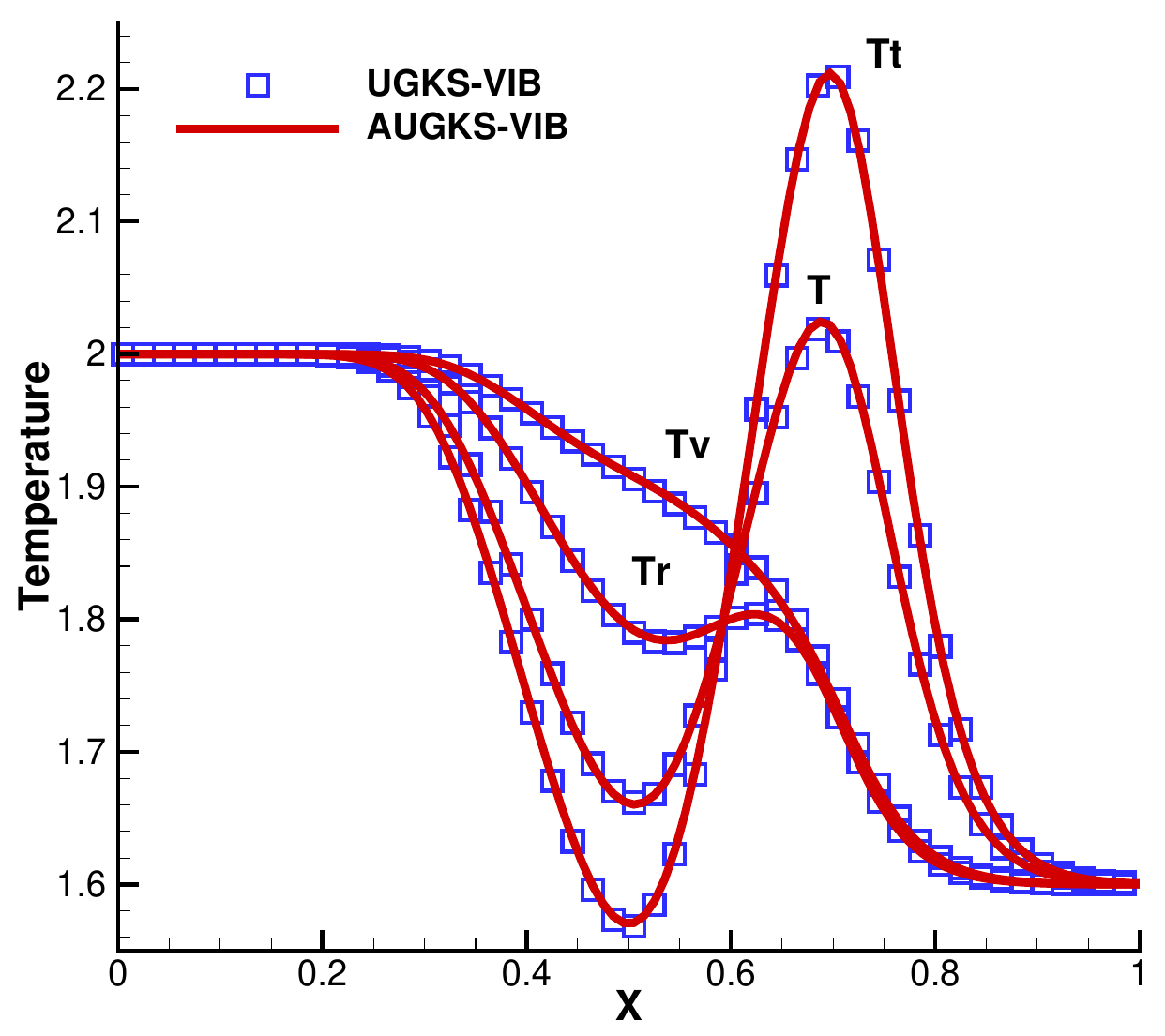}}
	\caption{Sod tube at ${\rm Kn}_\infty = 0.01$. (a) Density,
	(b) velocity, and (c) temperatures, compared with UGKS results using the same vibrational relaxation model.}
	\label{fig:sod-kn1e-2}
\end{figure}

\begin{figure}[H]
	\centering
	\subfloat[]{\includegraphics[width=0.33\textwidth]{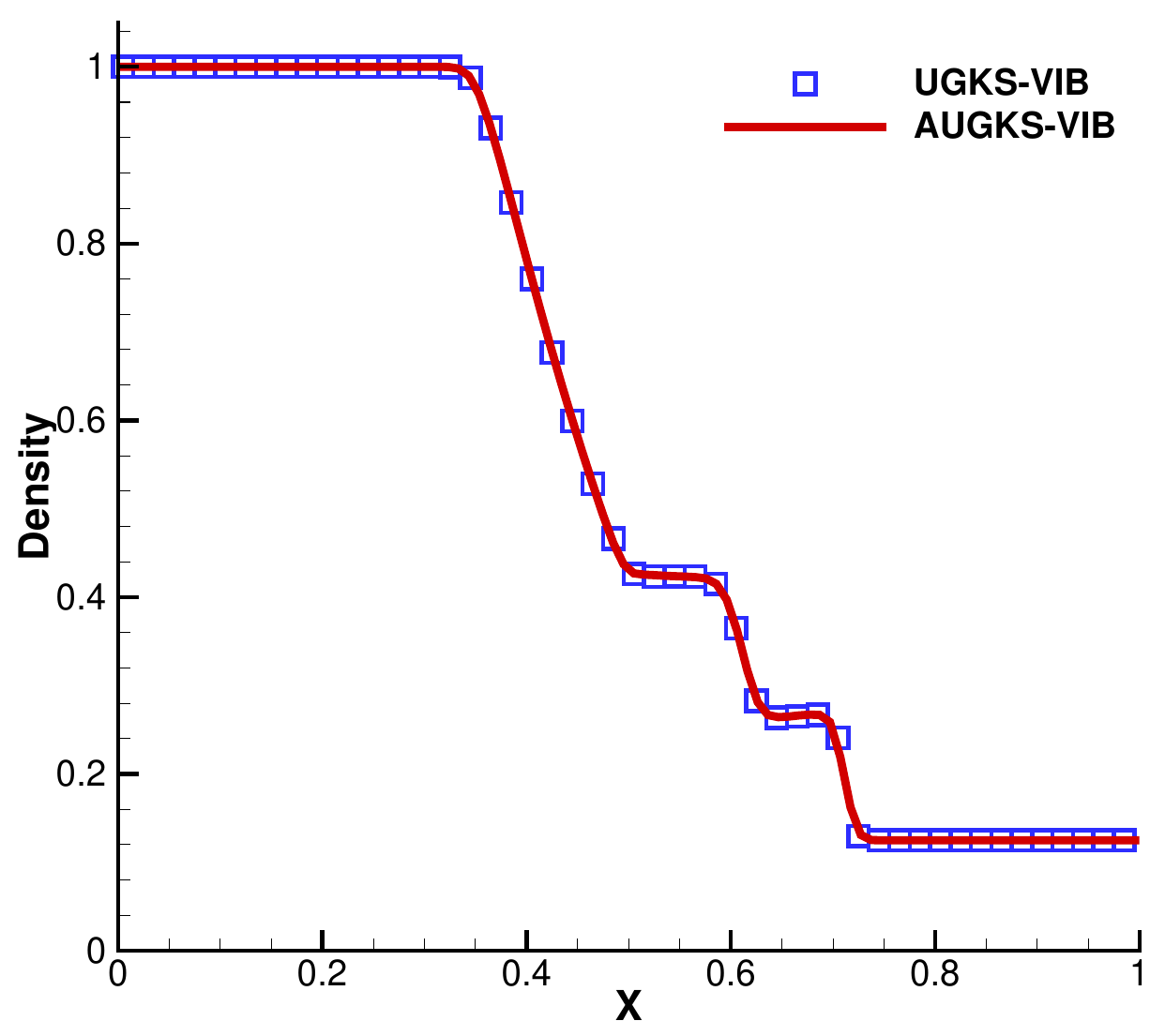}}
	\subfloat[]{\includegraphics[width=0.33\textwidth]{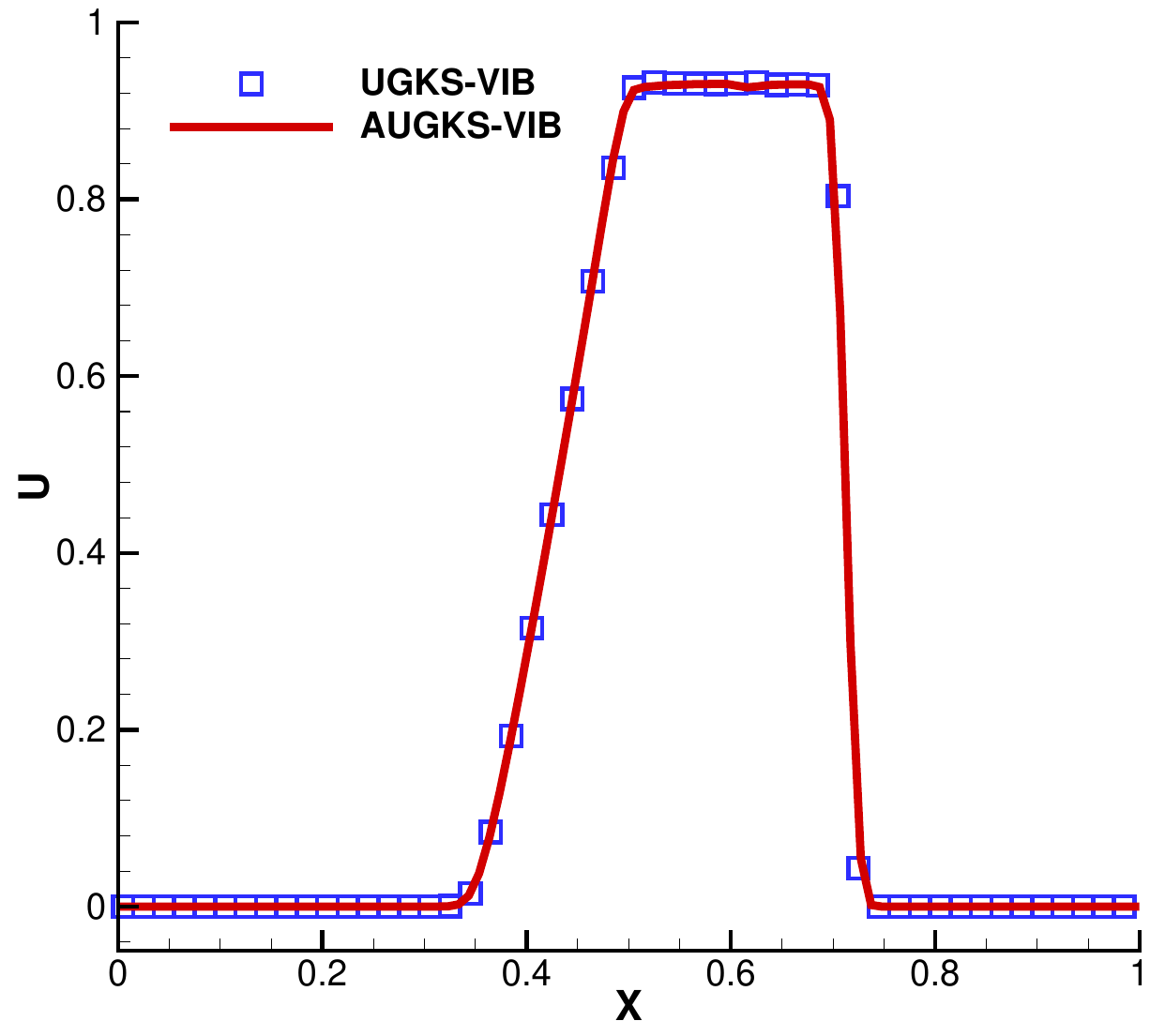}}
	\subfloat[]{\includegraphics[width=0.33\textwidth]{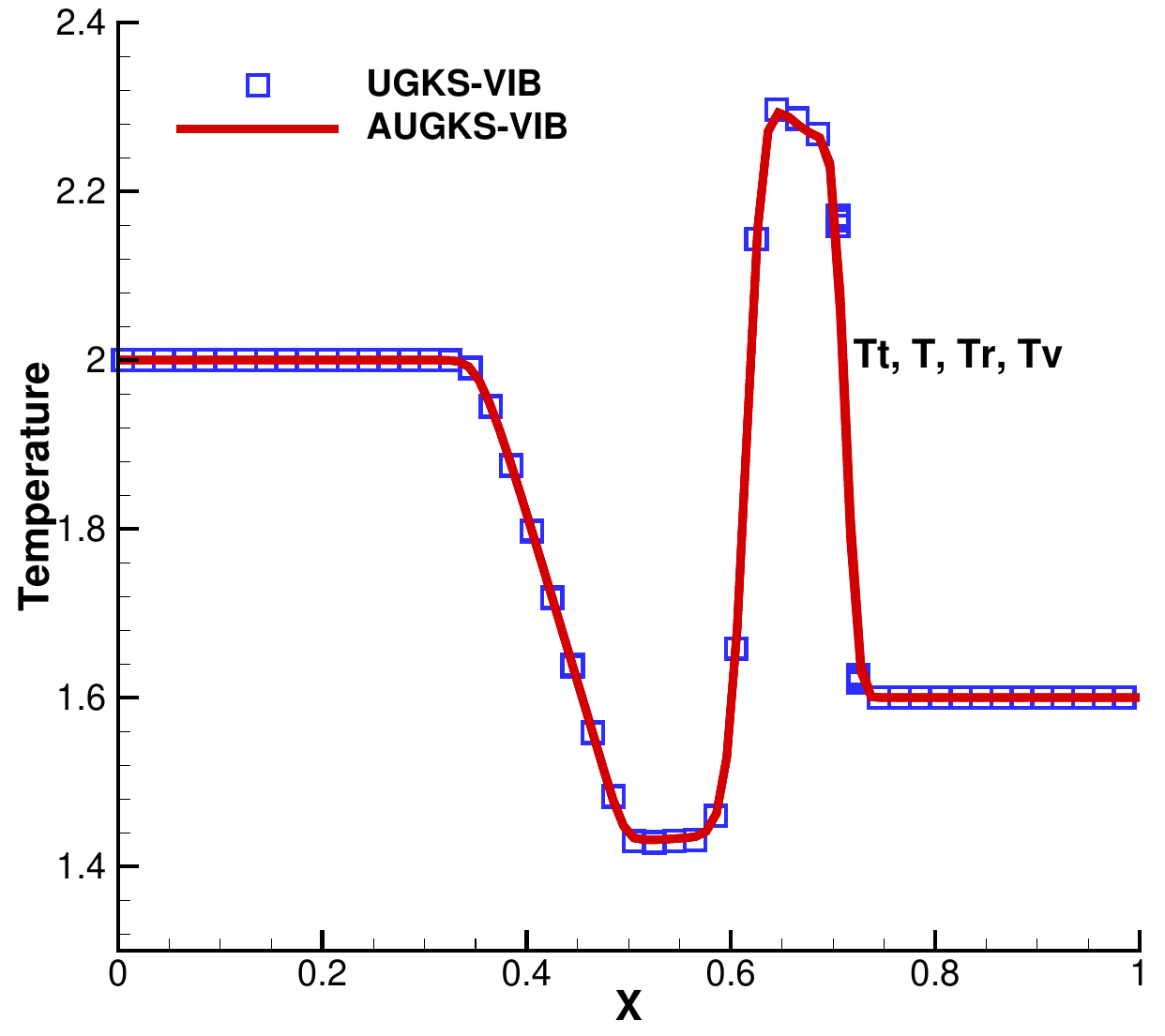}}
	\caption{Sod tube at ${\rm Kn}_\infty = 10^{-5}$. (a) Density,
	(b) velocity, and (c) temperatures, compared with UGKS results using the same vibrational relaxation model.}
	\label{fig:sod-kn1e-5}
\end{figure}

\begin{figure}[H]
	\centering
	\includegraphics[width=7cm]{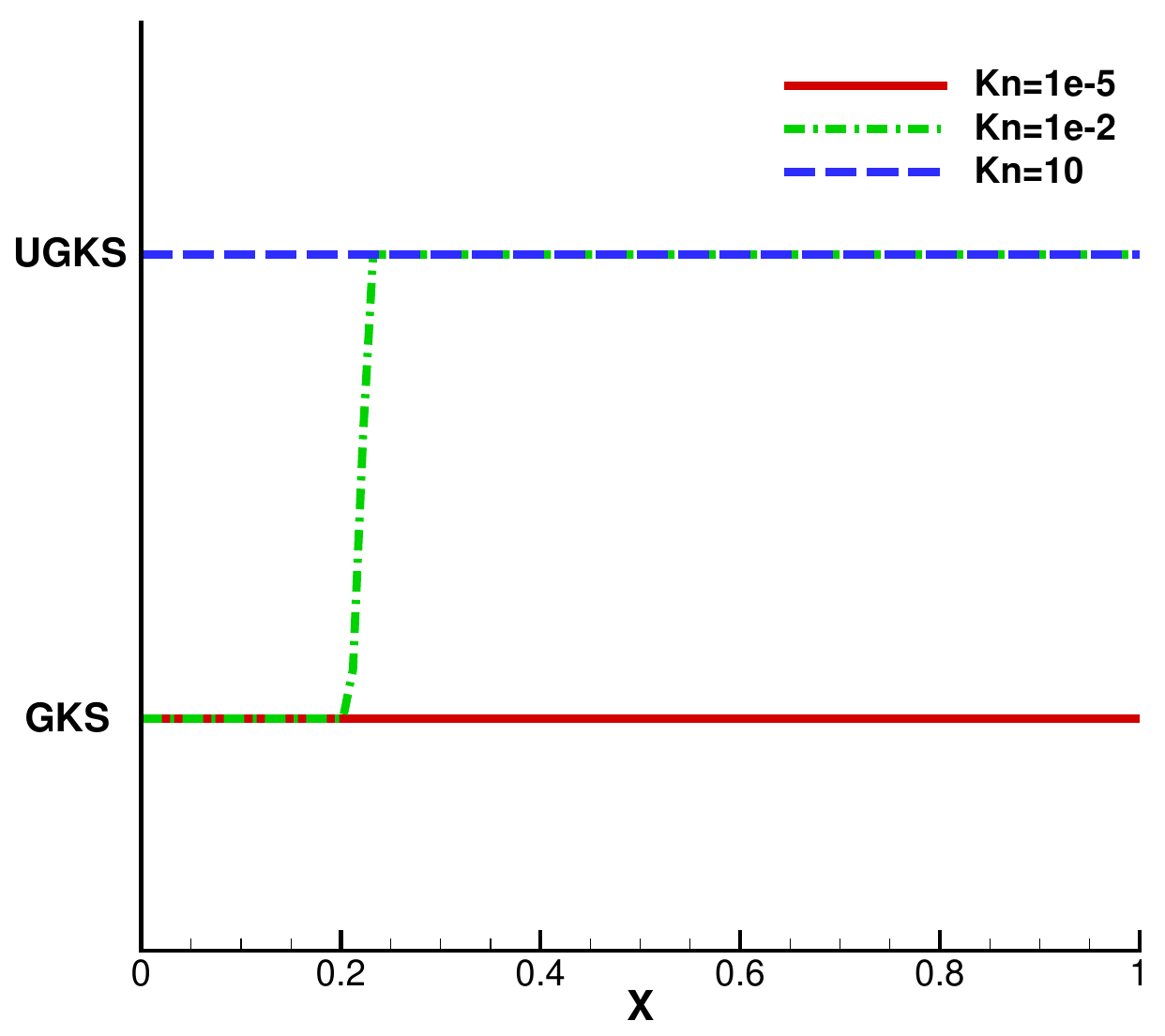}
	\caption{The velocity space adaptation for Sod tube at ${\rm Kn}_\infty = 10, 10^{-2}, 10^{-5}$.}
	\label{fig:sod-isDisc}
\end{figure}

The density, velocity as well as the temperatures including the translational, rotational, vibrational and the average temperatures obtained by AUGKS-vib and UGKS with the same vibrational model at different Knudsen numbers are plotted in Fig.~\ref{fig:sod-kn10}--\ref{fig:sod-kn1e-5}. In the calculation, the criteria for velocity space addition is set as $C_t = 0.01$, and the distribution of velocity space adaptation at different Knudsen number is plotted in Fig.~\ref{fig:sod-isDisc}. It shows that in the continuum and free molecular flow regimes, the continuous distribution function (GKS) and discretized velocity space (UGKS) are adopted respectively. At ${\rm Kn}_\infty = 0.01$, continuous and discretized distribution functions are applied adaptively to recover the real flow physics. For all the cases, the AUGKS-vib results agree well with the UGKS solutions with the same vibrational relaxation model, which shows the AUGKS-vib is capable of numerical simulations in both continuum and rarefied regimes with a complex physical model using the adaptive velocity space.

\subsection{Shock structure}
To verify the capability of capturing non-equilibrium effects with rotation and vibration modes, the normal shock wave at different upstream Mach number are tested. The initial conditions of the normal shock wave in the upstream and downstream with different specific heat ratios are determined by the conservation, which is given in \ref{sec:app-shock}. The computational domain $(-25, 25)$ has a length of $50$ times of the particle mean free path and is divided by $200$ cells uniformly. The discretized velocity space $(-U_1-15 \sqrt{ 2 R T_1}, U_1 + 15 \sqrt{2 R T_1})$ is discretized with 200 points based on midpoint rule. The left and right boundaries are treated as far field. The CFL number is taken as $0.95$.

\begin{figure}[H]
	\centering
	\subfloat[]{\includegraphics[width=0.33\textwidth]{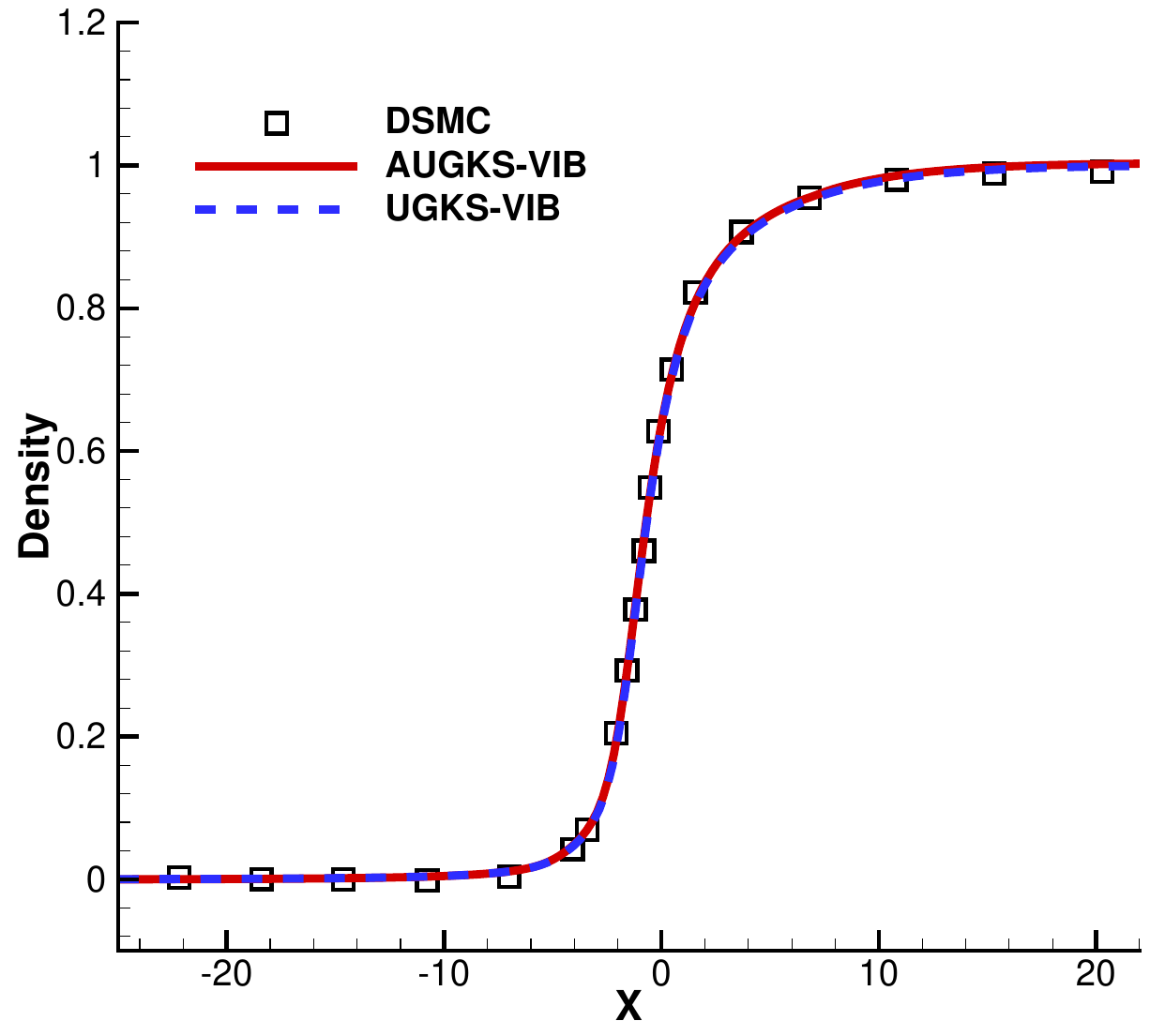}}
	\subfloat[]{\includegraphics[width=0.33\textwidth]{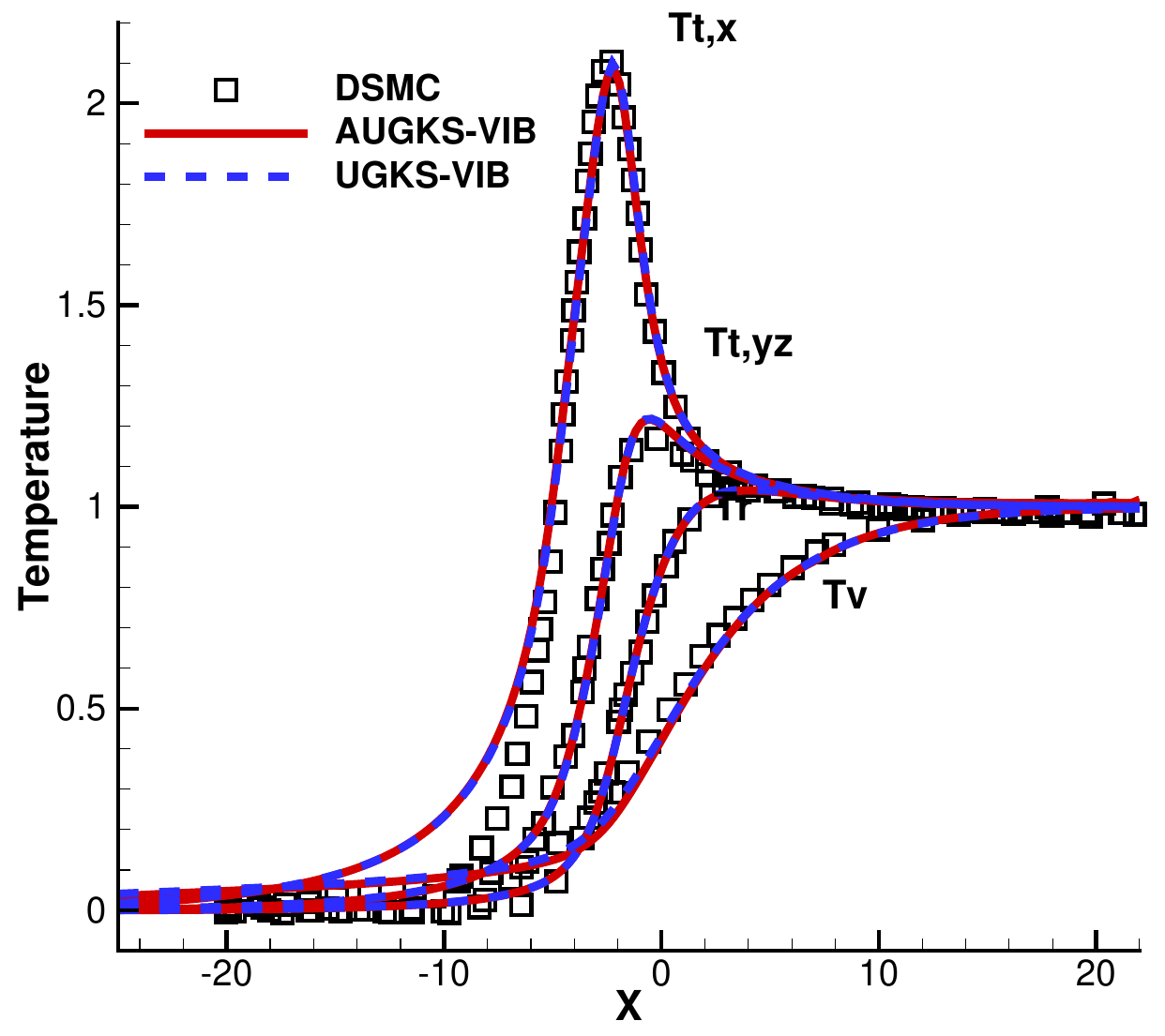}}
	\subfloat[]{\includegraphics[width=0.33\textwidth]{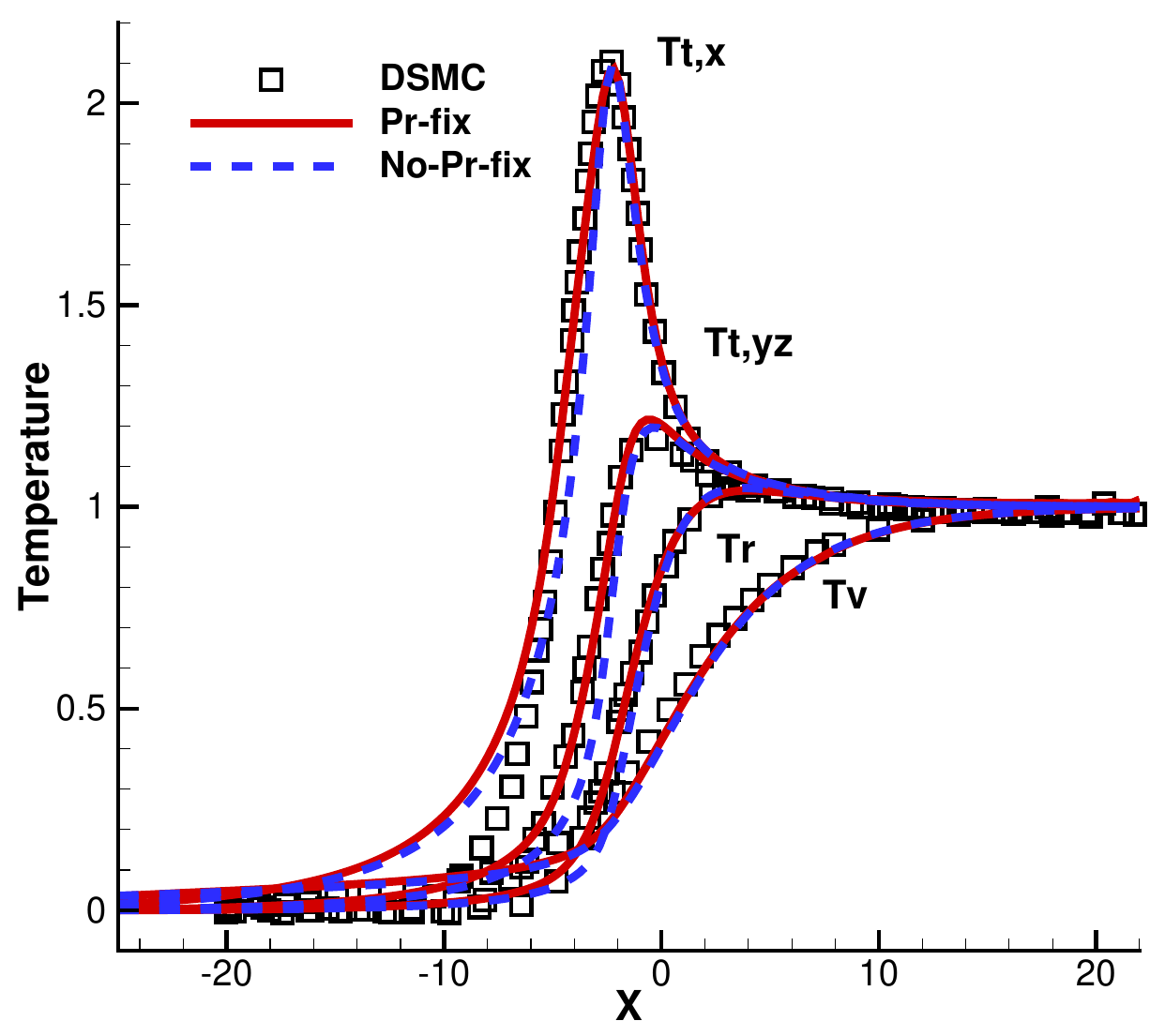}}
	\caption{Shock structure at ${\rm Ma}_\infty = 10$. (a) Density, and (b) temperatures compared
	with UGKS method and DSMC. (c) Comparison of temperatures between Pr-fixed model
	and original BGK-type vibrational model by AUGKS and DSMC.}
	\label{fig:shock}
\end{figure}

In this study, a strong shock wave at upstream Mach number ${\rm Ma}_1 = 10$ is investigated, and the upstream temperature is $T_1 = 226.149$ K. The rest parameters could be obtained from the non-dimensional initial condition
\begin{equation*}
	\begin{cases}\rho_1 = 1,  \qquad U_1=8.3666, \lambda_1=1, & x<0, \\
		         \rho_2 = 6.9294, U_2=1.2074, \lambda_2=0.05736, & x \geq 0.\end{cases}
\end{equation*}
The rotational and vibrational collision numbers keep constant as $Z_r = 5$ and $Z_v = 28$. The comparison with the DSMC \cite{cai2008one} simulation is plotted in Fig.~\ref{fig:shock}, which shows the good agreement between AUGKS-vib and DSMC data. In Fig.~\ref{fig:shock}(b), $T_r$ and $T_v$ denote rotational and vibrational temperature respectively. $T_{t,x}$ denotes the translational temperature in $x$ direction, and $T_{t,yz}$ is the average translational temperature in $y$ and $z$ directions, which are obtained from
\begin{equation*}\label{eq:Tt-x}
	T_{t,x} = \frac{1}{ \rho R } \int { ( u-U )^2 } f {\rm d}{\vec \Xi},
\end{equation*}
and
\begin{equation*}
	T_{t,yz} = \frac{1}{ 2 \rho R } \int { \left[  ( v-V )^2 + ( w-W )^2  \right] f {\rm d}{\vec \Xi}}.
\end{equation*}

For the relaxation type kinetic models, the early rising of the temperature occurs at high Mach number. As analyzed before, the reason is that the single relaxation time in the model equally re-distributes the energy to all molecules, which is inconsistent with the physical reality that the high speed particles should have shorter relaxation time. Therefore, the high speed particles transport longer distance and thermalize more efficiently the energy among all particles leading to the early rising of the temperature. But the density distributions in the shock structures from the kinetic relaxation models can match with the DSMC solutions very well. Fig.~\ref{fig:shock}(c) plots the effectiveness of Prandtl number fix on the shock structure which can better match the results of DSMC.

\begin{figure}[H]
	\centering
	\subfloat[Density]{\includegraphics[width=0.35\textwidth]{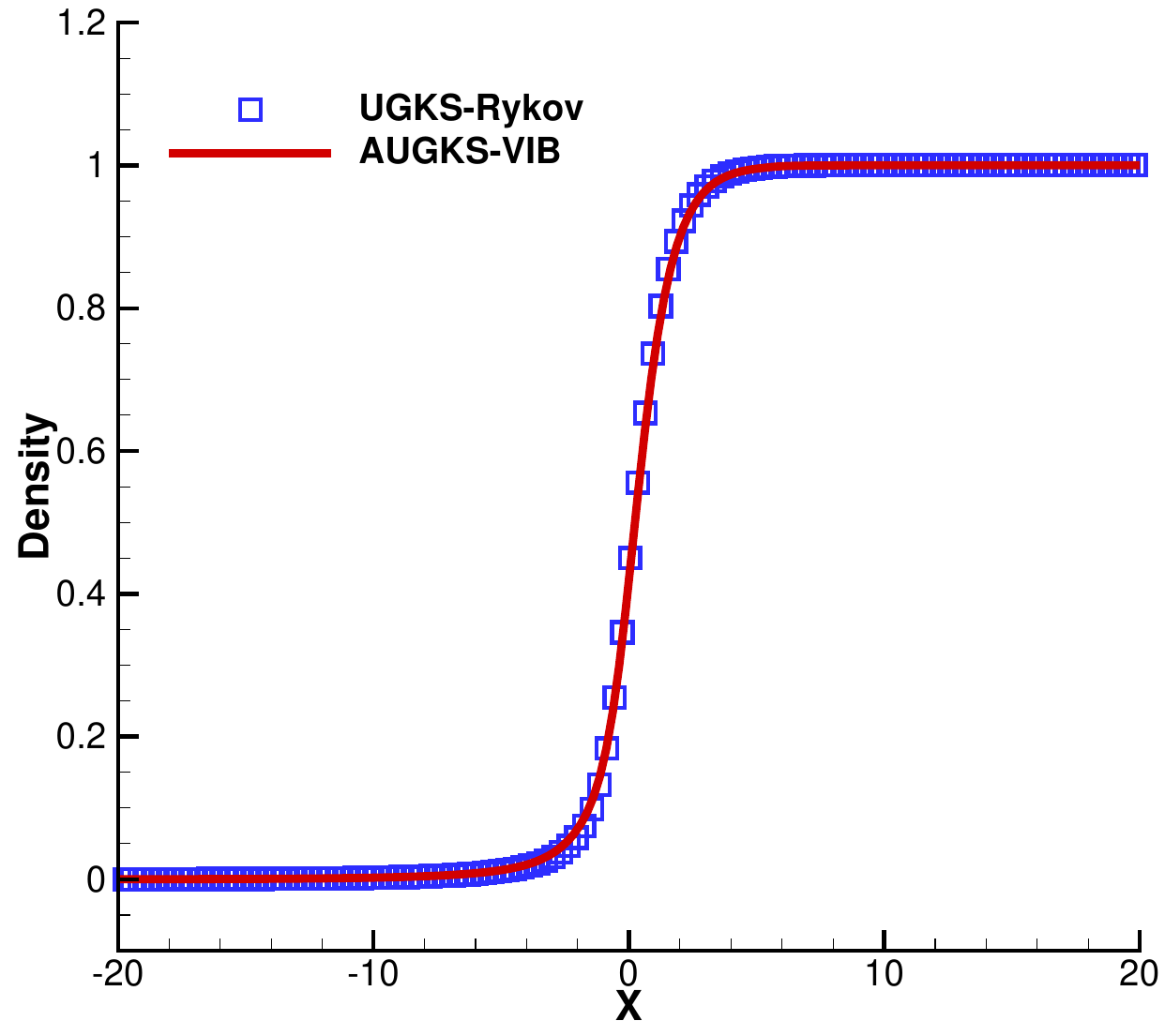}}
	\subfloat[Temperatures]{\includegraphics[width=0.35\textwidth]{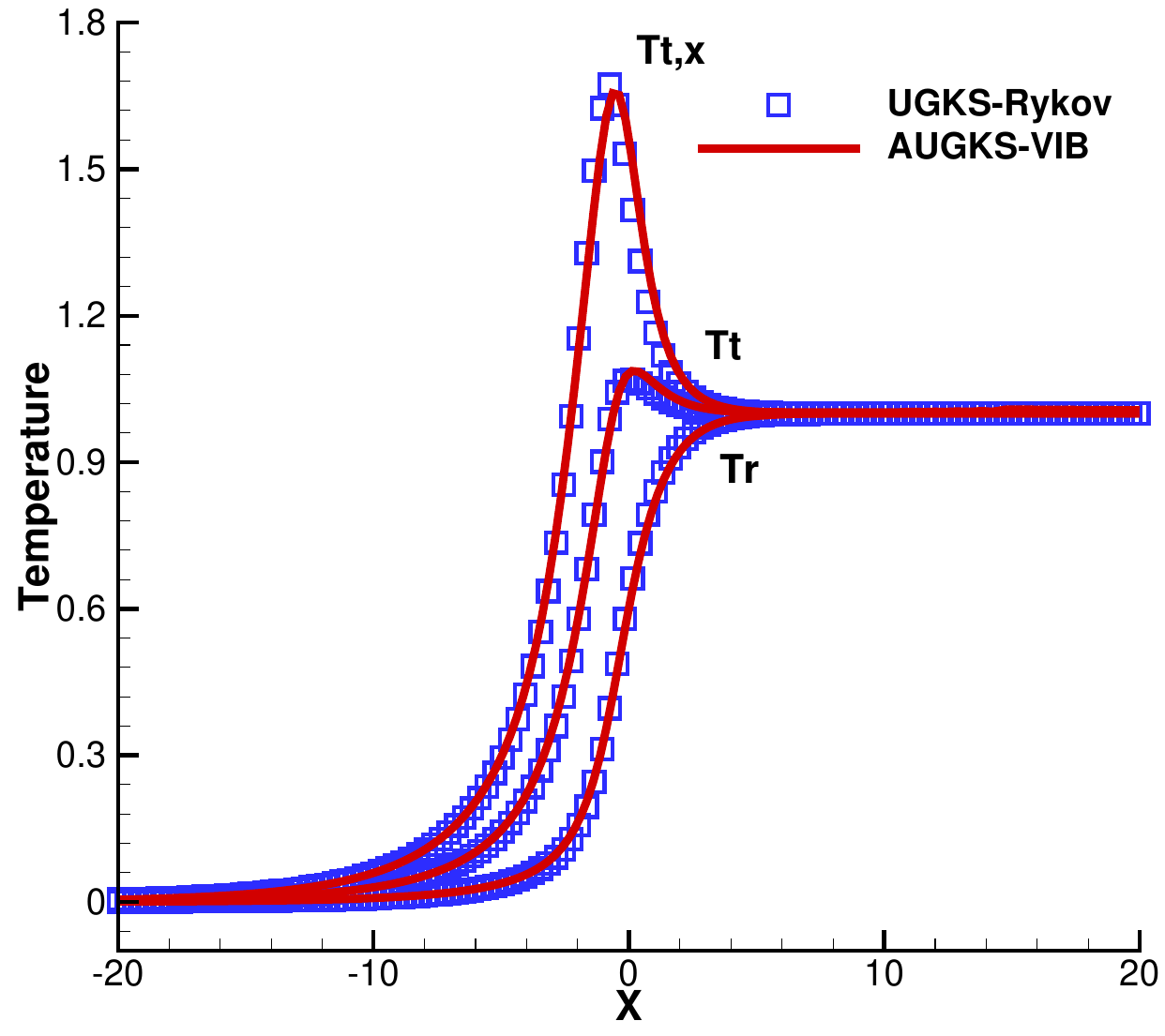}}
	\caption{Shock structure at ${\rm Ma}_\infty = 4$, $Z_r = 2.4$, $Z_v \to \infty$. (a) Density, and (b) temperatures compared with UGKS using Rykov model.}
	\label{fig:shock-Ma4-rot}
\end{figure}

\begin{figure}[H]
	\centering
	\subfloat[Density]{\includegraphics[width=0.35\textwidth]{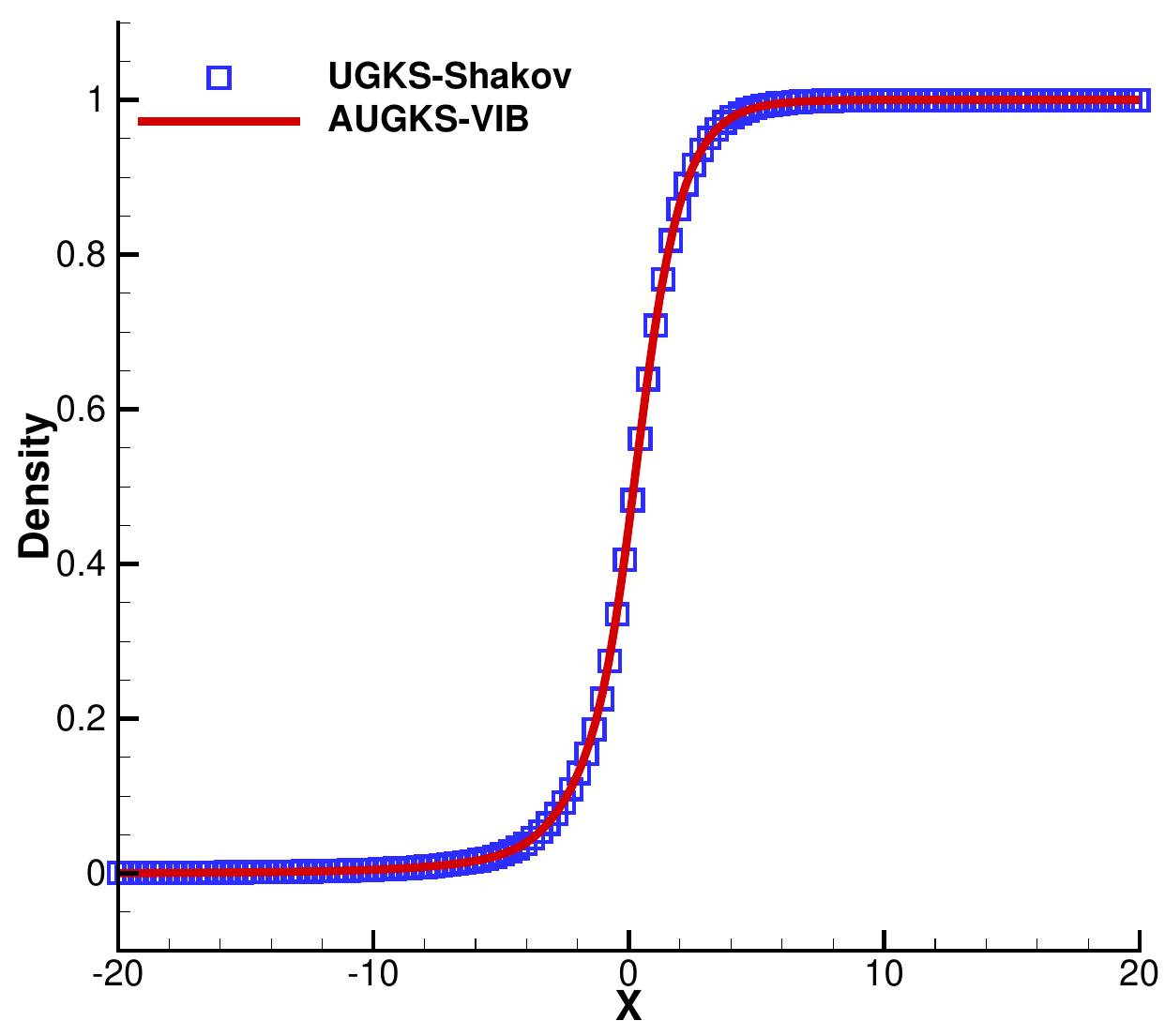}}
	\subfloat[Temperatures]{\includegraphics[width=0.35\textwidth]{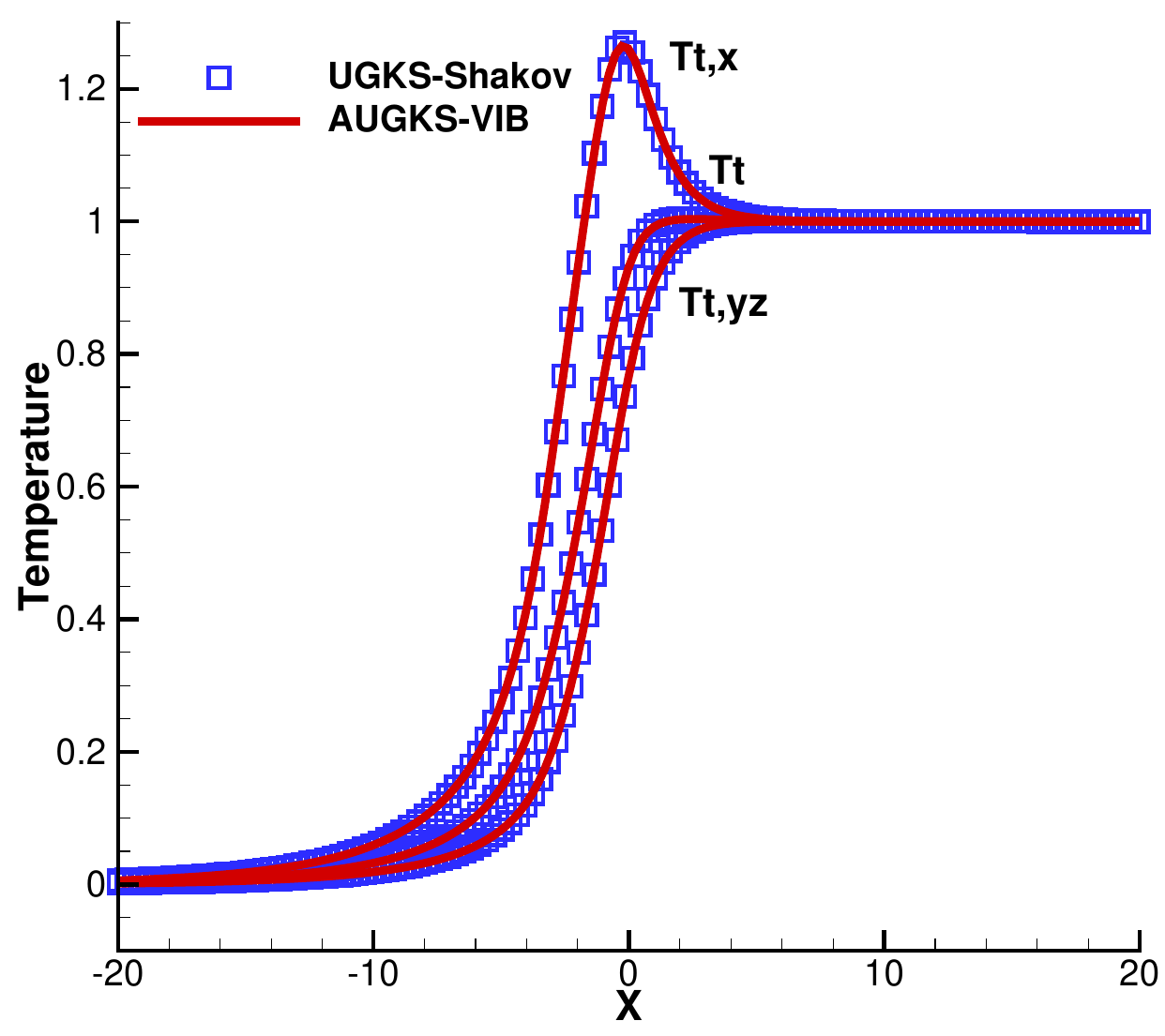}}
	\caption{Shock structure at ${\rm Ma}_\infty = 4$, $Z_r \to \infty$, $Z_v \to \infty$. (a) Density, and (b) temperatures compared with UGKS using Shakov model.}
	\label{fig:shock-Ma4-mono}
\end{figure}

The shock structure at ${\rm Ma} = 4.0$ is tested to further assess the accuracy of the model. In this case, the vibrational collision number $Z_v$ is set to infinity and the vibrational degrees of freedom are frozen. The solution obtained by AUGKS-vib will return to the Rykov model solved by UGKS, as shown in Fig.~\ref{fig:shock-Ma4-rot}. Furthermore, by setting the rotational collision number to $Z_r \to \infty$, AUGKS-vib will recover the solution of the Shakov model, as shown in Fig.~\ref{fig:shock-Ma4-mono}. This demonstrates the consistency of the current vibrational model with the Rykov and Shakov models in dealing with rotation and translation degrees of freedom for diatomic gas.

\subsection{Hypersonic Flow around a circular cylinder}
\begin{figure}[H]
	\centering
	\includegraphics[width=0.4\textwidth]
		{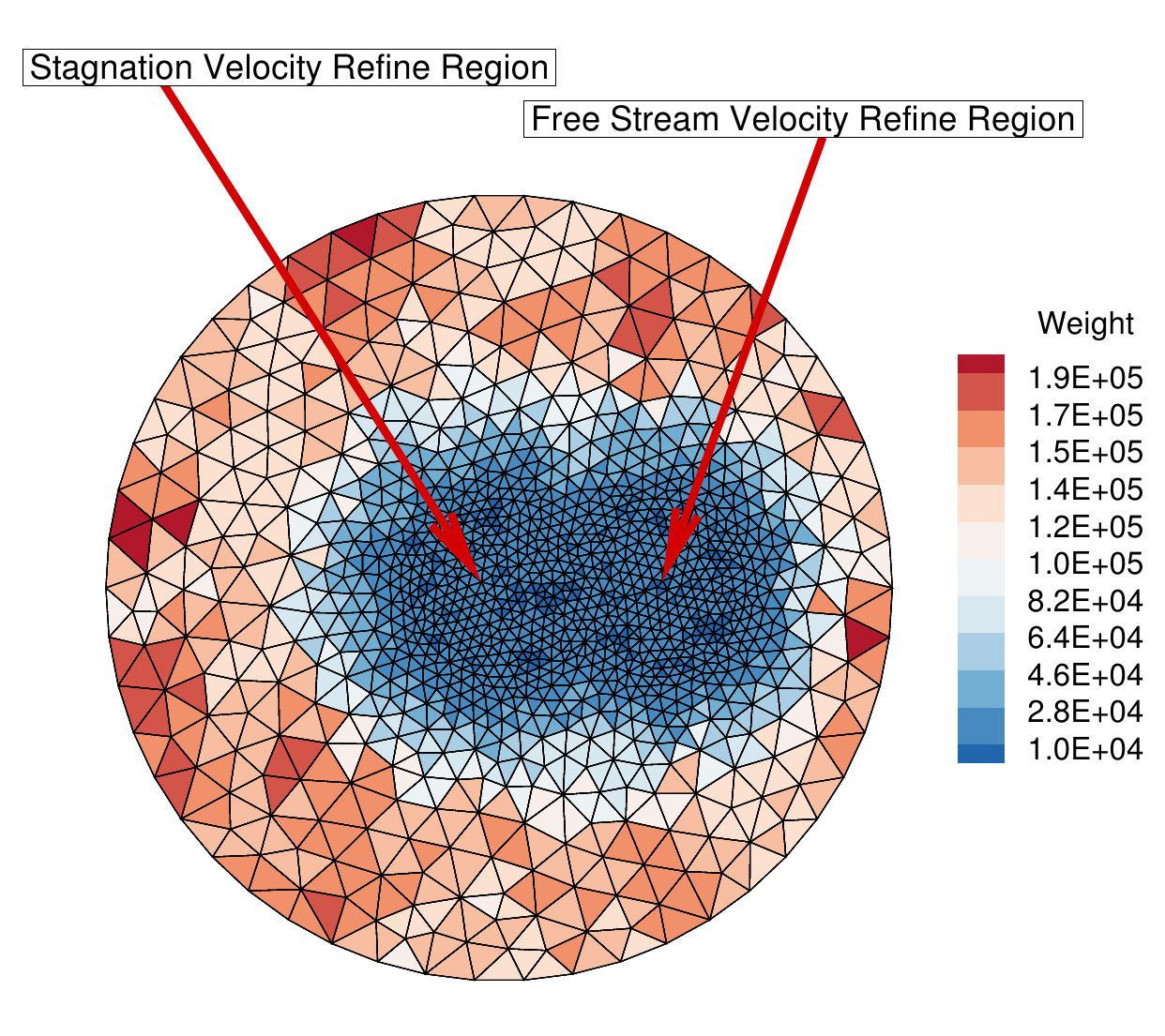}
	\caption{Unstructured discrete velocity space mesh adopted for hypersonic flow at ${\rm Kn} = 0.01$ and ${\rm Ma} = 5$ passing over a cylinder by the AUGKS method.}
	\label{fig:cylinder-Ma5-DVS}
\end{figure}

Hypersonic gas flow of nitrogen around a circular cylinder has been computed at ${\rm Ma}_\infty = 5$ and ${\rm Kn}_\infty = 0.01$. The Knudsen number is defined with respect to the diameter $D = 1$ m. The temperature in the free stream is $T_\infty = 500$ K. An isothermal wall with a fixed temperature of $T_w = 500$ K is applied for the solid surface. The physical domain is discretized by $180 \times 88 \times 1$ quadrilateral cells. Figure~\ref{fig:cylinder-Ma5-DVS} shows the unstructured DVS mesh consists of 2,112 cells. The DVS is discretized in a circle region with the center $0.4\times(U_\infty,V_\infty,W_\infty)$, and the radius is $6\sqrt{R T_s}$ where $T_s$ is the stagnation temperature of the free stream flow. The unstructured DVS mesh is refined at zero velocity point with a radius of $3\sqrt{R T_w}$, and the free stream velocity point with a radius of $3\sqrt{R T_\infty}$. The rotational collision number is evaluated by Eq.~\eqref{eq:Zr-tumuklu} with $Z_r^\infty = 23.5$, $T^\ast = 91.5$ K, and a constant vibrational collision number $Z_v=50$ is taken. The CFL number is taken as 0.95.

\begin{figure}[H]
	\centering
	\subfloat[]{\includegraphics[width=0.3\textwidth]
	{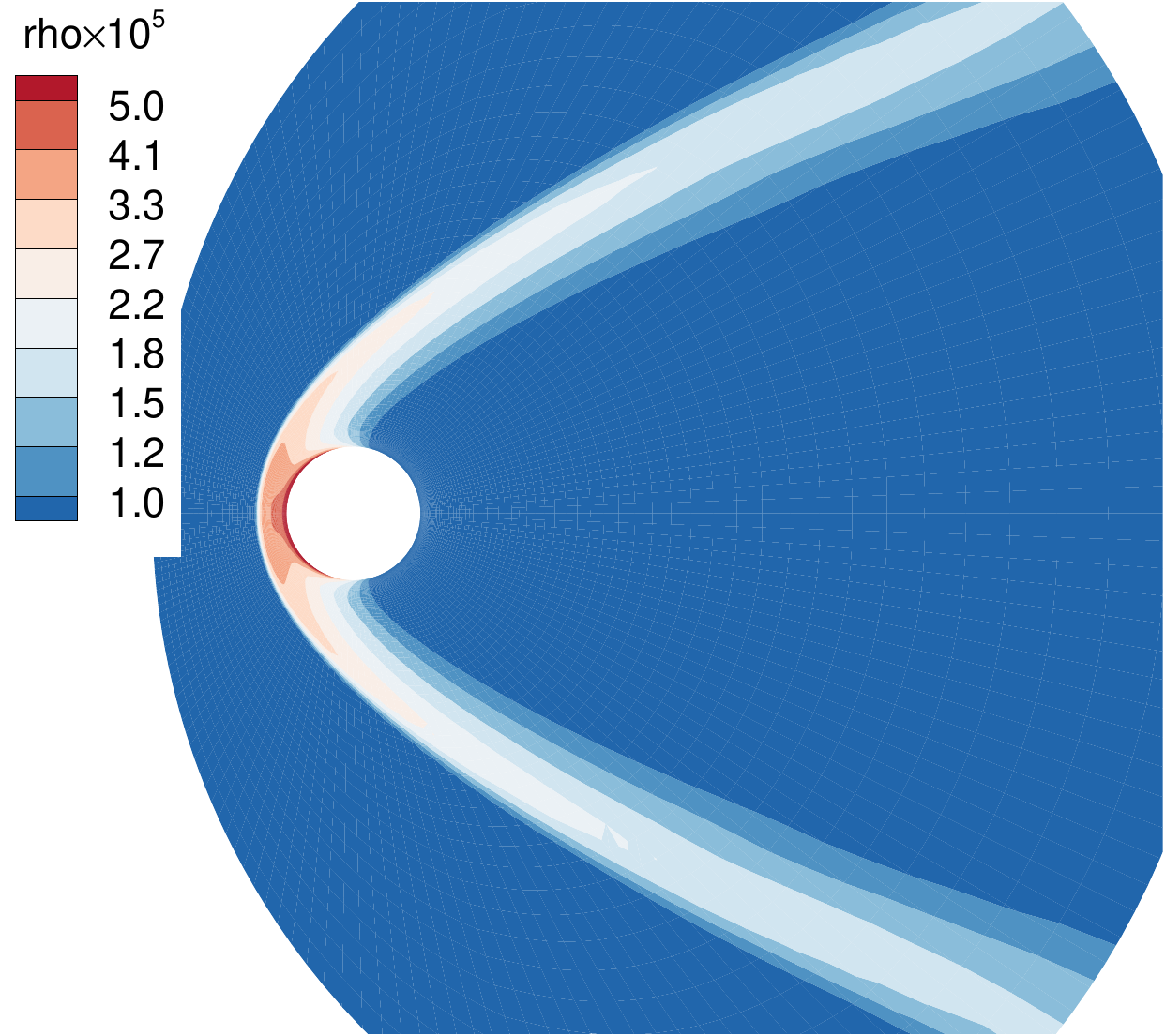}}
	\subfloat[]{\includegraphics[width=0.3\textwidth]
	{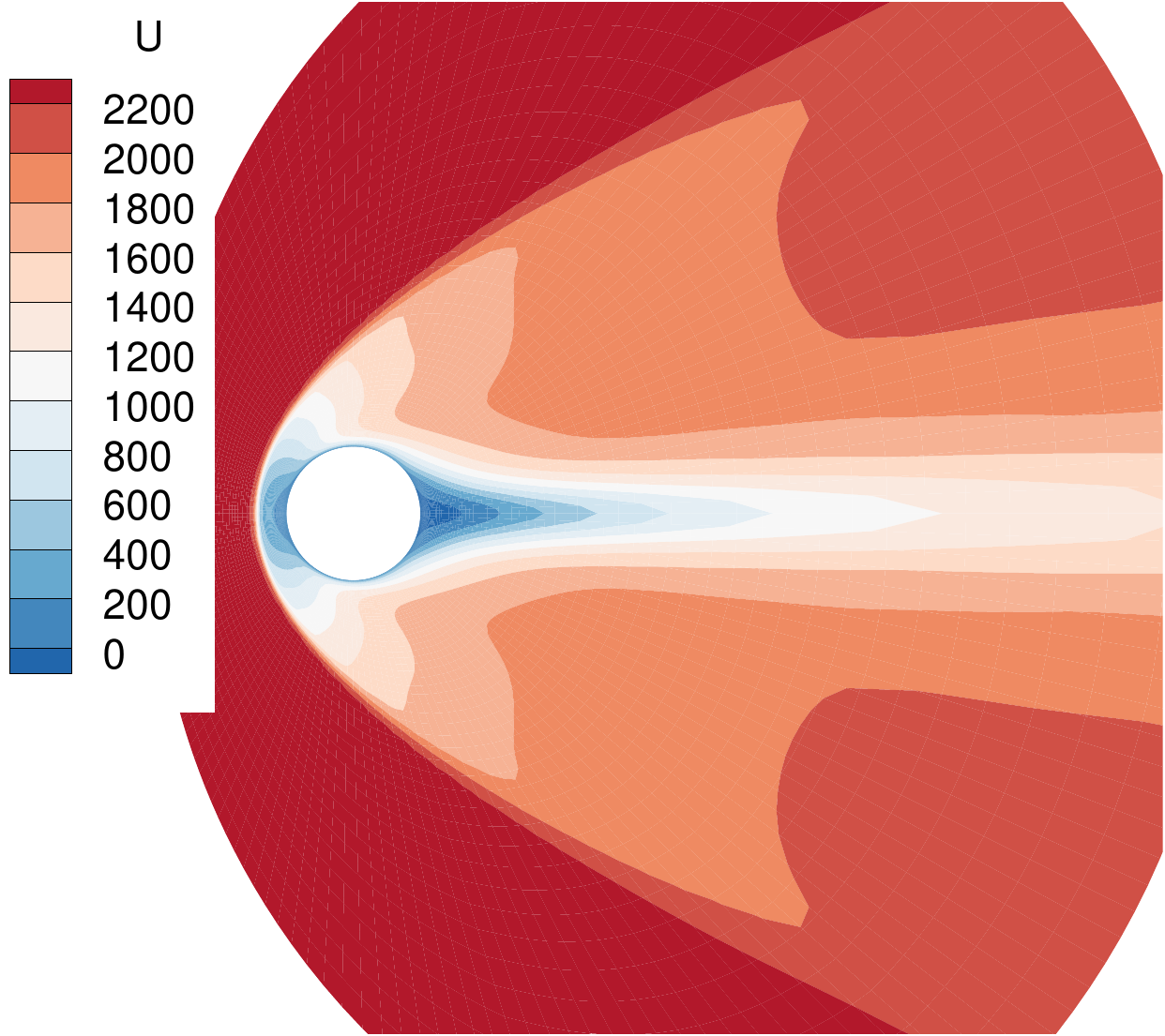}}
	\subfloat[]{\includegraphics[width=0.3\textwidth]
	{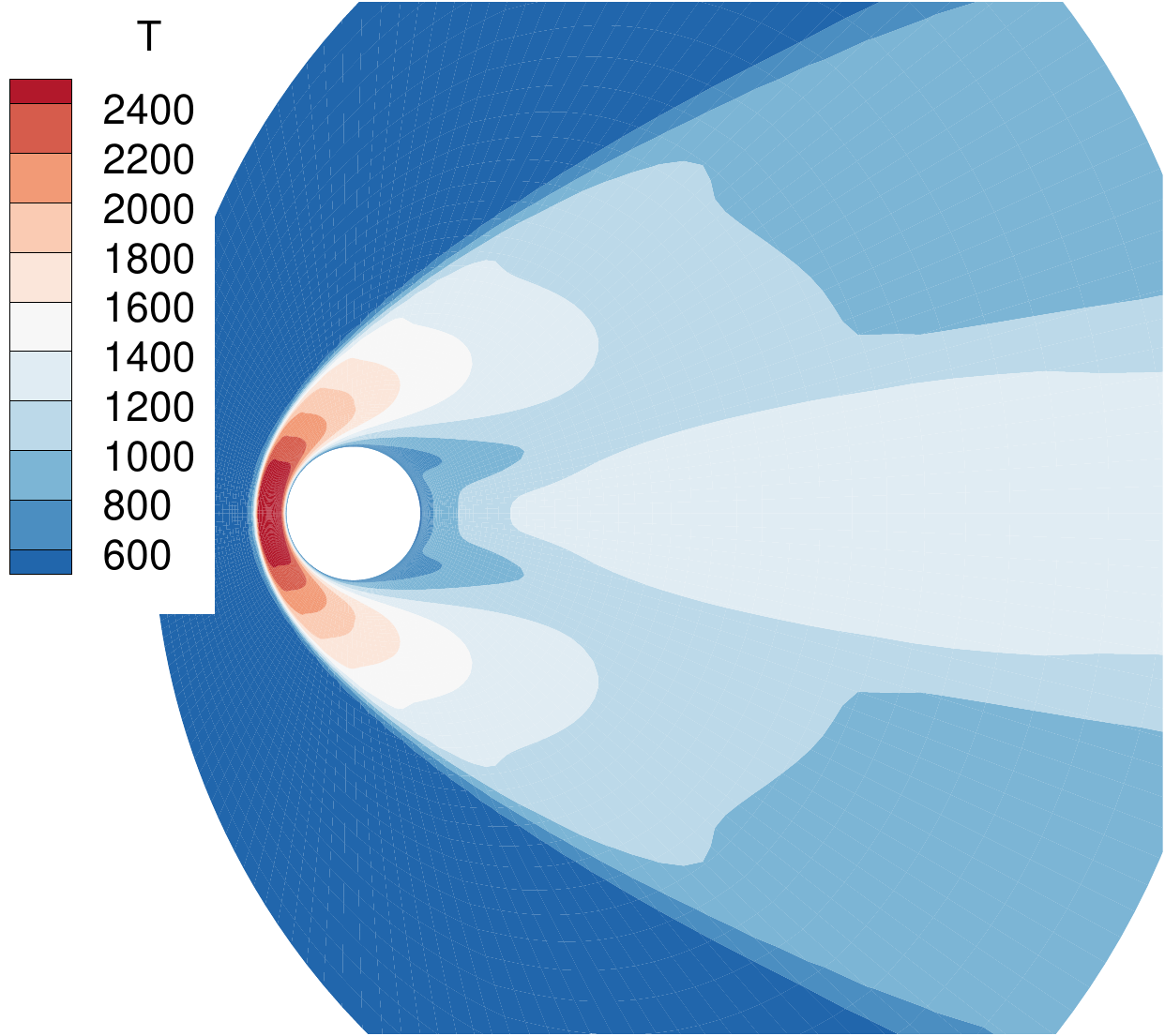}} \\
	\subfloat[]{\includegraphics[width=0.3\textwidth]
	{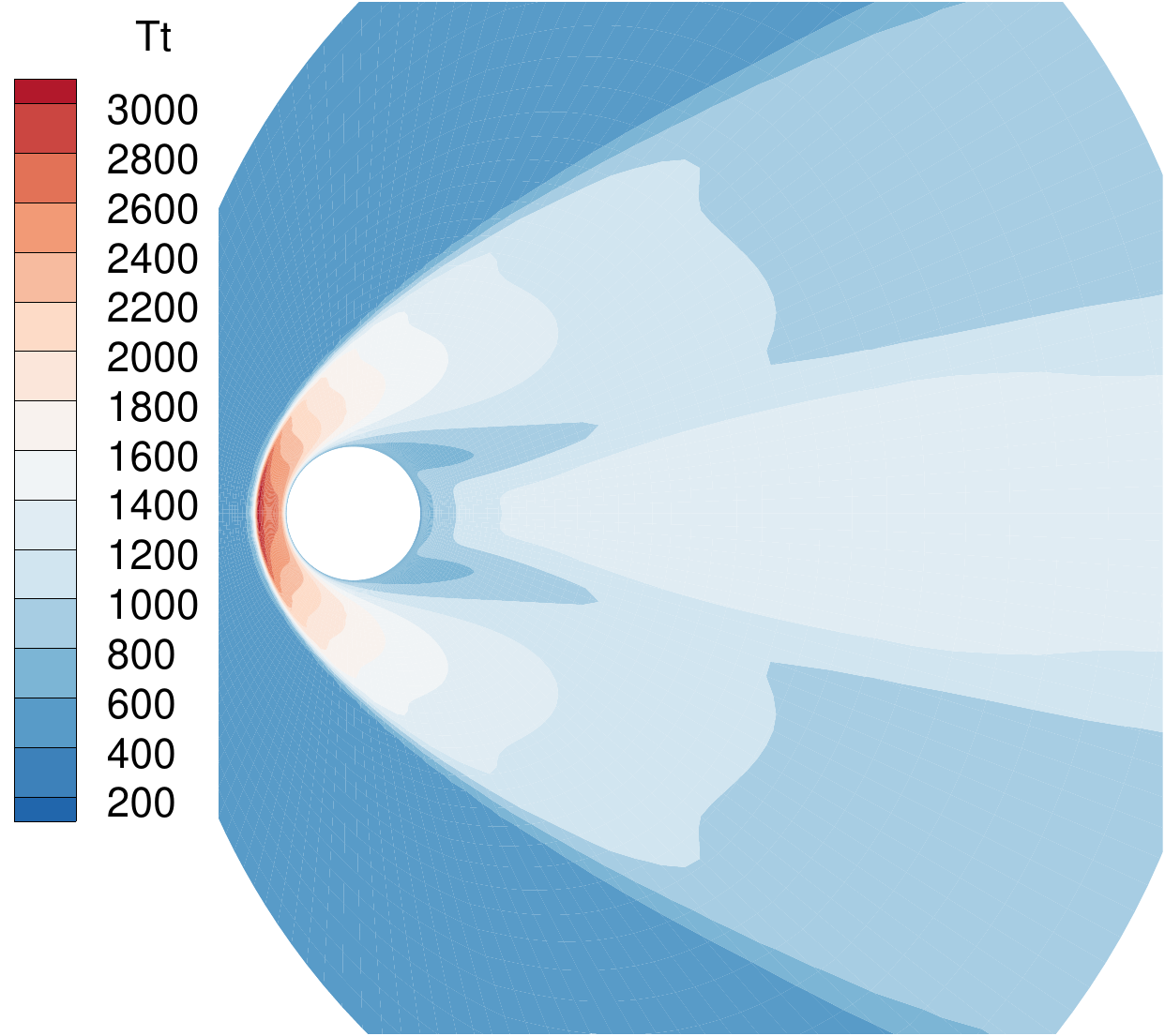}}
	\subfloat[]{\includegraphics[width=0.3\textwidth]
	{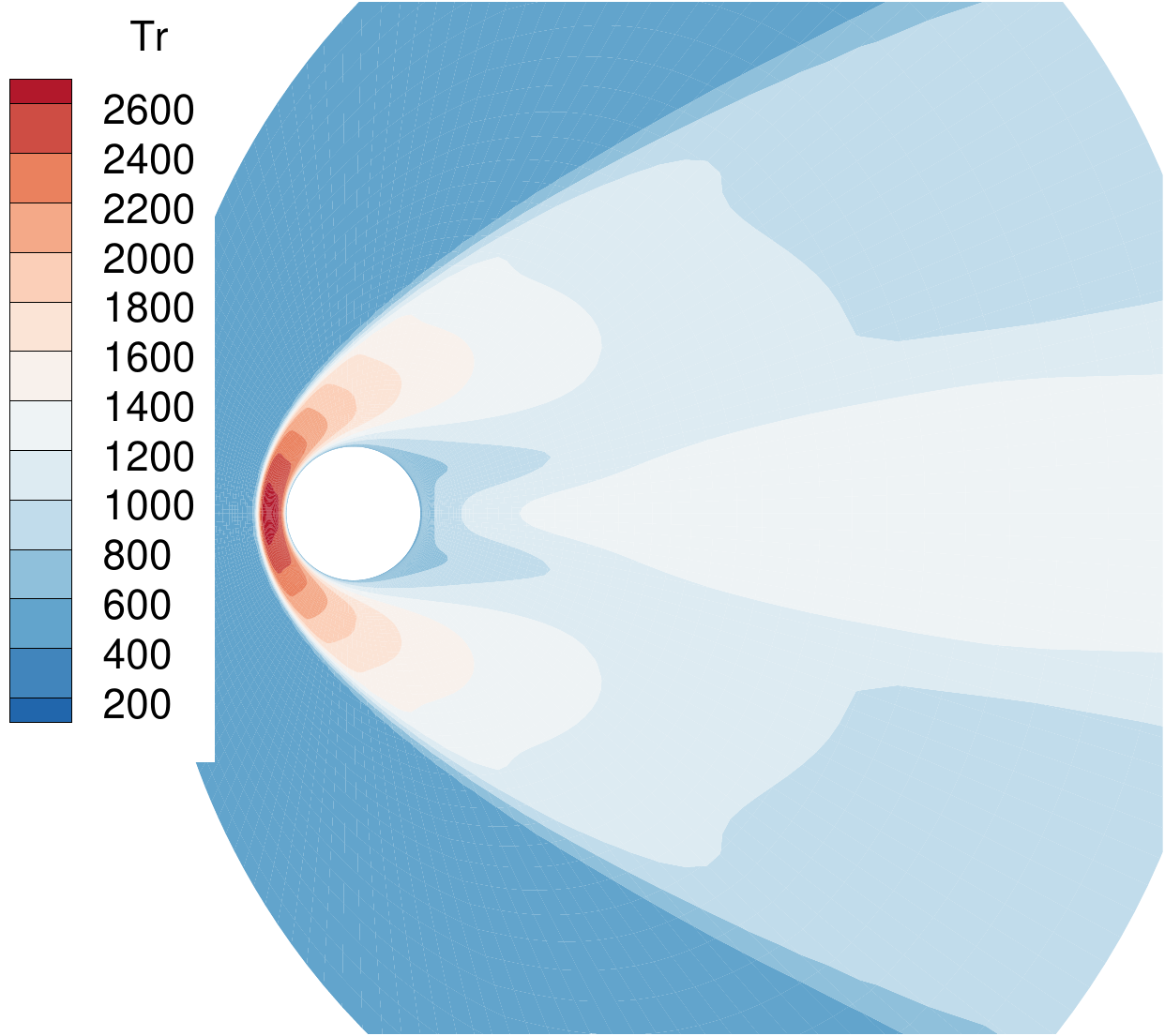}}
	\subfloat[]{\includegraphics[width=0.3\textwidth]
	{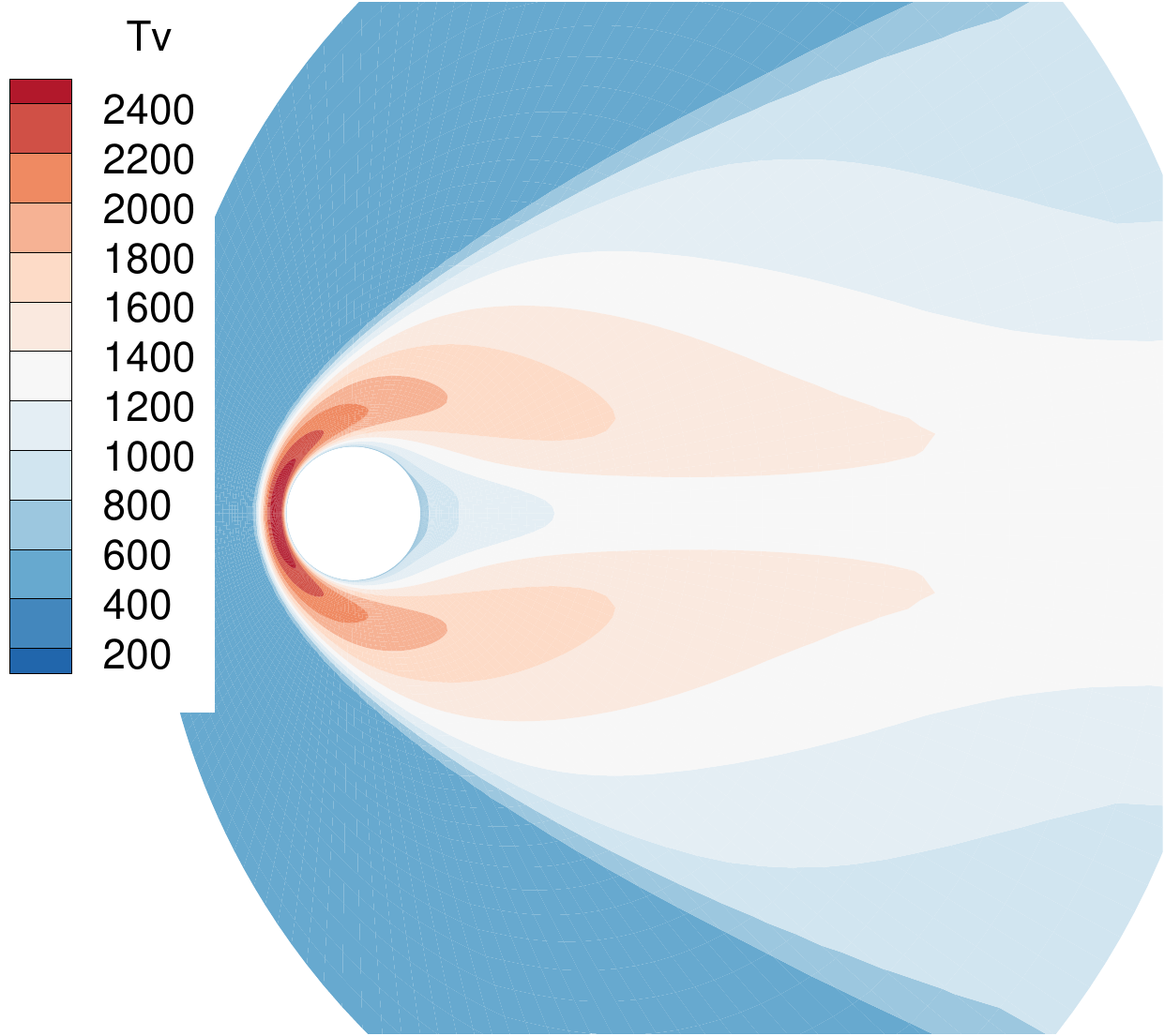}} \\
	\caption{Hypersonic flow at ${\rm Kn} = 0.01$ and ${\rm Ma} = 5$ passing over a circular cylinder by the AUGKS method. (a) Density, (b) $x$ direction velocity,
	(c) temperature, (d) translational temperature,
	(e) rotational temperature and (f) vibrational
	temperature contours.}
	\label{fig:cylinder-Ma5}
\end{figure}

\begin{figure}[H]
	\centering
	\includegraphics[width=7cm]{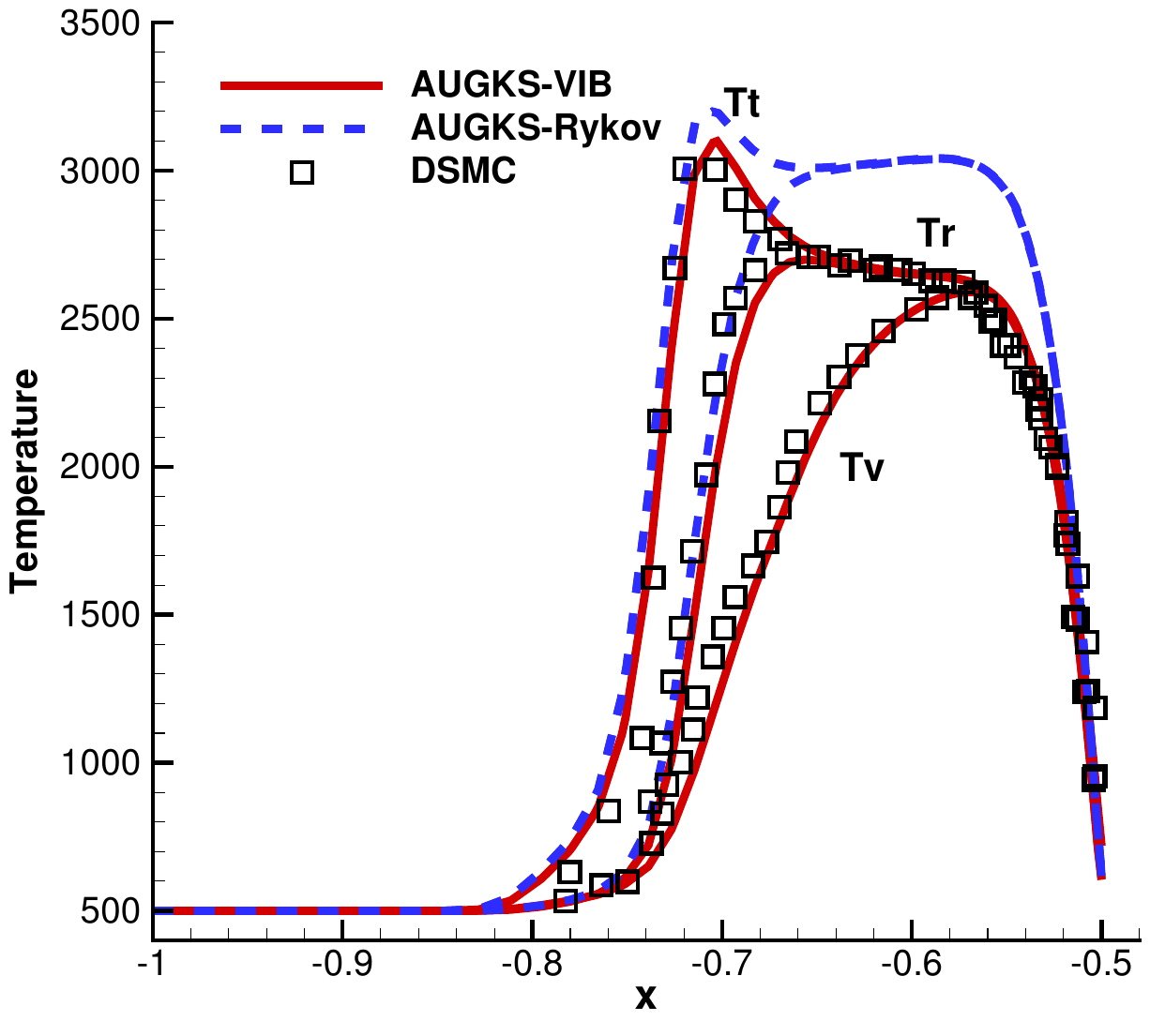}
	\caption{Hypersonic flow at ${\rm Kn} = 0.01$ and ${\rm Ma} = 5$ passing over a circular cylinder by the AUGKS method. Temperature distributions along the stagnation line.}
	\label{fig:cylinder-Ma5-T}
\end{figure}

\begin{figure}[H]
	\centering
	\includegraphics[width=7cm]{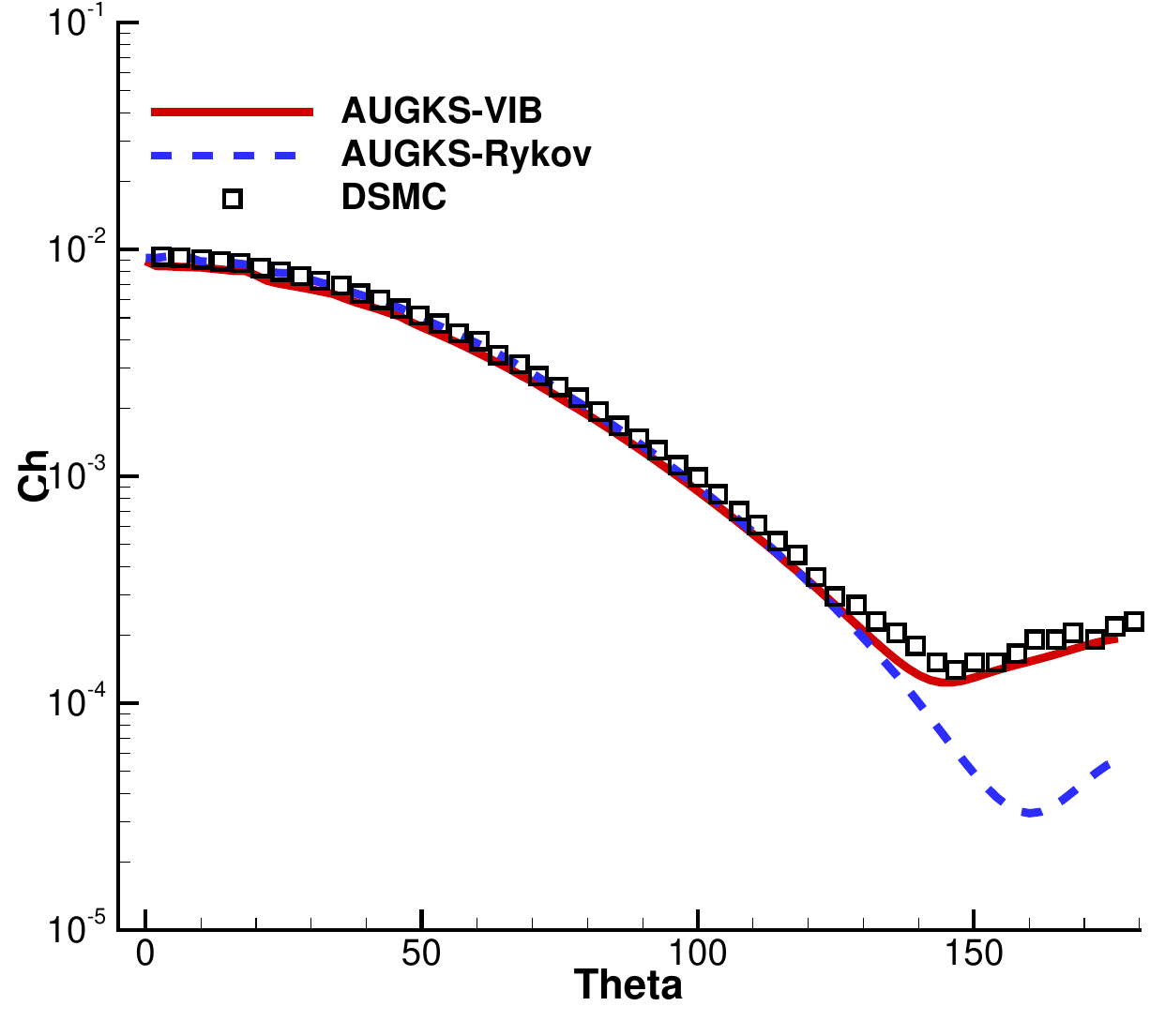}
	\caption{Hypersonic flow at ${\rm Kn}_\infty = 0.01$ and ${\rm Ma} = 5$ passing over a circular cylinder by the AUGKS method. Heat flux coefficient distributions on the cylinder surface.}
	\label{fig:cylinder-Ma5-cH-vib-rot}
\end{figure}

\begin{figure}[H]
	\centering
	\includegraphics[width=8cm]{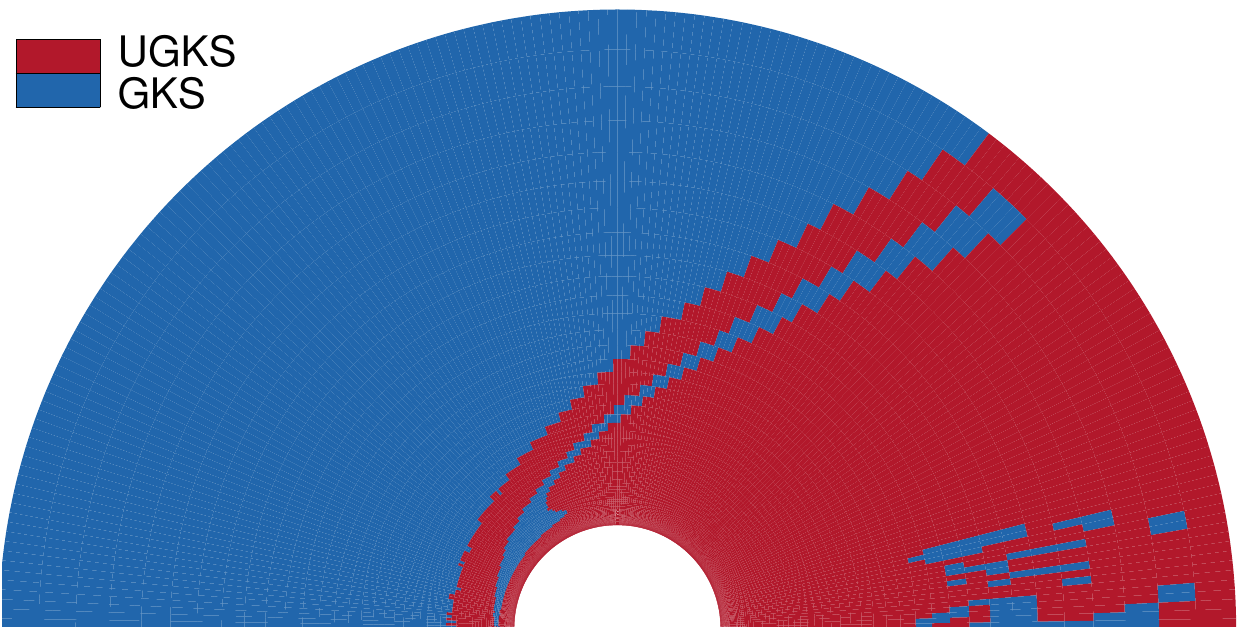}
	\caption{Hypersonic flow at ${\rm Ma}_\infty = 15$ and ${\rm Kn}_\infty = 0.01$ around a half-circular
		cylinder by the AUGKS method. Distributions of velocity space adaptation with $C_t = 0.01$ where the discretized velocity space (UGKS) is used in 78.78\% of physical domain.}
	\label{fig:cylinder-Ma15-isDisc}
\end{figure}

The contours of flow are plotted in Fig.~\ref{fig:cylinder-Ma5} where an initial flow field provided by 10000 steps of first order GKS calculation and another 15000 steps of AUGKS is adopted. Figure~\ref{fig:cylinder-Ma15-isDisc} depicts the distribution of velocity space adaptation where 63.72\% of the computational domain is covered by discretized velocity space (UGKS). The temperature distributions along the upstream central show good agreement with DSMC data and the difference of translational and rotational temperatures given by vibrational model and Rykov model using AUGKS can be observed in Fig.~\ref{fig:cylinder-Ma5-T}. The influence of vibration mode of diatomic gas on the high speed nonequilibrium flow is also illustrated in the comparison of surface heat flux coefficient computed by AUGKS with vibrational model, AUGKS with Rykov model, and DSMC, which is shown in Fig.~\ref{fig:cylinder-Ma5-cH-vib-rot}.

\begin{figure}[H]
	\centering
	\includegraphics[width=8cm]{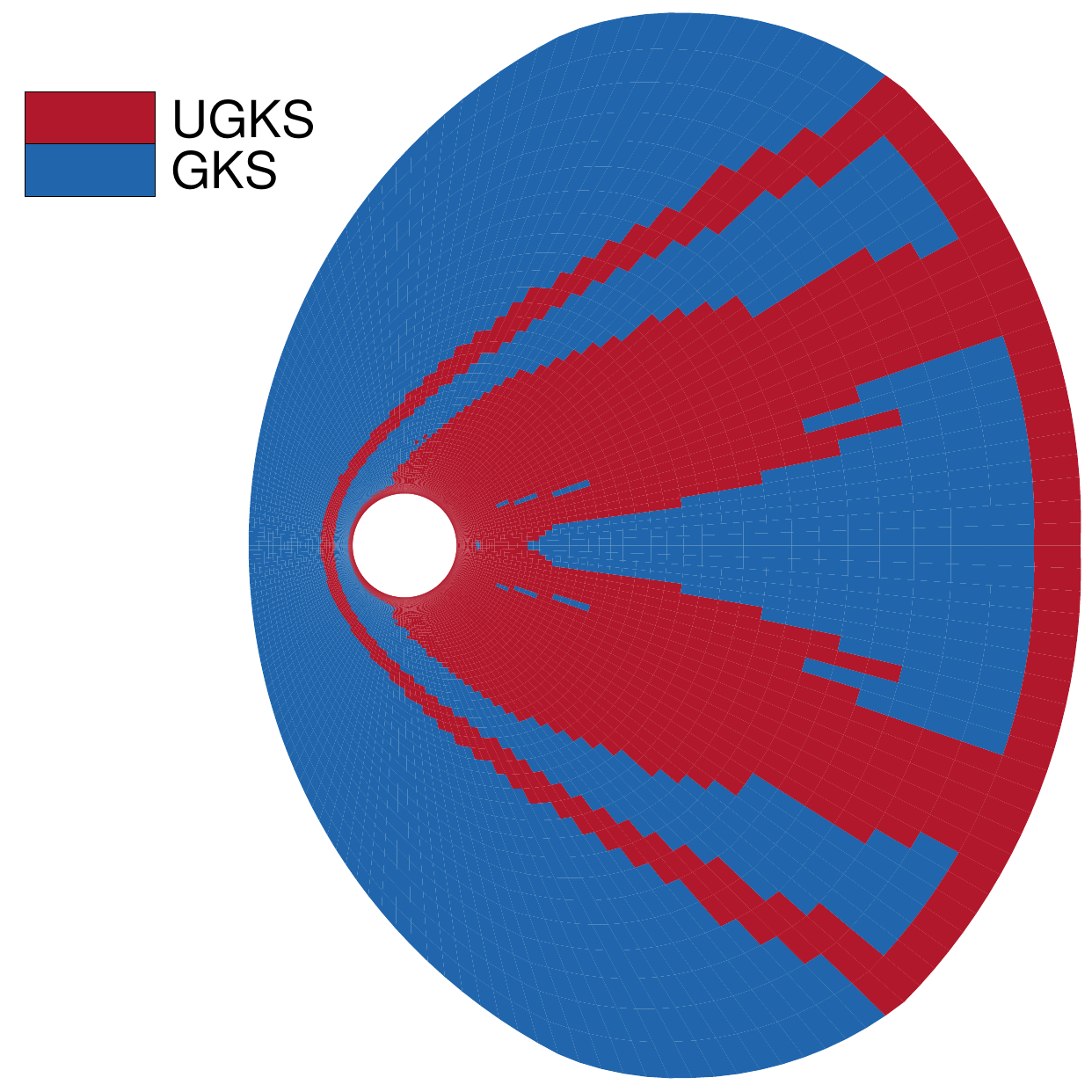}
	\caption{Hypersonic flow at ${\rm Kn} = 0.01$ and ${\rm Ma} = 5$ passing over a circular cylinder by the AUGKS method. Distributions of velocity space adaptation with $C_t = 0.01$ where the discretized velocity space (UGKS) is used in 63.72\% of physical domain.}
	\label{fig:cylinder-Ma5-isDisc}
\end{figure}

\begin{figure}[H]
	\centering
	\subfloat[]{\includegraphics[width=0.3\textwidth]
	{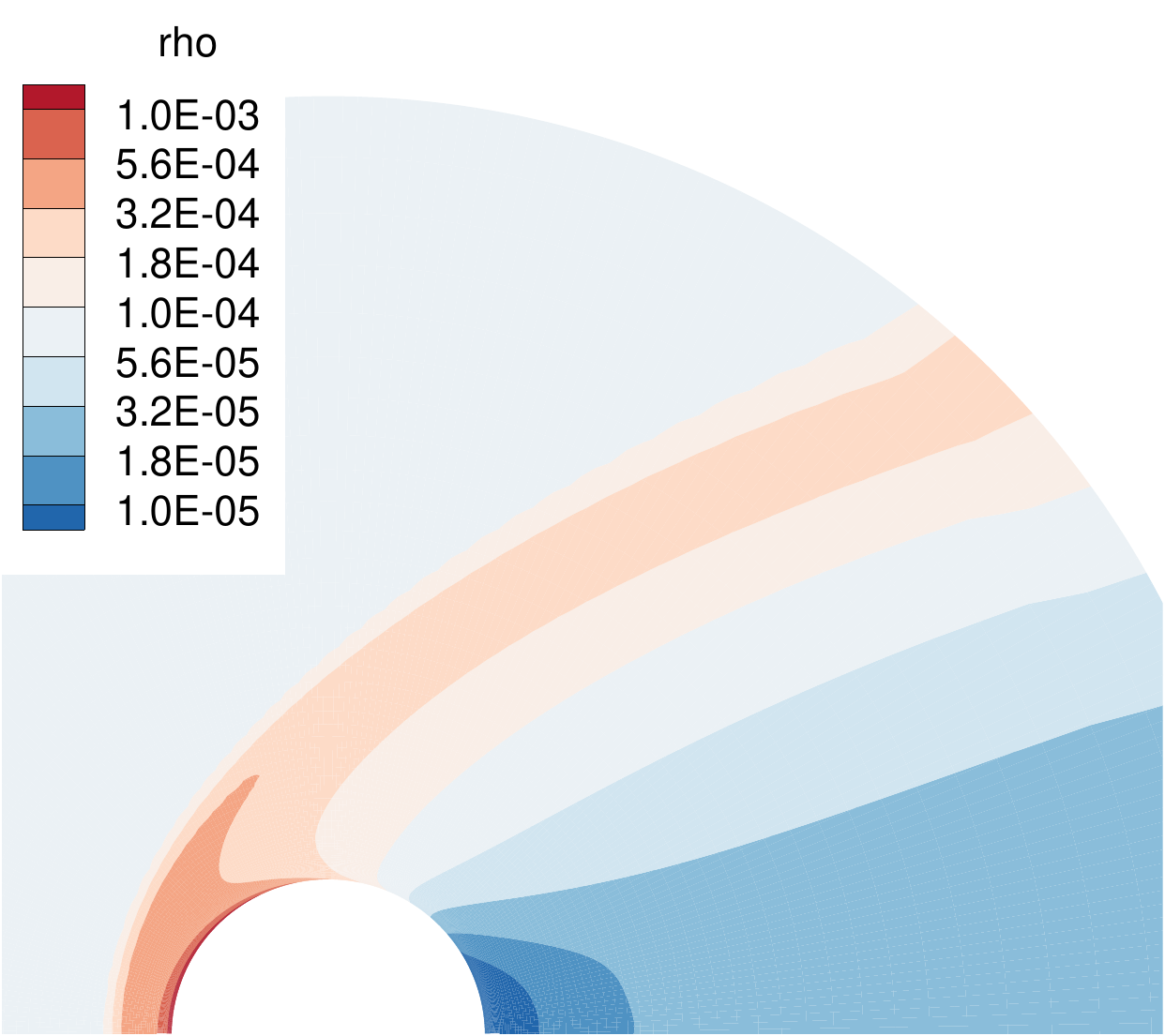}}
	\subfloat[]{\includegraphics[width=0.3\textwidth]
	{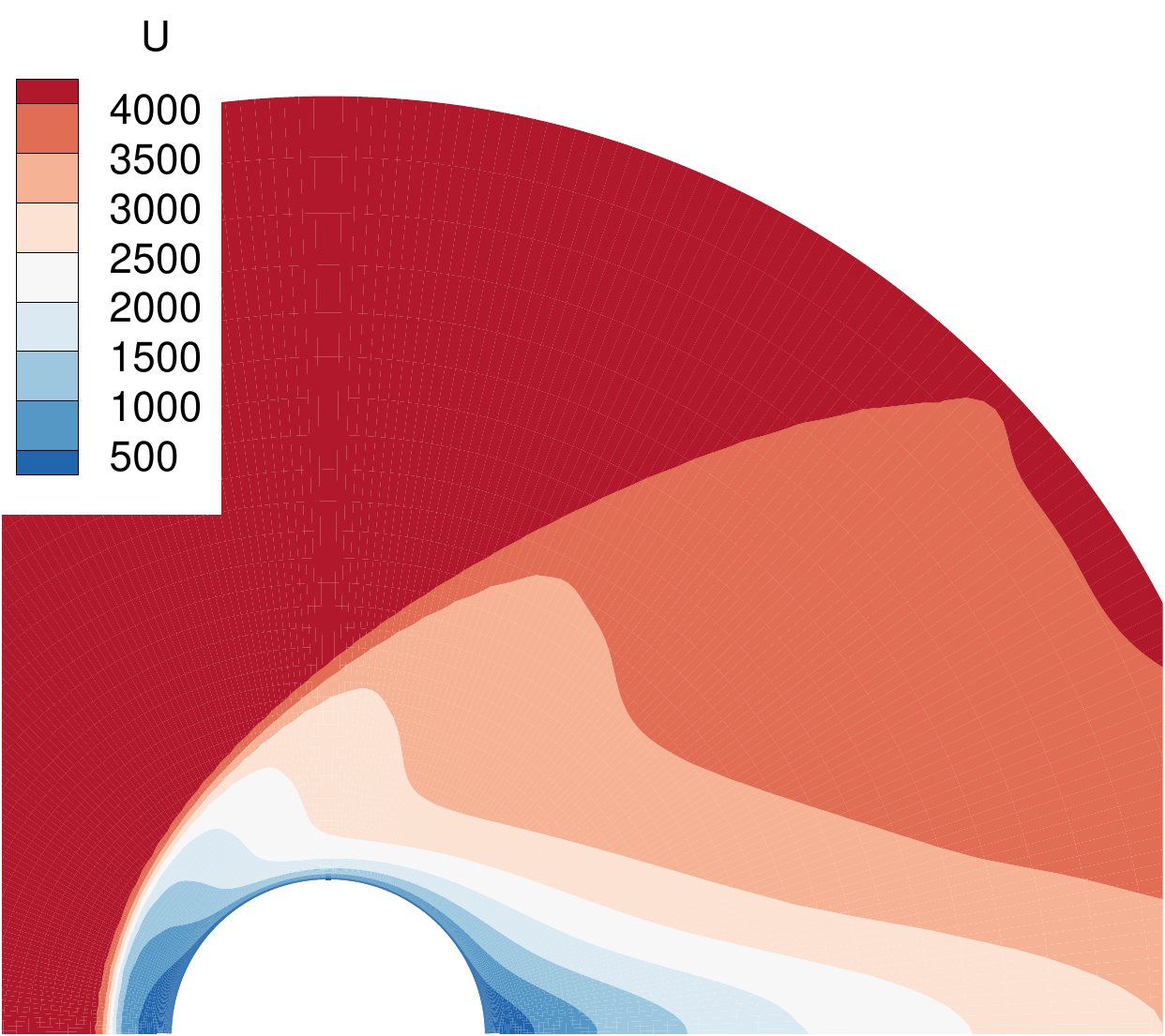}}
	\subfloat[]{\includegraphics[width=0.3\textwidth]
	{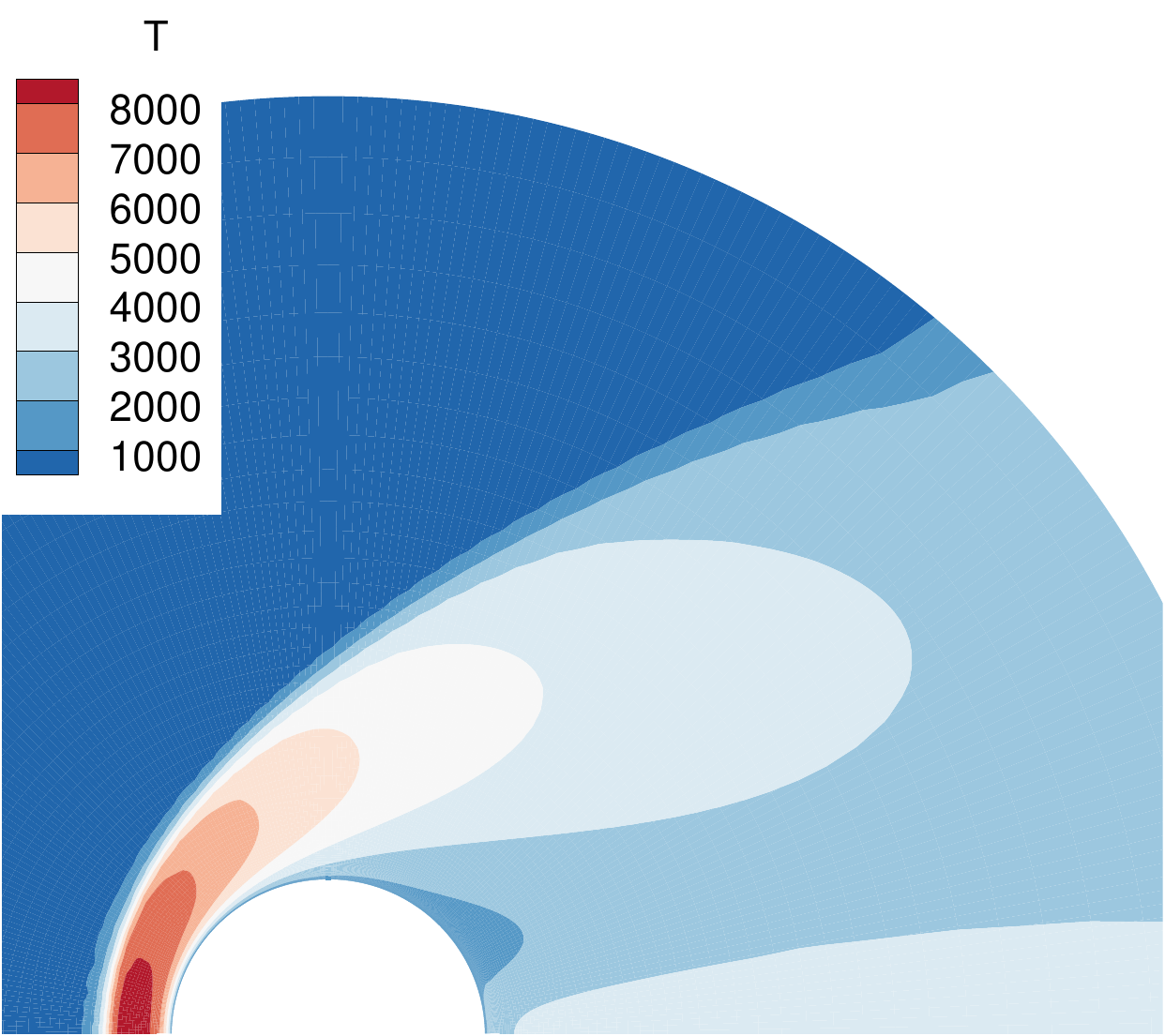}} \\
	\subfloat[]{\includegraphics[width=0.3\textwidth]
	{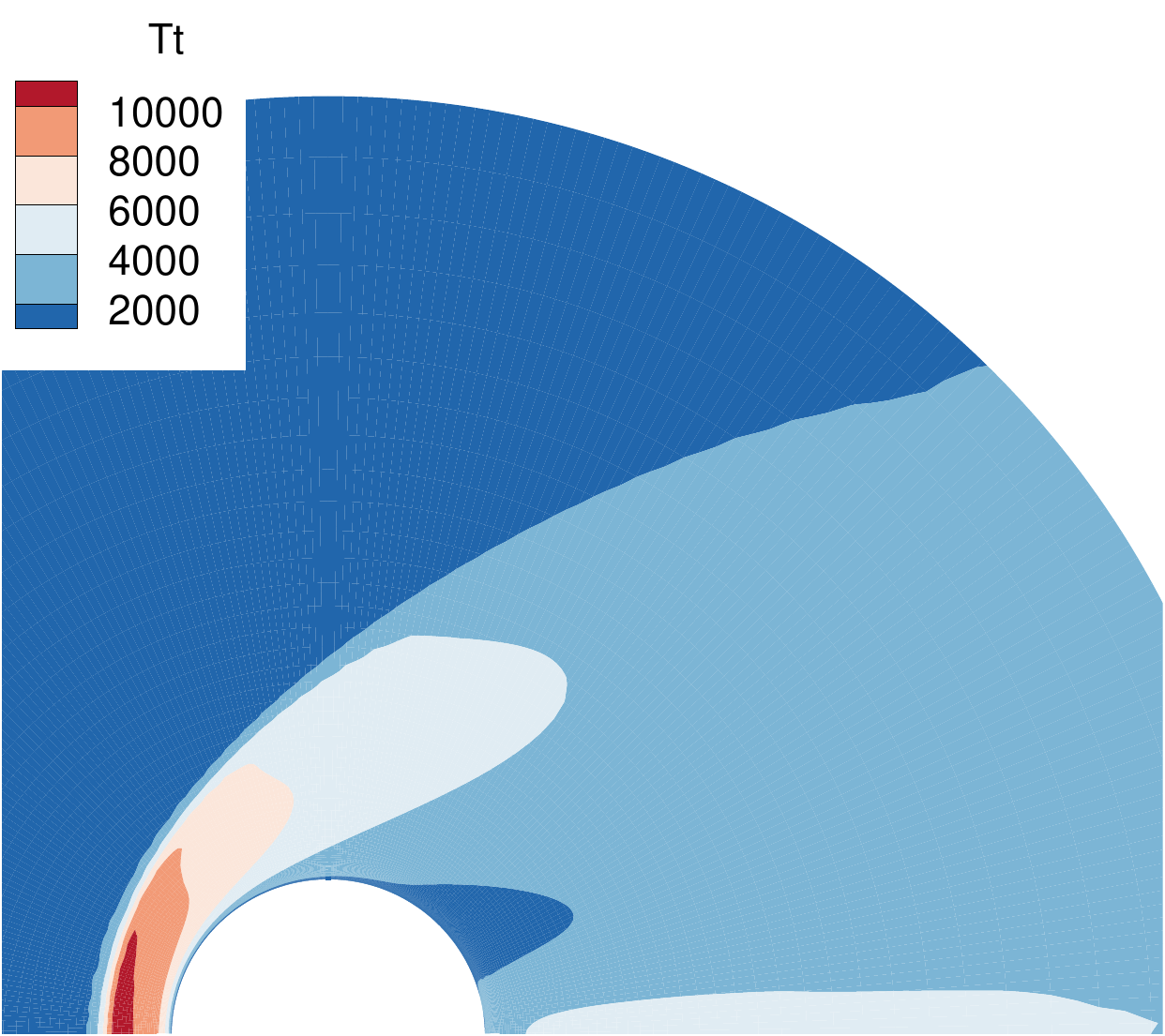}}
	\subfloat[]{\includegraphics[width=0.3\textwidth]
	{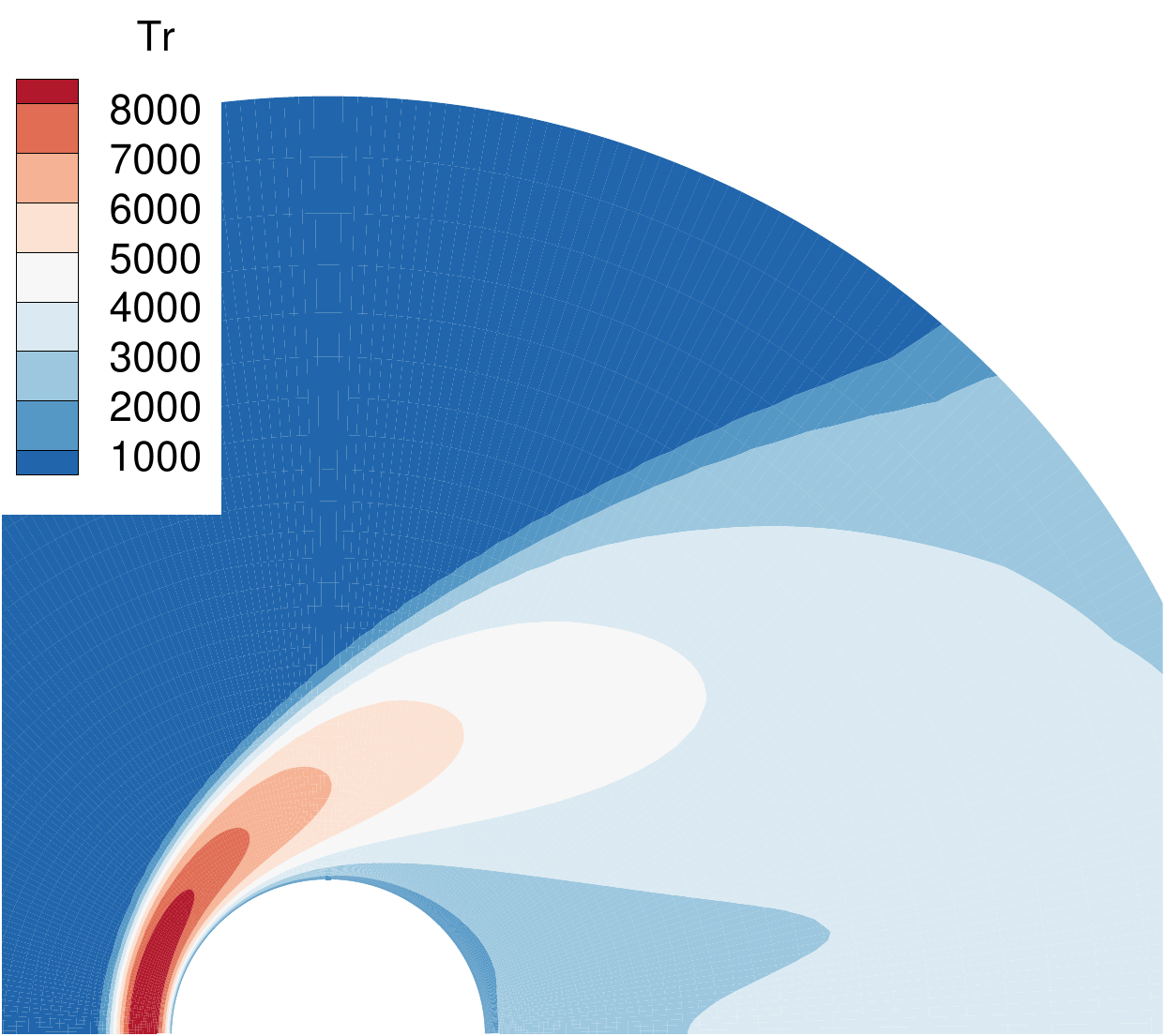}}
	\subfloat[]{\includegraphics[width=0.3\textwidth]
	{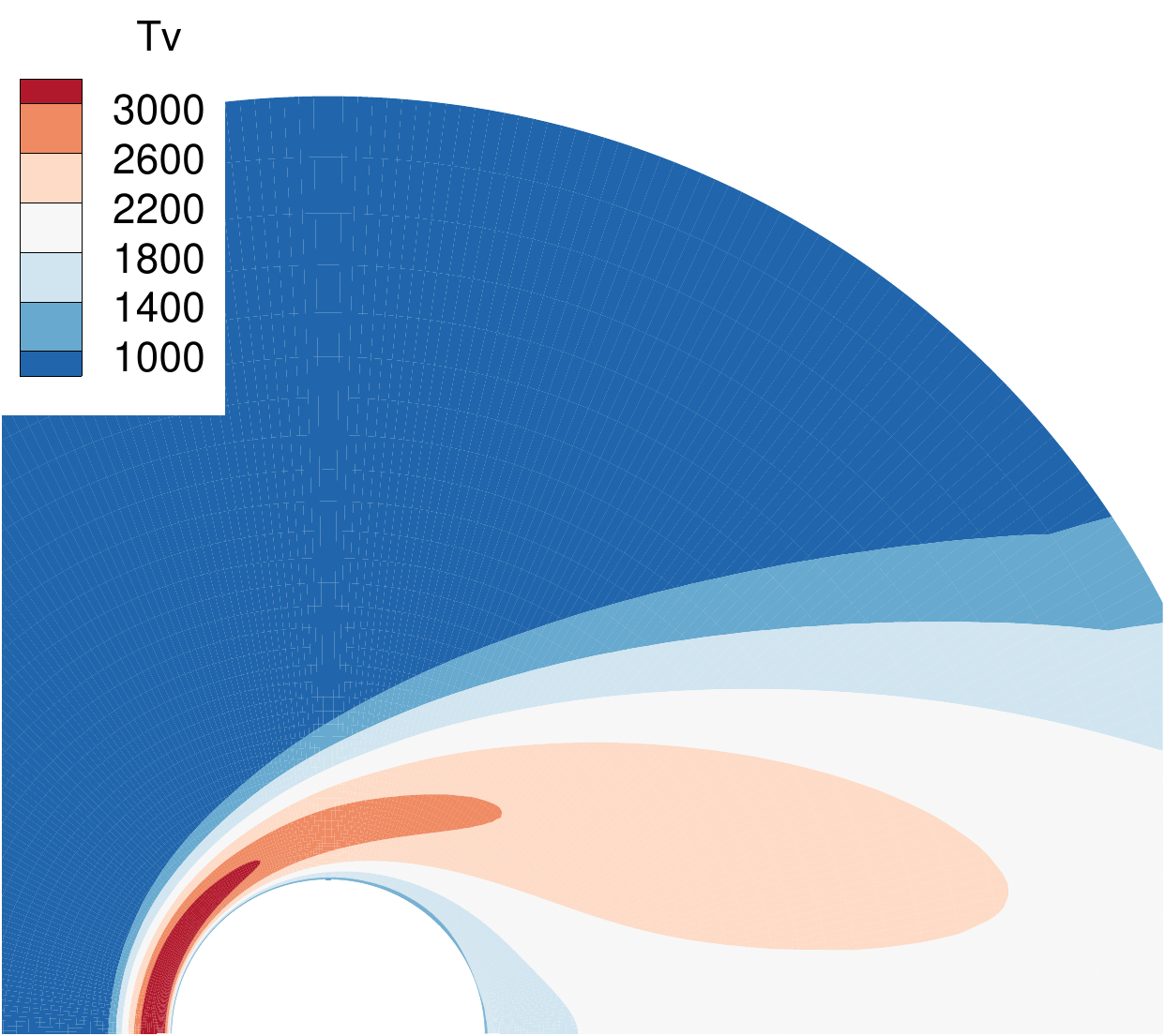}}
	\caption{Hypersonic flow at ${\rm Ma}_\infty = 15$ and ${\rm Kn}_\infty = 0.01$ around a half-circular
	cylinder by the AUGKS method.
	(a) Density, (b) $x$ direction velocity,
	(c) temperature, (d) translational temperature,
	(e) rotational temperature, and (f) vibrational
	temperature contours.}
	\label{fig:cylinder}
\end{figure}

\begin{figure}[H]
	\centering
	\includegraphics[width=7cm]{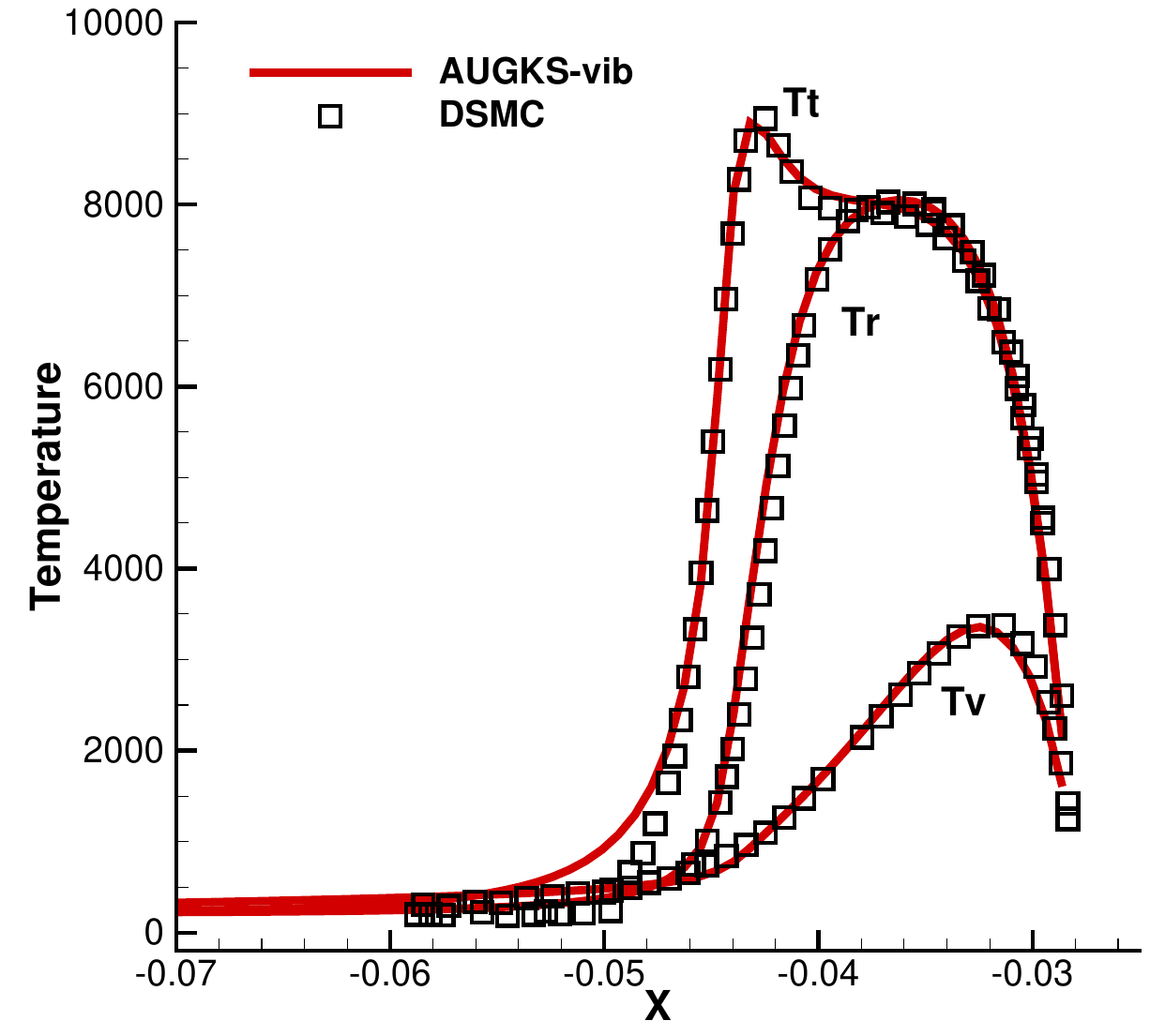}
	\caption{Hypersonic flow at ${\rm Ma}_\infty = 15$ and ${\rm Kn}_\infty = 0.01$ around a half-circular
		cylinder by the AUGKS method. Temperature distributions along the ${45^{\rm{\circ }}}$
	extraction line.}
	\label{fig:cylinder-Ma15-T}
\end{figure}

\begin{figure}[H]
	\centering
	\includegraphics[width=7cm]{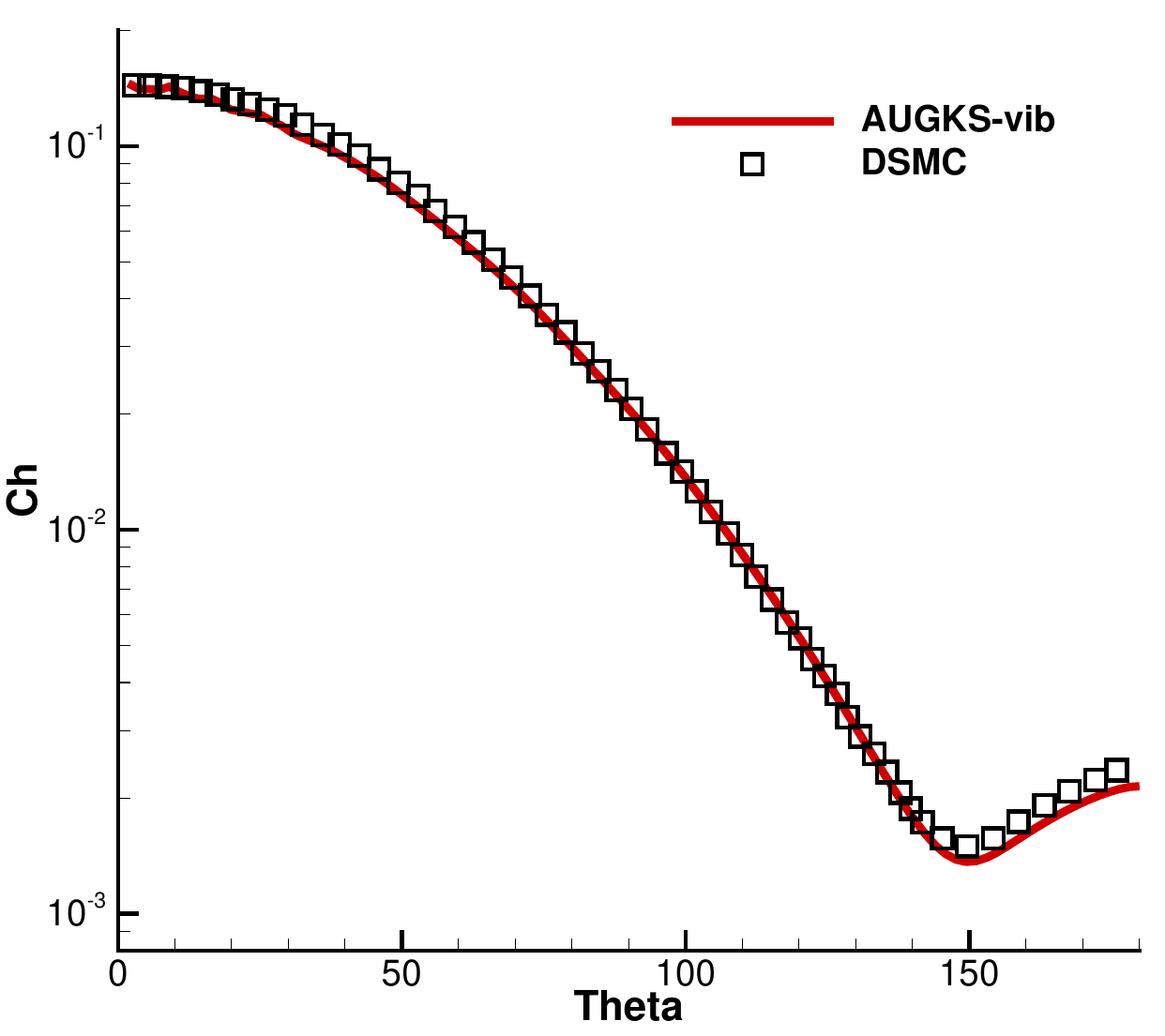}
	\caption{Hypersonic flow at ${\rm Ma}_\infty = 15$ and ${\rm Kn}_\infty = 0.01$ around a half-circular
		cylinder by the AUGKS method. Heat flux coefficient distributions at the cylinder surface.}
	\label{fig:cylinder-Ma15-Ch}
\end{figure}

To further verify the computational accuracy and robustness of the current scheme with a complex kinetic model, hypersonic flow passing over a semi-circular cylinder at a very large Mach number $15$ for ${\rm Kn}_\infty = 0.01$ is simulated. The diameter of the cylinder $D = 0.08$ m. The physical domain is discretized by $100 \times 160 \times 1$ quadrilateral cells. The structured discrete velocity space with a range $(-30\sqrt{RT_\infty}, 30 \sqrt{RT_\infty})$ consists of $89 \times 89 \times 1$ cells. The temperature of free stream gives $T_\infty = 217.5$ K, and the isothermal wall temperature is fixed at $T_w = 1000$ K. The rotational and vibrational collision numbers are evaluated by Eq.~\eqref{eq:Zr-tumuklu} and Eq.~\eqref{eq:Zv-tumuklu} with $Z_r^\infty = 12.5$ and $T^\ast = 91.5$ K. The CFL number is taken as 0.95. Fig.~\ref{fig:cylinder} plots the contours of flow field simulated by the AUGKS-vib method, where a fast approximation of initial flow field provided by $5000$ steps of GKS calculation \cite{xu2001} is adopted, and another $22000$ AUGKS-vib steps. The velocity space adaptation is illustrated in Fig.~\ref{fig:cylinder-Ma15-isDisc}. As the Mach number of the incoming flow increases, the nonequilibrium effects intensify, leading to a higher percentage (78.78\%) of the discrete velocity space used in the computational domain. Fig.~\ref{fig:cylinder-Ma15-T} shows the results from the AUGKS-vib and DSMC method for the translational, rotational, and vibrational temperatures extracted along the $45^{\circ}$ line in the upstream. The deviation of vibrational temperature from translational and rotational temperatures shows the strong thermodynamic non-equilibrium effects in high Mach number flow. By comparing the heat flux coefficient distribution with the DSMC method, as shown in Fig.~\ref{fig:cylinder-Ma15-Ch}, the accuracy of AUGKS-vib in aerodynamic heating has been further validated.

\subsection{Supersonic flow around a sphere}
\begin{figure}[H]
 	\centering
 	 	\subfloat[]{\includegraphics[width=0.4\textwidth]
 		{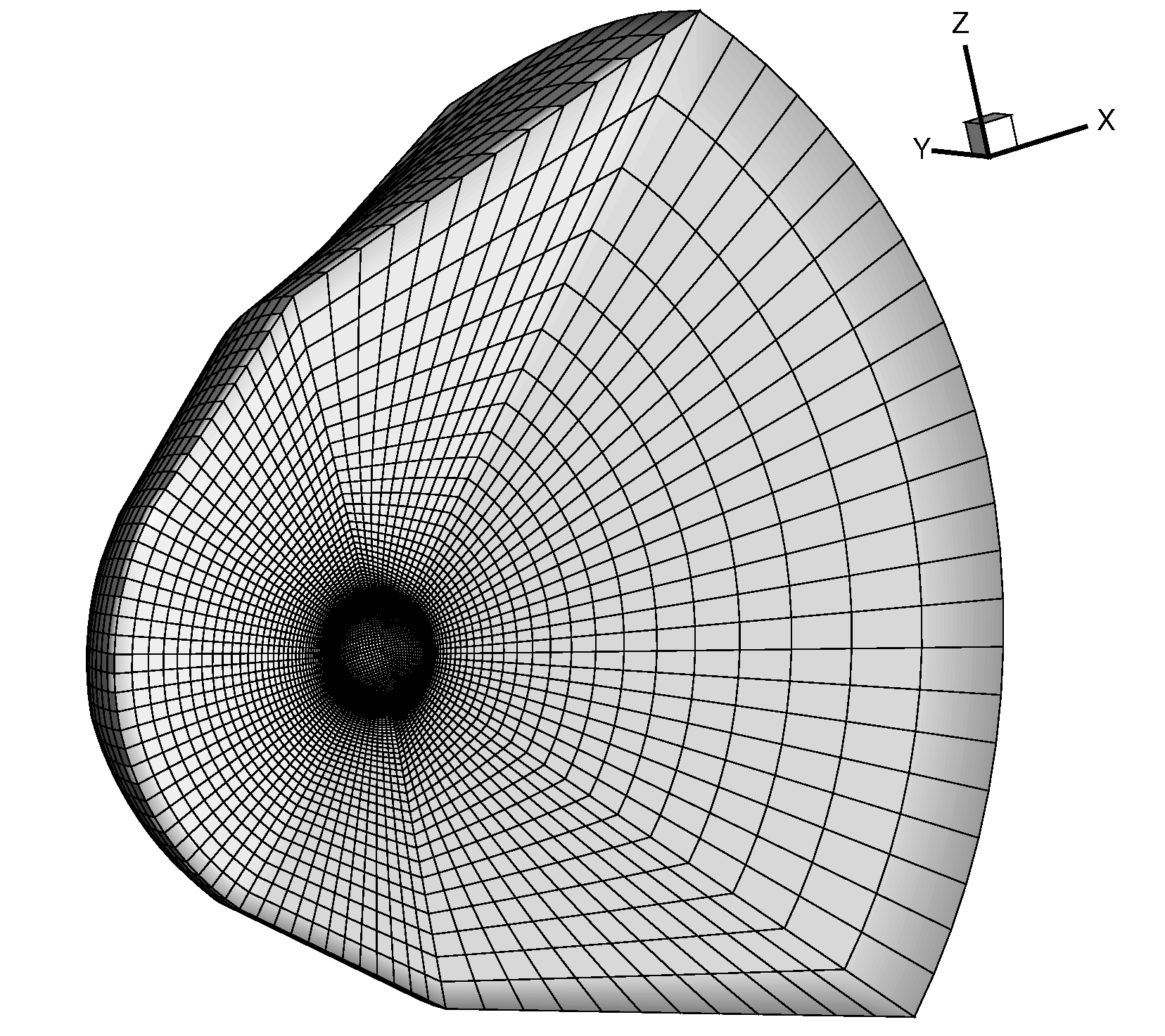}}
 	\subfloat[]{\includegraphics[width=0.4\textwidth]
 		{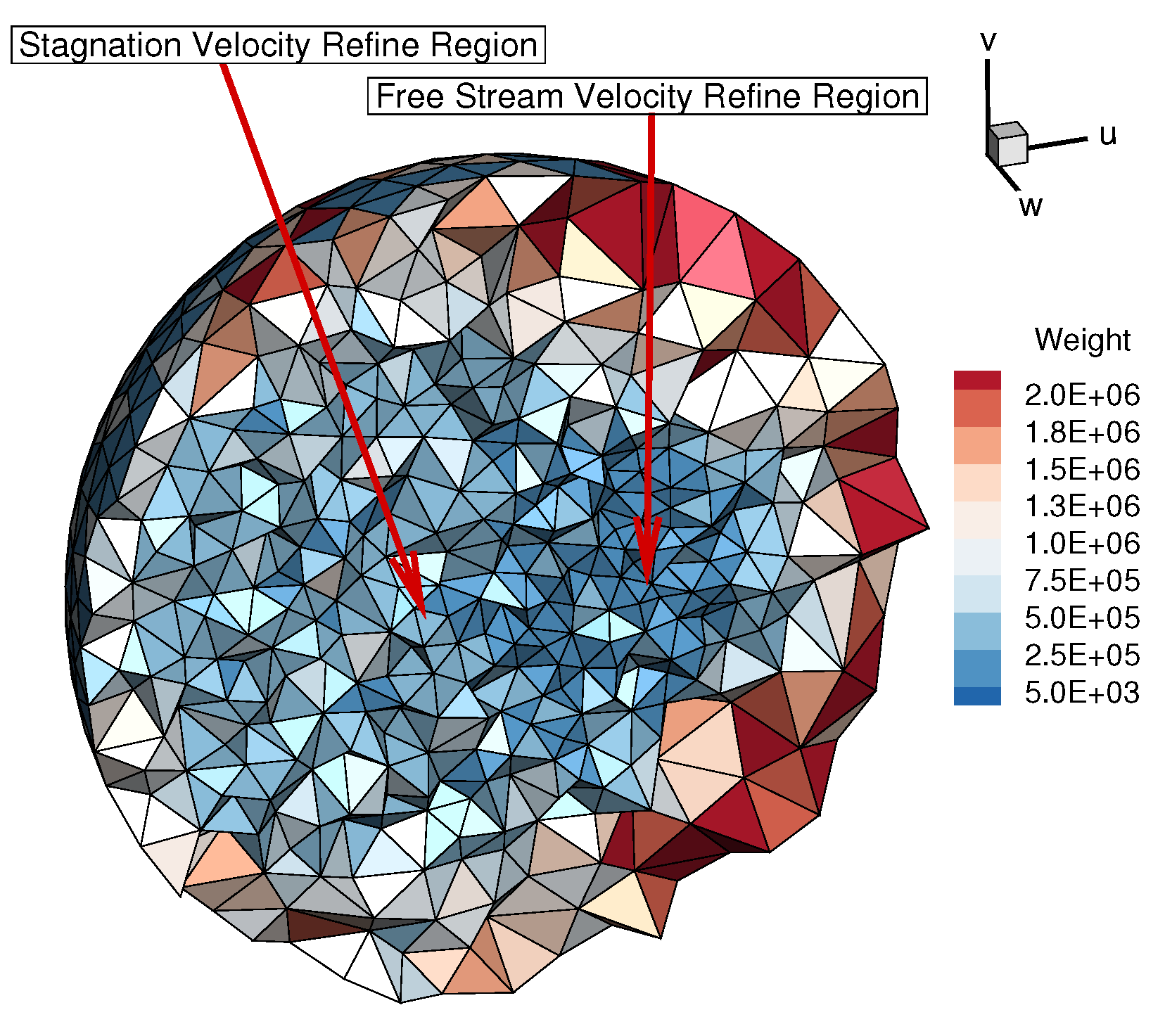}}
 	\caption{Supersonic flow at ${\rm Ma}_\infty = 4.25$ passing over sphere by the AUGKS method. (a) Physical mesh consisting 138,240 cells, and (b) unstructured discrete velocity space mesh consisting 18,802 cells.}
 	\label{fig:sphere-mesh}
\end{figure}

Supersonic flow passing over a three-dimensional sphere at Mach number $4.25$ for different ${\rm Kn}_\infty$ number is computed for nitrogen gas. To define the Knudsen number, the reference length is chosen as the diameter of the sphere, i.e., $D = 0.002$ m. The physical domain is discretized by $3,456 \times 40$ hexahedron cells, where the surface mesh is divided into 6 domains with $24 \times 24$ cells in each domain. Figure~\ref{fig:sphere-mesh} illustrated the section view of unstructured discrete velocity space mesh with 18,802 cells. The DVS mesh is discretized into a sphere with center coordinates at $0.4\times(U_\infty,V_\infty,W_\infty)$, and a radius of $6\sqrt{R T_s}$. To capture the non-equilibrium flow characteristics, the velocity space near the zero velocity point and free stream velocity point are refined within a spherical region of radius $r=3\sqrt{R T_w}$ and $r=3\sqrt{R T_\infty}$ respectively. Table ~\ref{table:spherecondition} gives the initial conditions of free stream, i.e., $\rho_\infty$ and $T_\infty$ under different ${\rm Kn}_\infty$, and the temperature of isothermal wall boundary condition $T_w$. The rotational and vibrational collision numbers are constant values of $Z_r = 3.5$ and $Z_v = 10$.

\begin{table}[H]
	\caption{Free stream flow parameters of supersonic flow at ${\rm Ma}_\infty=4.25$ around a sphere.}
	\centering
	\begin{tabular}{ccccc}
		\hline
		 ${\rm Kn}_\infty$ & Altitude, km & $\rho_\infty$, kg/($\rm{m}^3 \cdot \rm{s}$) & $T_\infty$, K & $T_w$, K \\
		\hline
		0.672 & 89.1 & $3.173\times 10^{-5}$ & $65.04$ & $302$  \\
		0.338 & 83.3 & $6.313\times 10^{-5}$ & $65.04$ & $302$  \\
		0.080 & 71.1 & $2.675\times 10^{-4}$ & $65.04$ & $302$  \\
		0.031 & 63.2 & $6.879\times 10^{-4}$ & $65.04$ & $302$  \\
		0.0031& 41.5 & $6.879\times 10^{-3}$ & $65.04$ & $302$  \\
		\hline
	\end{tabular}
	\label{table:spherecondition}
\end{table}
The drag coefficient computed by AUGKS-vib is compared in Table~\ref{table:spherecd}, with those obtained from experiment (Air) \cite{wendtJF} and UGKS calculation without vibrational model \cite{jiang2019implicit}. Accurate results have been obtained with all relative errors less than 2.5\%.

To further verify the computational accuracy in terms of aerodynamic heating and force, the pressure, shear stress, and heat flux coefficients distribution are compared with DSMC method for diatomic gas \cite{zhang2023conservative} under ${\rm Kn}_\infty = 0.031$ by setting ${\rm Pr} = 0.69$, $\sigma = 0.5$ in AUGKS-vib with the non-dimensionalized surface coefficients
\begin{equation*}
	C_p     = \dfrac{p_s-p_\infty} {\frac12 \rho_\infty U_\infty^2},~
	C_\tau  = \dfrac{f_s}{\frac12 \rho_\infty U_\infty^2},~
	C_h     = \dfrac{h_s}{\frac12 \rho_\infty U_\infty^3},
\end{equation*}
where velocity $U_\infty$ can be calculated by free-stream Mach number ${\rm Ma}_\infty$, $p_s$ is the surface pressure, $p_{\infty}$ is the pressure in freestream flow, $f_s$ is the surface friction and $h_s$ is the surface heat flux.

\begin{figure}[H]
	\centering
	\subfloat[]{\includegraphics[width=0.3\textwidth]
		{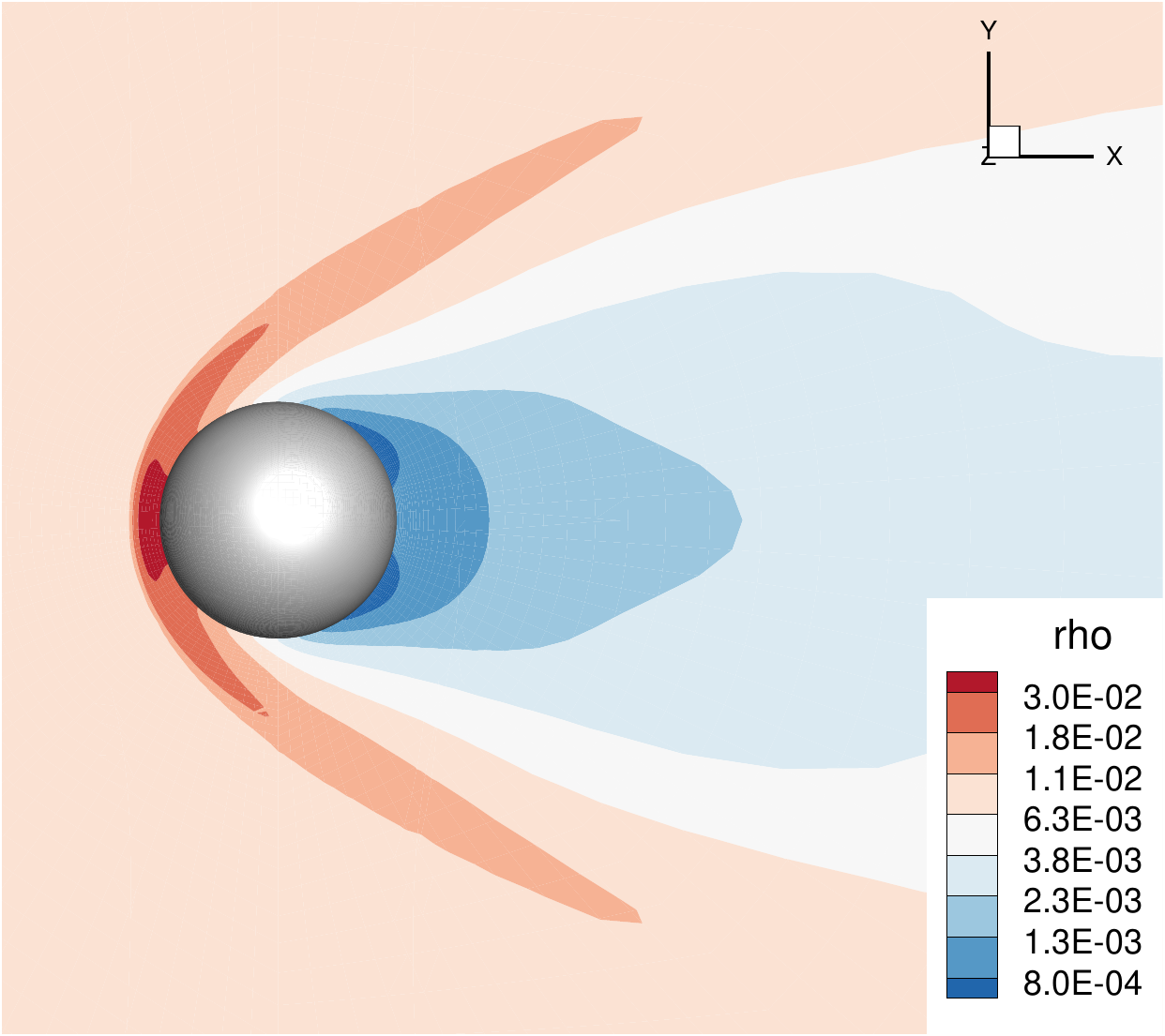}}
	\subfloat[]{\includegraphics[width=0.3\textwidth]
		{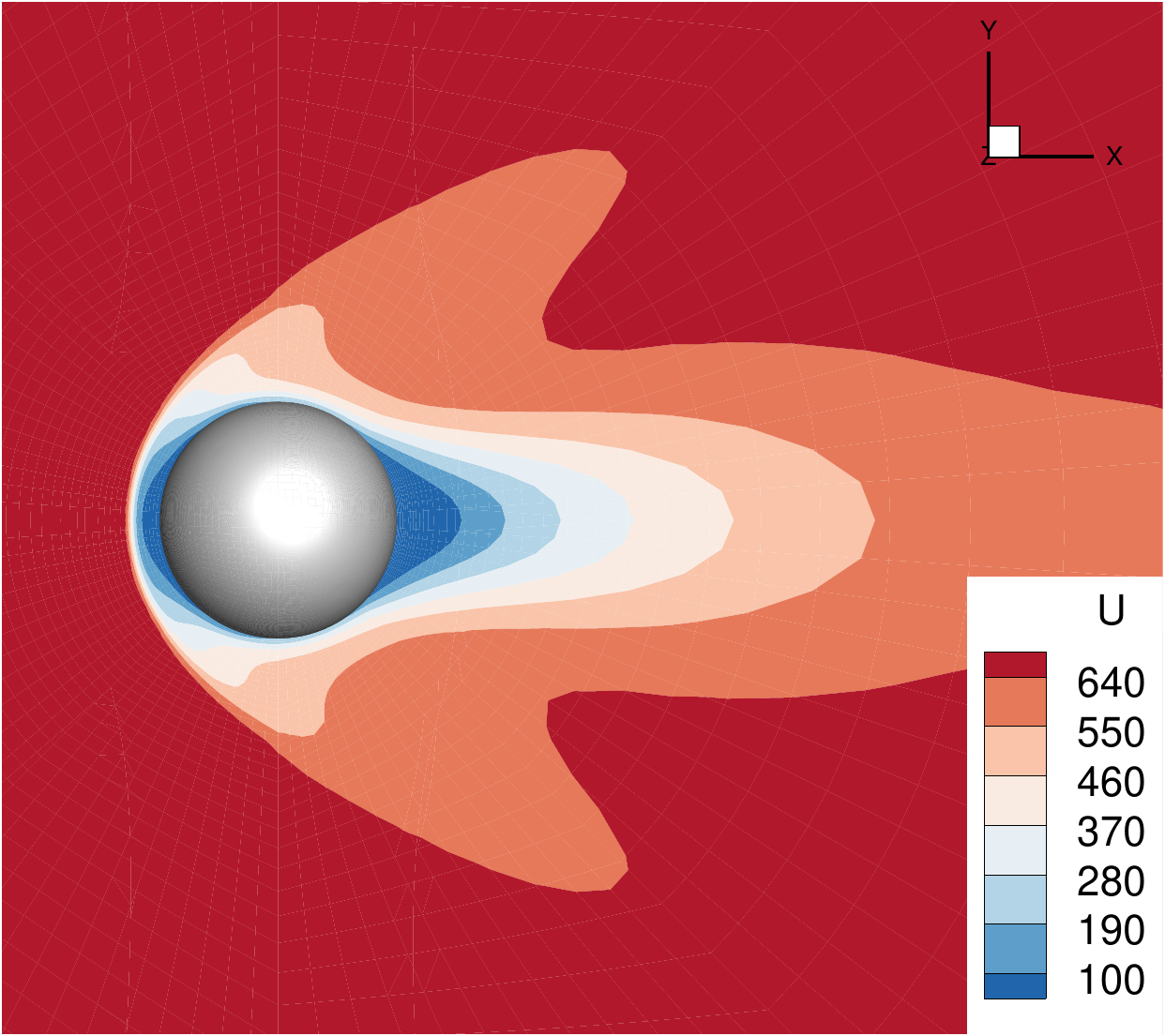}}
	\subfloat[]{\includegraphics[width=0.3\textwidth]
		{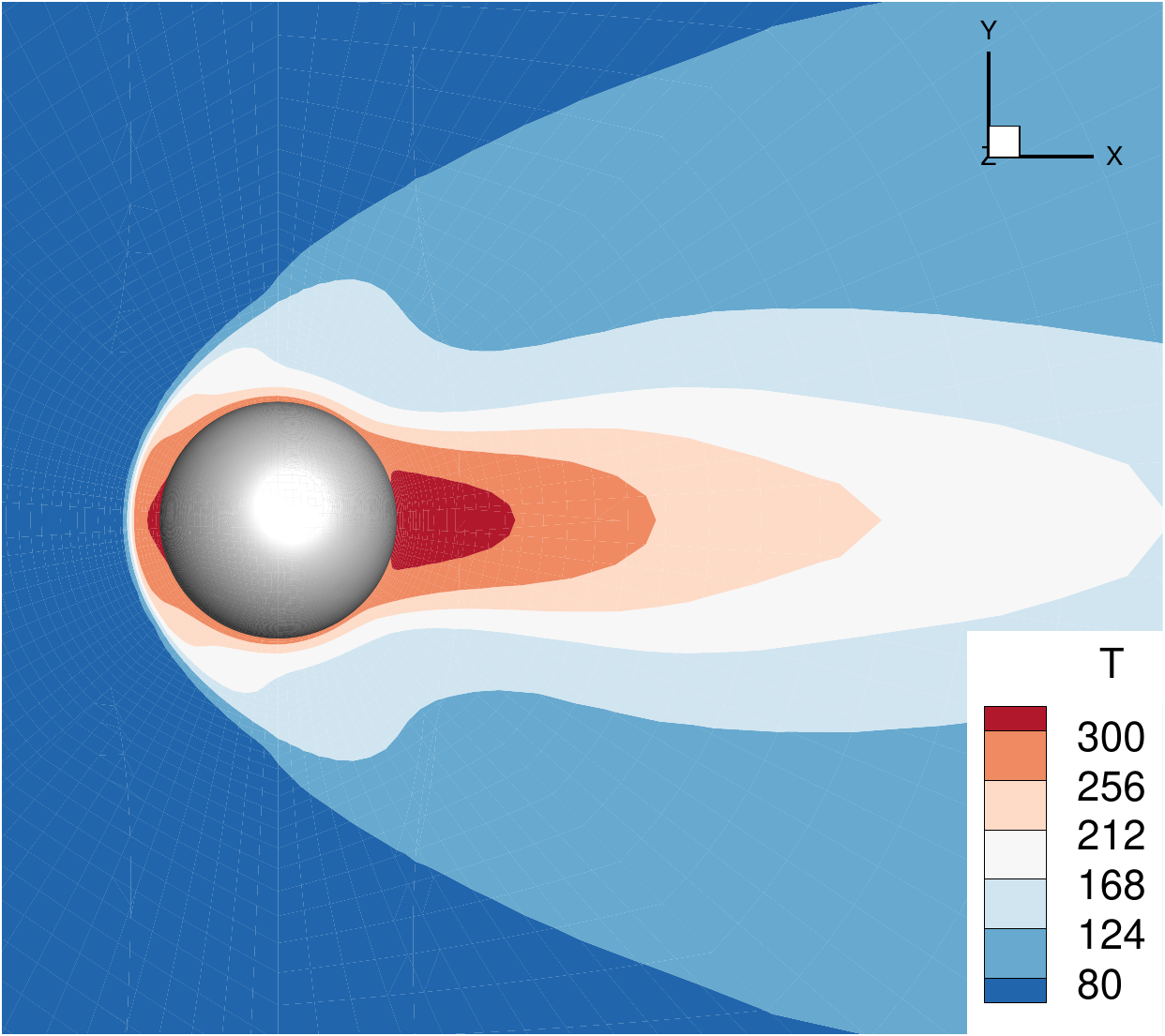}} \\
	\subfloat[]{\includegraphics[width=0.3\textwidth]
		{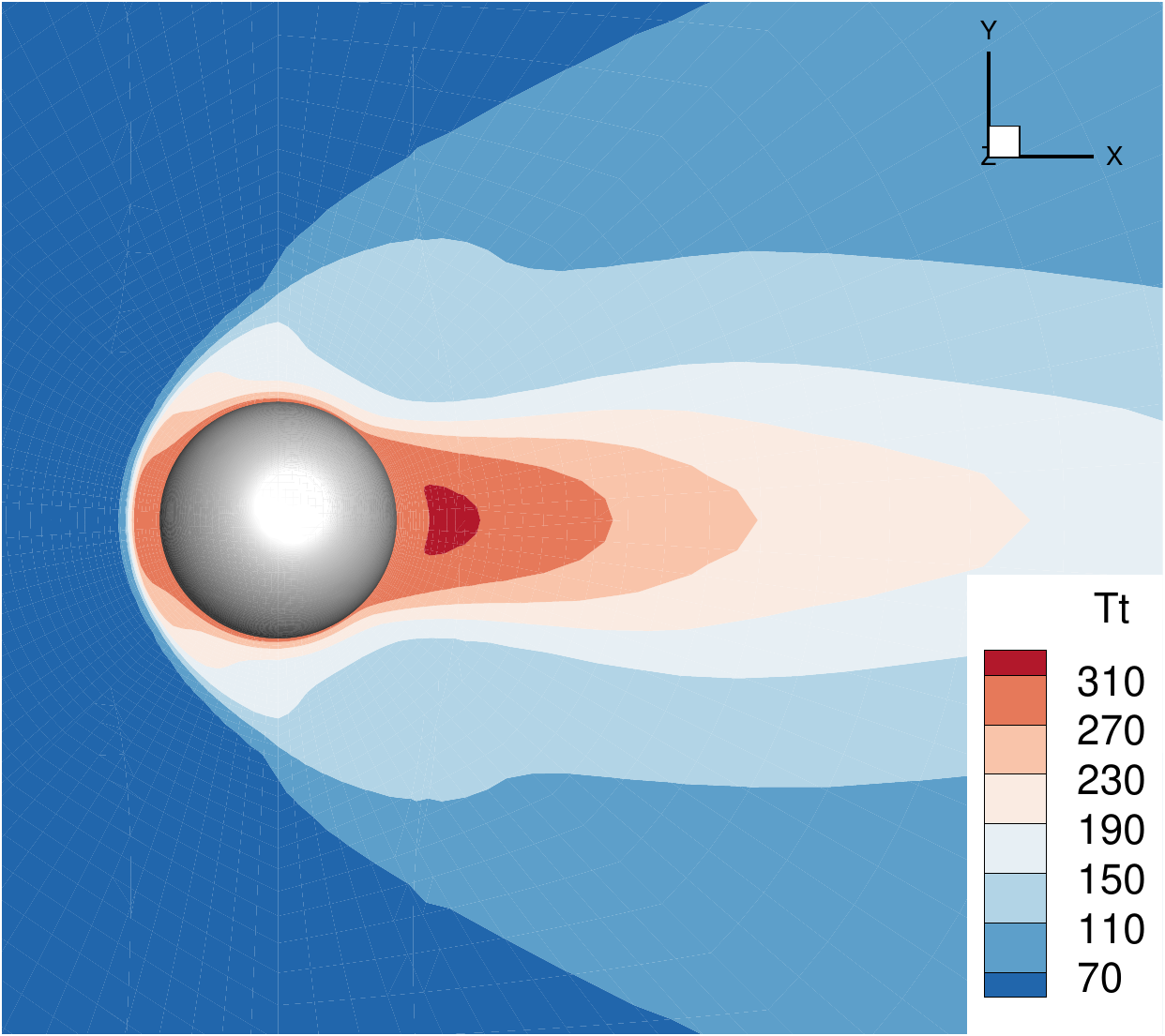}}
	\subfloat[]{\includegraphics[width=0.3\textwidth]
		{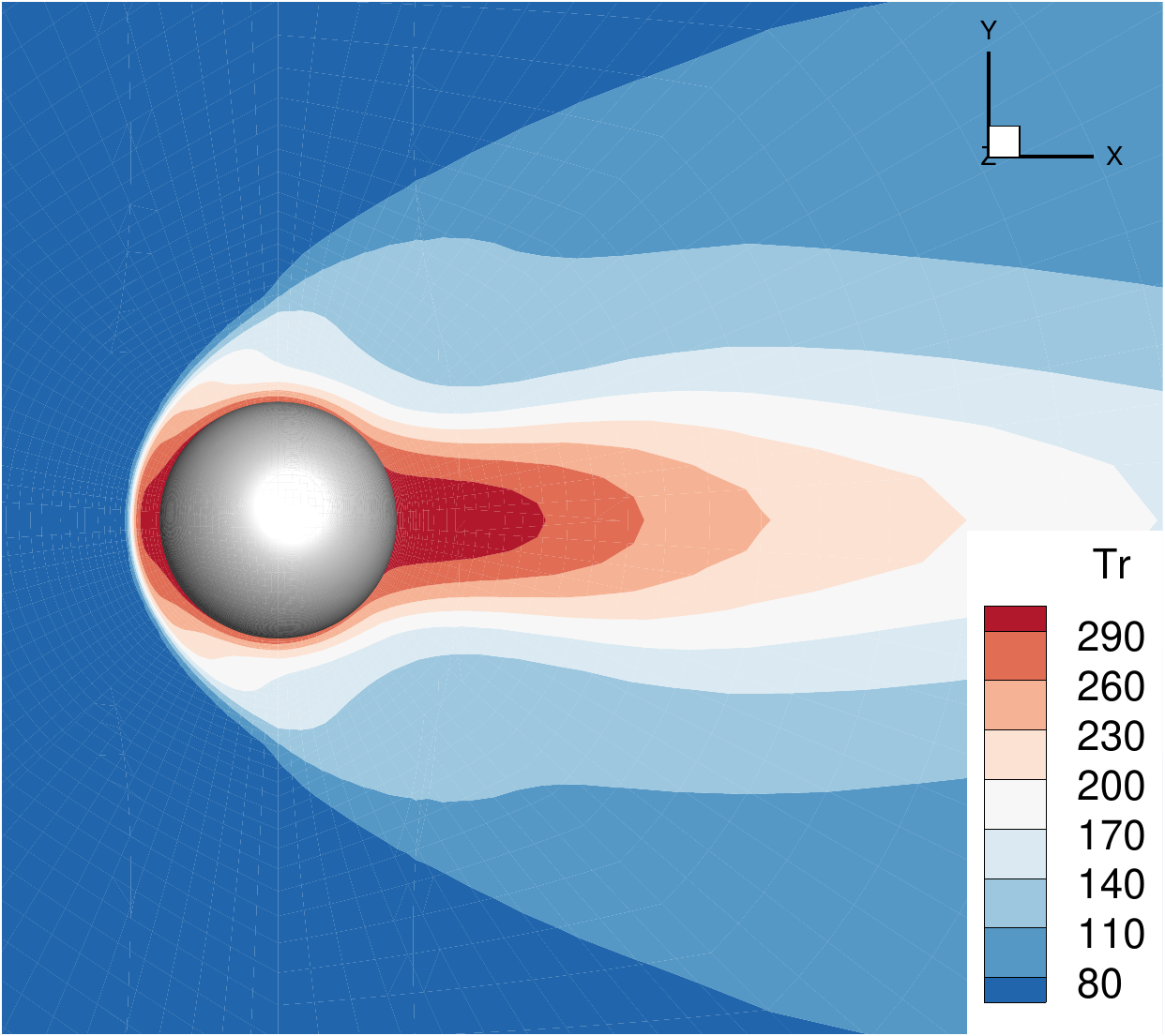}}
	\subfloat[]{\includegraphics[width=0.3\textwidth]
		{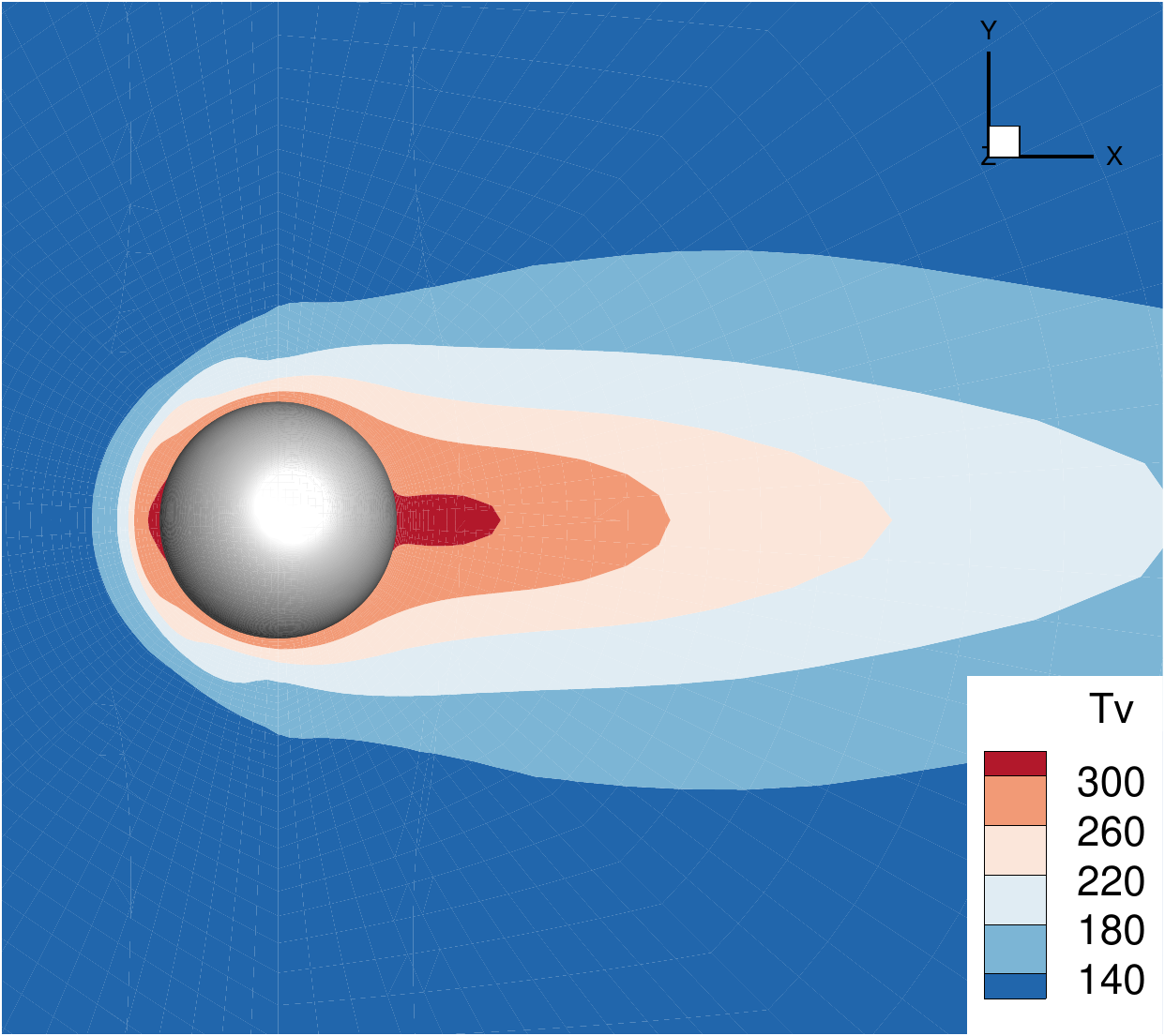}} \\
	\caption{Supersonic flow
		around a sphere at ${\rm Ma}_\infty = 4.25$ for ${\rm Kn}_\infty = 0.0031$. (a) Density, (b) $x$ direction velocity,
		(c) temperature, (d) translational temperature,
		(e) rotational temperature and (f) vibrational
		temperature contours.}
	\label{fig:sphere-Ma4.25}
\end{figure}

\begin{figure}[H]
	\centering
	\includegraphics[width=0.4\textwidth]
		{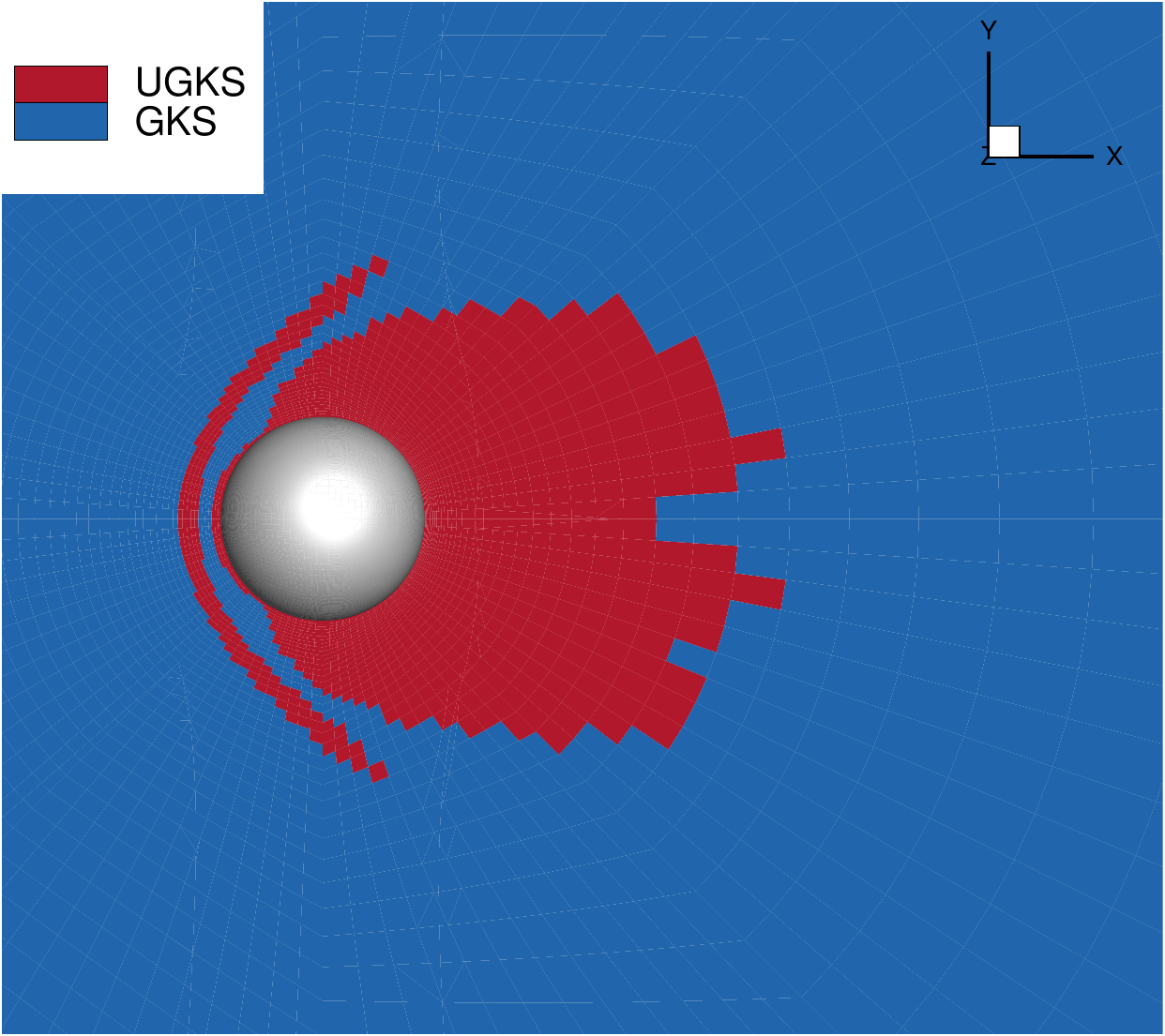}
	\caption{Supersonic flow around a sphere at ${\rm Ma}_\infty = 4.25$ for ${\rm Kn}_\infty = 0.0031$. Distributions of velocity space adaptation with $C_t = 0.01$ where the discretized velocity space (UGKS) is used in 43.75\% of physical domain.}
	\label{fig:sphere-isDisc}
\end{figure}

Figure~\ref{fig:sphere-Ma4.25} plots the contours of density, $x$ direction velocity, temperature, translational, rotational and vibrational temperature simulated by the AUGKS-vib under ${\rm Kn}_\infty = 0.0031$. Figure~\ref{fig:sphere-isDisc} illustrates the distribution of velocity space adaptation where 43.75\% of the computational domain is simulated by discretized velocity space (UGKS).

\begin{table}[H]
	\caption{Drag coefficients of supersonic flow around a sphere at ${\rm Ma_\infty} = 4.25$ by the AUGKS method.}
	\centering
	\begin{tabular}{ccccc}
		\toprule
		\multirow{2}{*}{${\rm Ma}_\infty$} & \multirow{2}{*}{${\rm Kn}_\infty$} & \multicolumn{3}{c}{Drag Coefficient (Error)} \\
		\cline{3-5} &  &
		\begin{tabular}[c]{@{}c@{}}Experiment (Air)\end{tabular} &
		\begin{tabular}[c]{@{}c@{}}UGKS      (${\rm N}_2$)\end{tabular} &
		\begin{tabular}[c]{@{}c@{}}AUGKS-vib (${\rm N}_2$)\end{tabular}
		\\ \midrule
		4.25 & 0.672 & 2.42 & 2.356 (-2.64\%) & 2.443 (1.22\%)  \\
		4.25 & 0.338 & 2.12 & 2.101 (-0.87\%) & 2.161 (2.14\%)	\\
		4.25 & 0.080  & 1.53 & 1.558 (1.80\%) & 1.527 (-0.14\%) \\
		4.25 & 0.031 & 1.35 & 1.355 (0.39\%) & 1.348 (-0.12\%)\\
		4.25 & 0.0031 & - &  - & {1.162} (-)\\
		\bottomrule
	\end{tabular}
	\label{table:spherecd}
\end{table}

\begin{figure}[H]
	\centering
\subfloat[]{\includegraphics[width=0.33\textwidth]
	{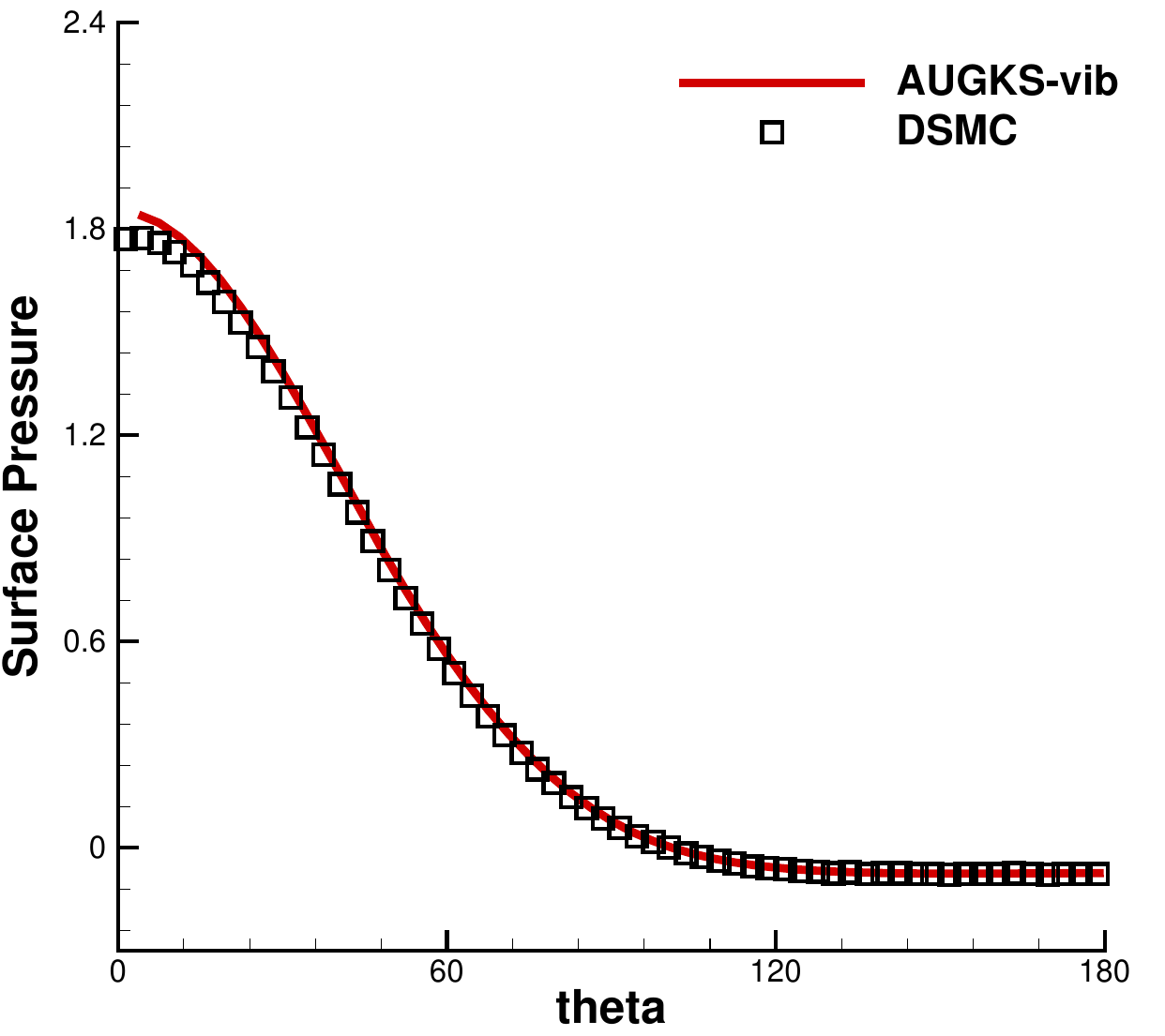}}
\subfloat[]{\includegraphics[width=0.33\textwidth]
	{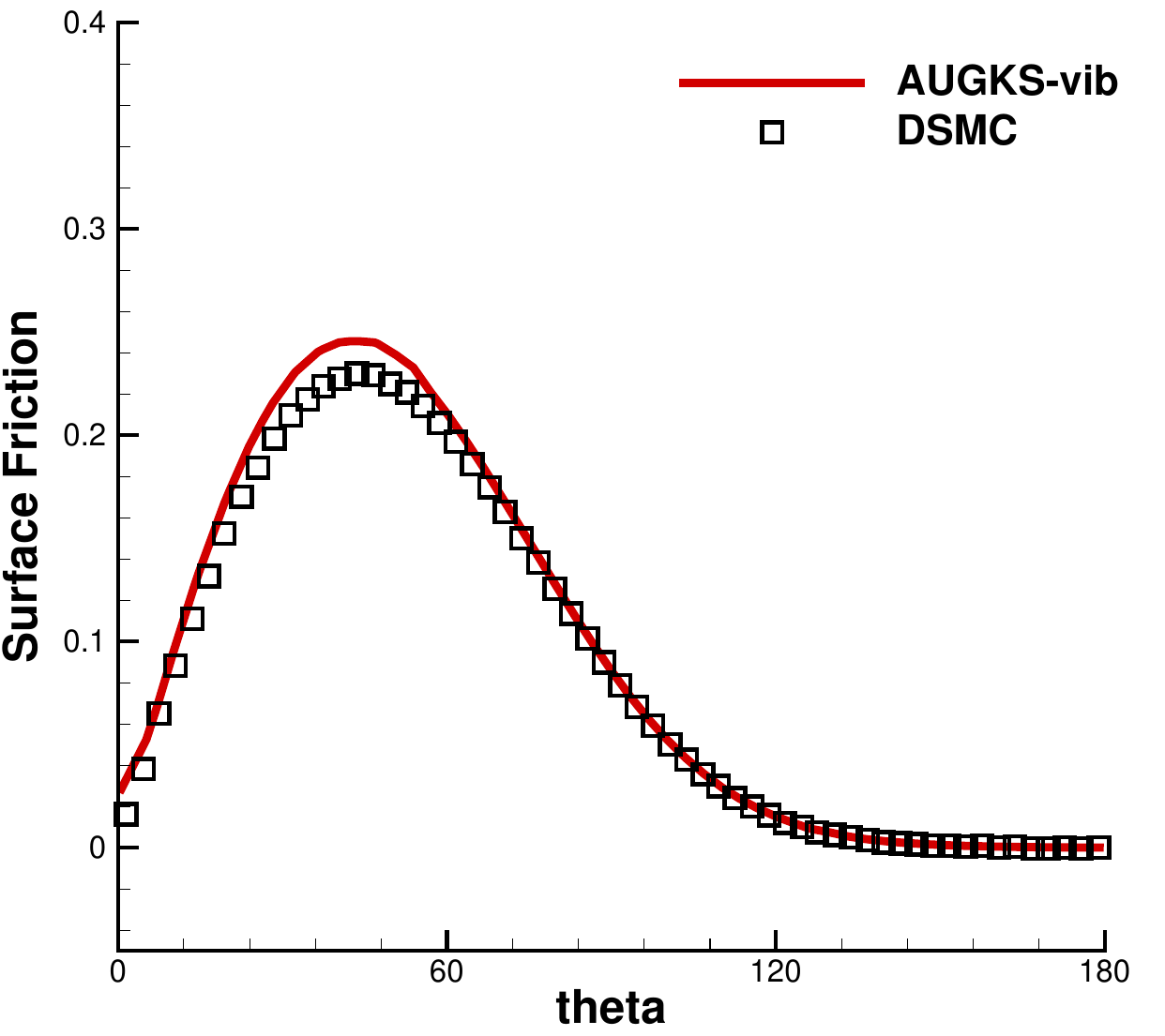}}
\subfloat[]{\includegraphics[width=0.33\textwidth]
	{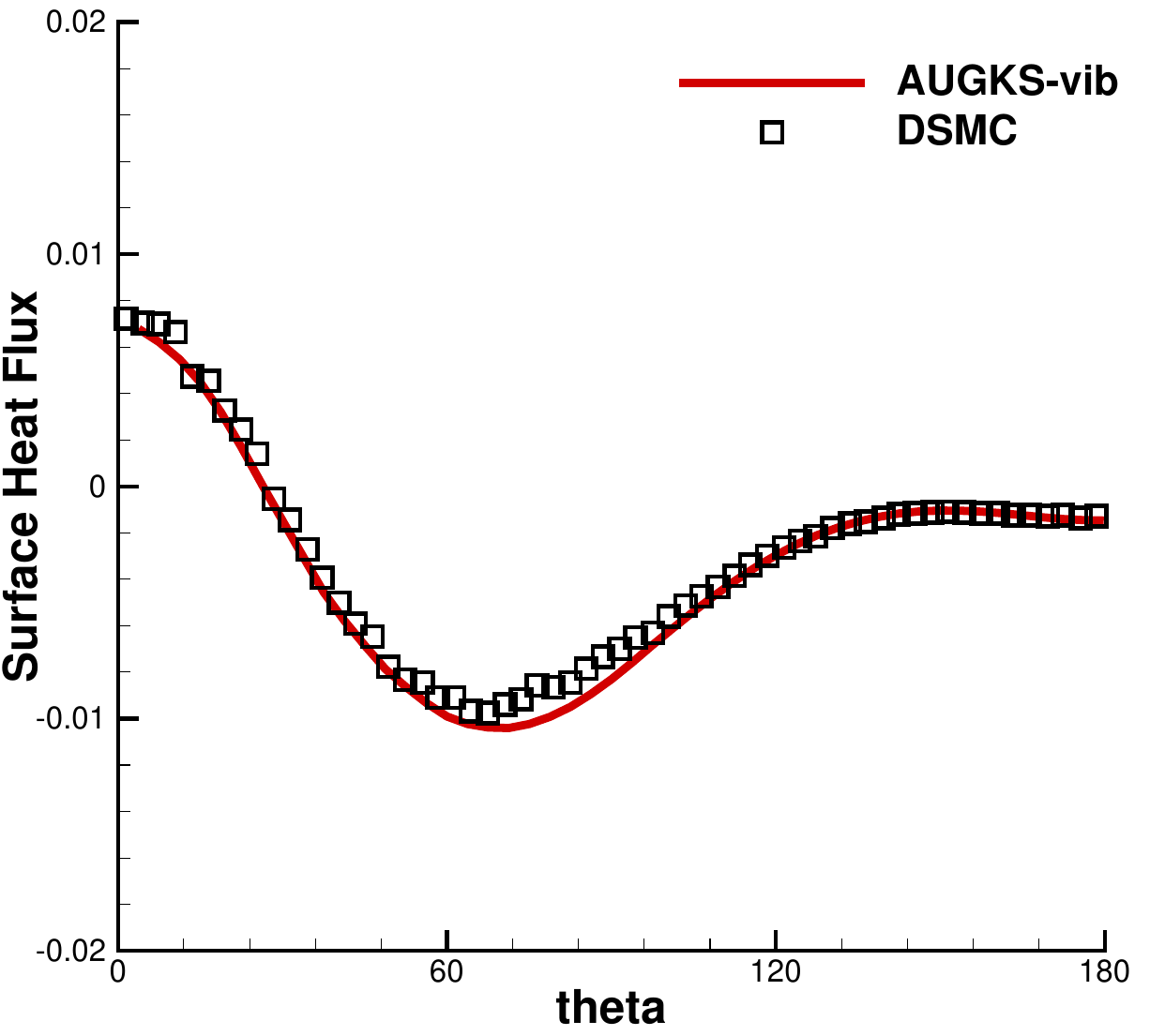}}\\
\caption{Surface quantities of supersonic flow
	around a sphere at ${\rm Ma}_\infty = 4.25$ for ${\rm Kn}_\infty = 0.031$ by the AUGKS method. (a) Pressure coefficient, (b) shear stress coefficient,
	and (c) heat flux coefficient.}
\label{fig:spheresurface}
\end{figure}

Table~\ref{table:spheretime} shows the computational efficiency and resource consumption. All the simulations are conducted on the SUGON computation platform with a CPU model of 7285 32C 2.0GHz.
\begin{table}[H]
	\caption{The computational cost for simulations of the supersonic flow around a sphere at ${\rm Ma_\infty} = 4.25$ by the AUGKS method. The physical domain consists of 138,240 cells, and the unstructured DVS mesh is discretized by 18,802 cells.}
	\centering
	\begin{threeparttable}
		\begin{tabular}{cccc}
			\hline
			${\rm Kn}_\infty$ & Cores & Steps & Wall Clock Time, h   \\
			\hline
			$0.672$ & 1920 & $6400\tnote{1}+1600$ & 6.60 \\
			$0.338$ & 1920 & $6400\tnote{1}+1600$ & 6.66 \\
			$0.080$ & 1920 & $4000\tnote{1}+6000$ & 14.82 \\
			$0.031$ & 1920 & $4000\tnote{1}+3500$ & 9.97 \\
			$0.0031$ & 1920 & $2500\tnote{1}+3100$ & 7.96 \\
			\hline
		\end{tabular}
		
		\begin{tablenotes}
			\item[1] Steps of first order GKS simulations.
		\end{tablenotes}
	\end{threeparttable}
	\label{table:spheretime}
\end{table}

\subsection{Hypersonic flow around a X38-like space vehicle}
Hypersonic flow at ${\rm Ma}_\infty = 8.0$ passing over a X38-like space vehicle for ${\rm Kn}_\infty = 0.0025$ at angles of attack of ${\rm AoA} = 0^\circ$ is simulated. According to the particle mean free path and the normal size of the space vehicle (5 m), the above Knudsen number corresponds to the flight at 78.6 km altitude. At the hypersonic speed, all flow regimes can emerge at different parts of the flying vehicle. This case can test the efficiency and capability of AUGKS-vib for simulating three-dimensional hypersonic flow over complex geometry in the transition regime.

\begin{figure}[H]
	\centering
	\subfloat[]{\includegraphics[height=3.5cm]
		{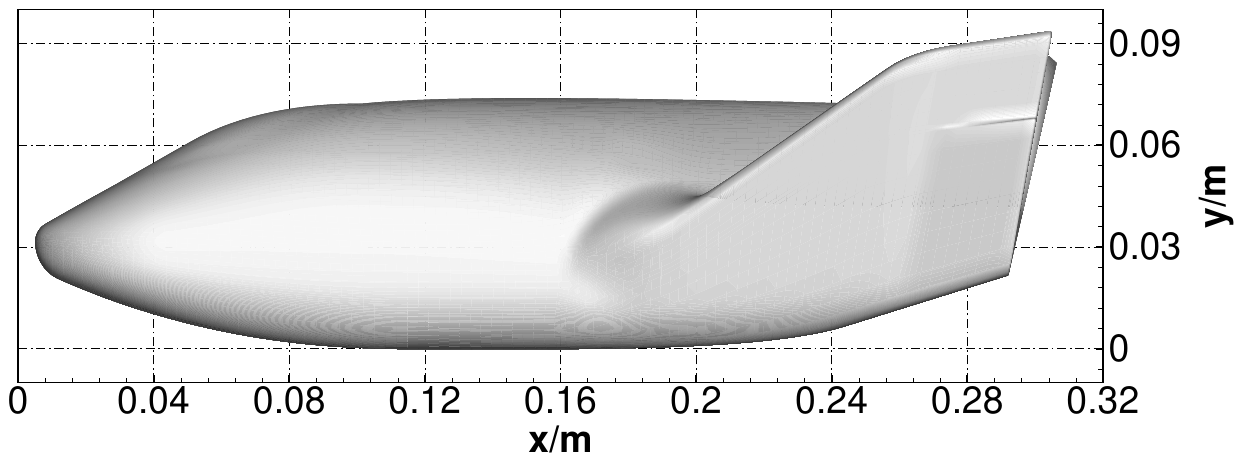}} \\
	\subfloat[]{\includegraphics[height=3.5cm]
		{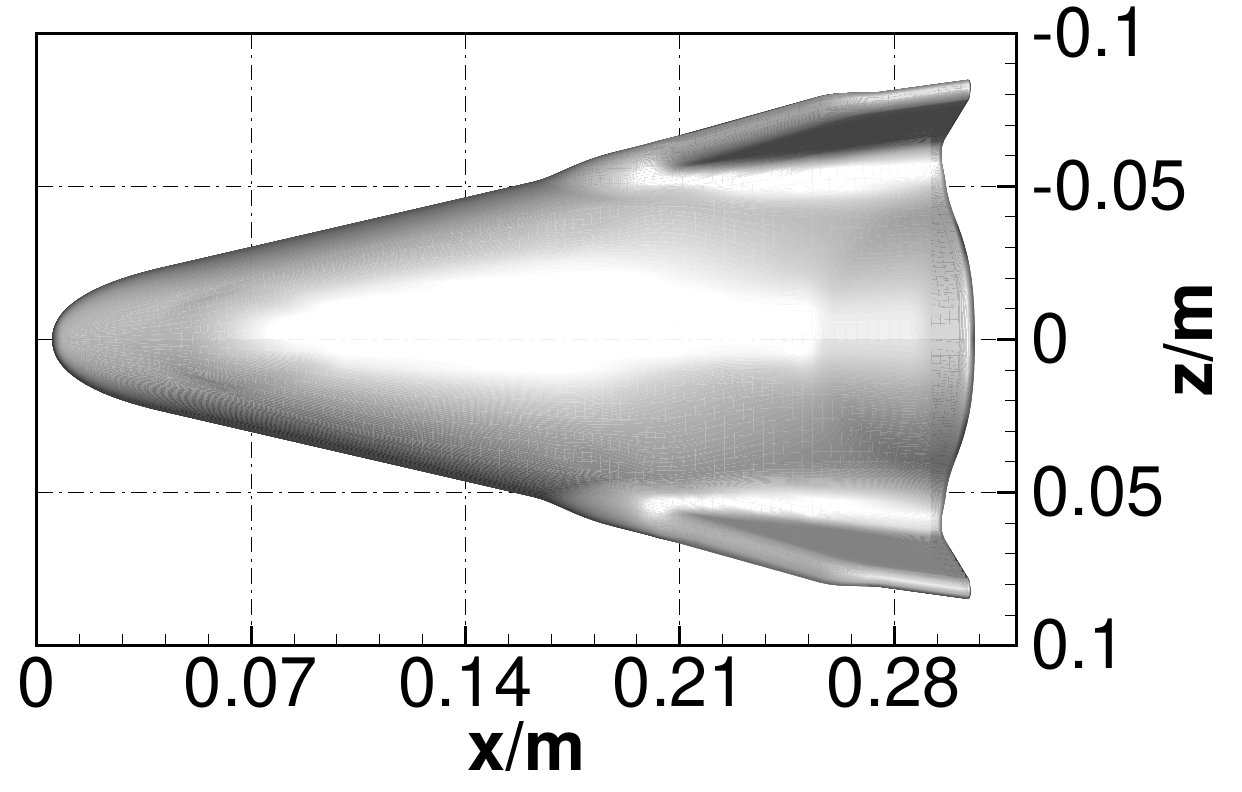}}
	\subfloat[]{\includegraphics[height=3.5cm]
		{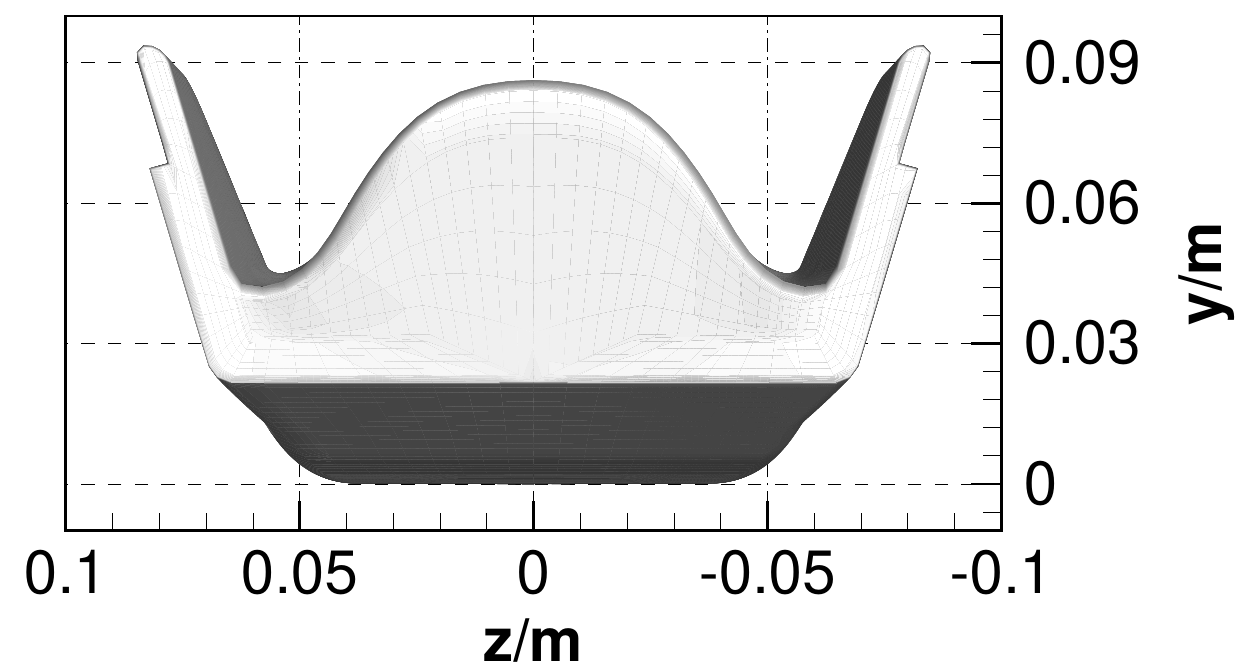}}
	\caption{Sketch of the X38-like space vehicle.}
	\label{fig:x38-geo}
\end{figure}

\begin{figure}[H]
	\centering
	\includegraphics[width=0.45\textwidth]{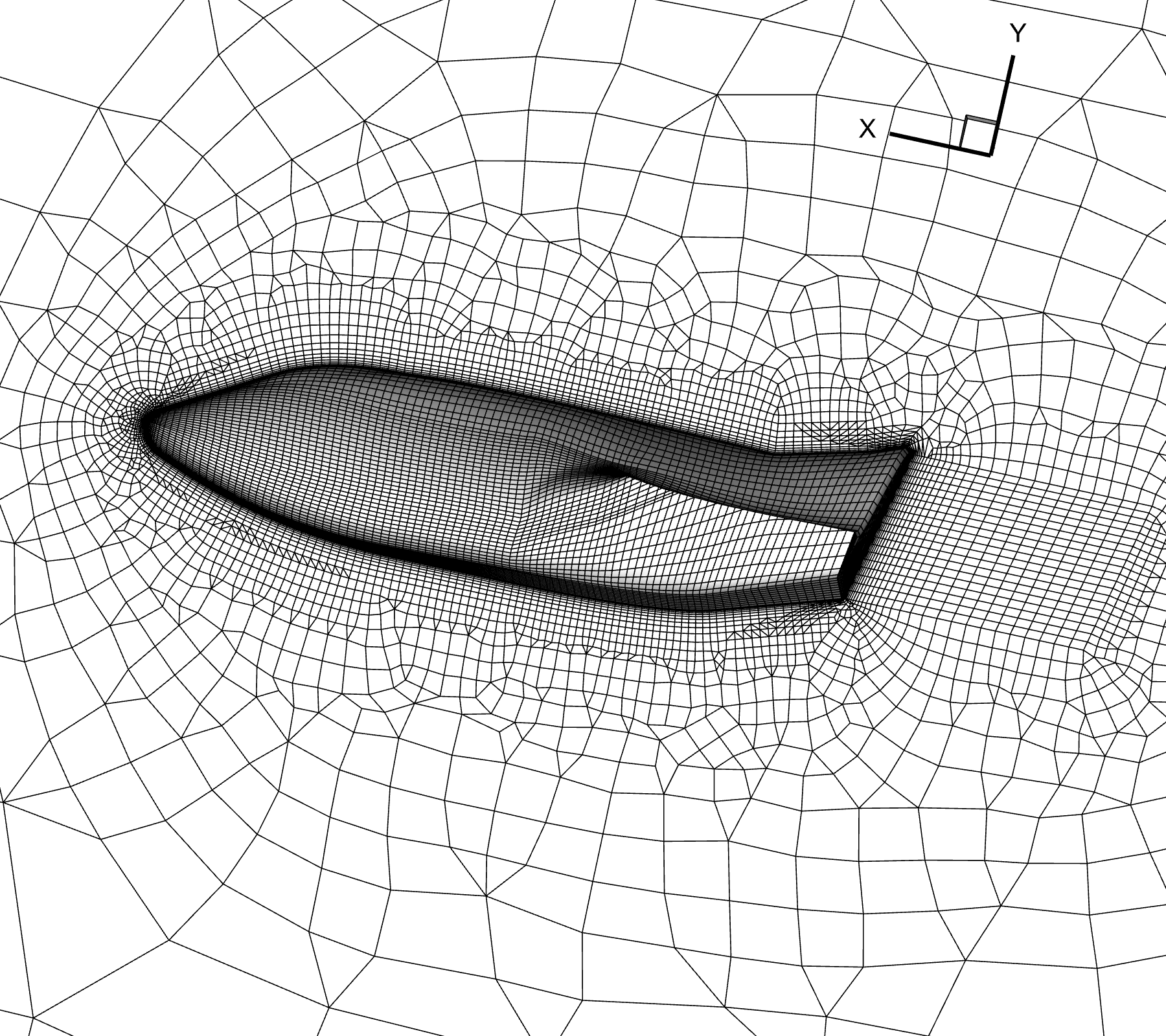}
	\caption{Three-dimensional meshes of hypersonic flow at ${\rm Ma}_\infty = 8.0$ around a X38-like space vehicle. The computational domain consists of 246,558 cells.}
	\label{fig:x38-mesh}
\end{figure}

The sketch of the space vehicle is depicted in Fig.~\ref{fig:x38-geo}. The reference length for the definition of the Knudsen number is $L_{ref} = 0.28$ m. The unstructured symmetry mesh used is shown in Fig.~\ref{fig:x38-mesh}, consisting of 246,558 cells with the minimum cell height of $L_{ref} \times 10^{-4}$ near the front of the vehicle surface. The structured discrete velocity space with a range $(-U_\infty - 15 \sqrt{ 2 R T_1}, U_\infty + 5 \sqrt{2 R T_\infty})$ consists of $33\times33\times33$ velocity points. The free-stream temperature is $T_\infty = 56$ K, and the vehicle surface is treated as an isothermal wall with a constant temperature $T_w = 300$ K.

\begin{figure}[H]
	\centering
\subfloat[]{\includegraphics[width=0.33\textwidth]
	{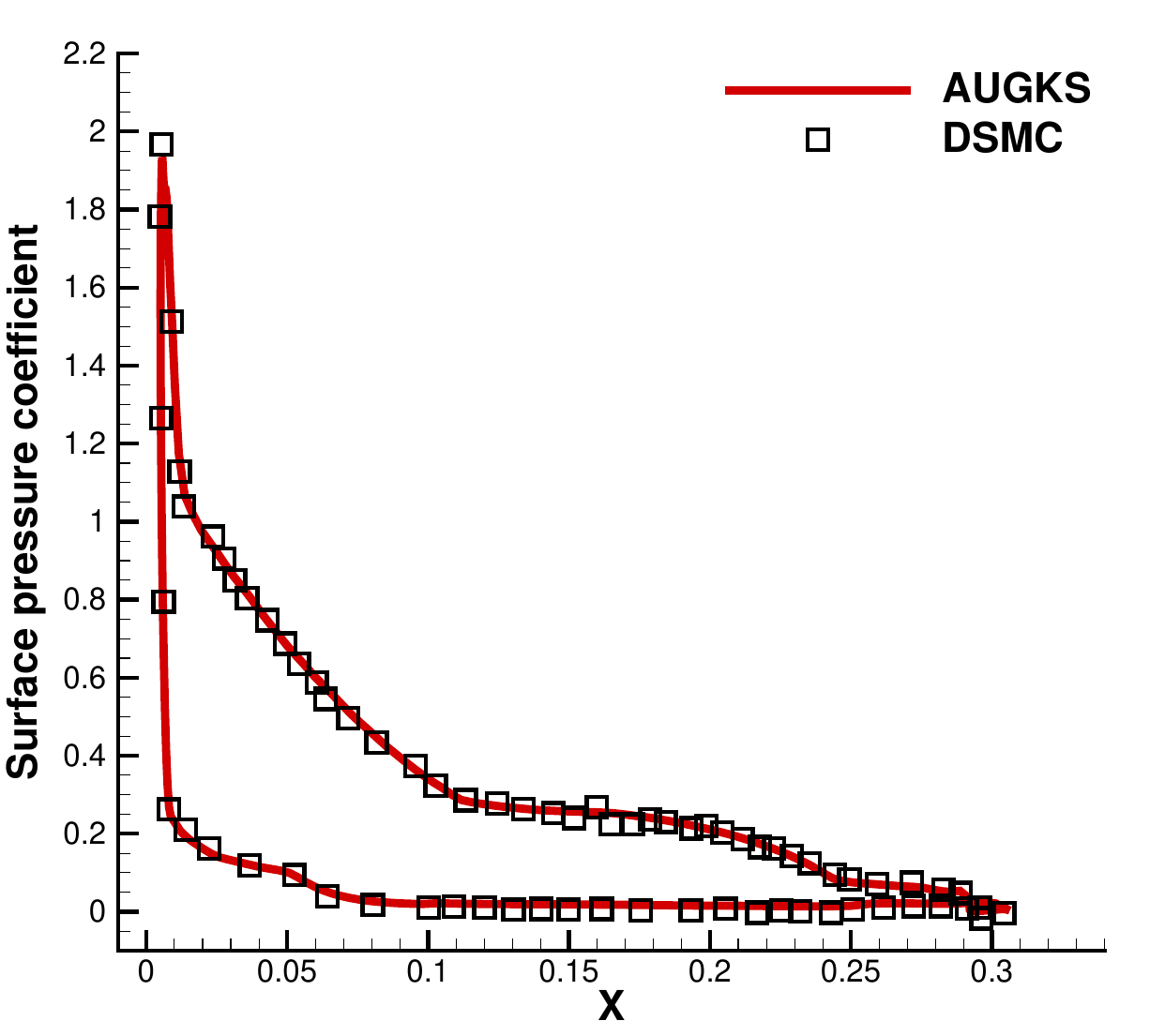}}
\subfloat[]{\includegraphics[width=0.33\textwidth]
	{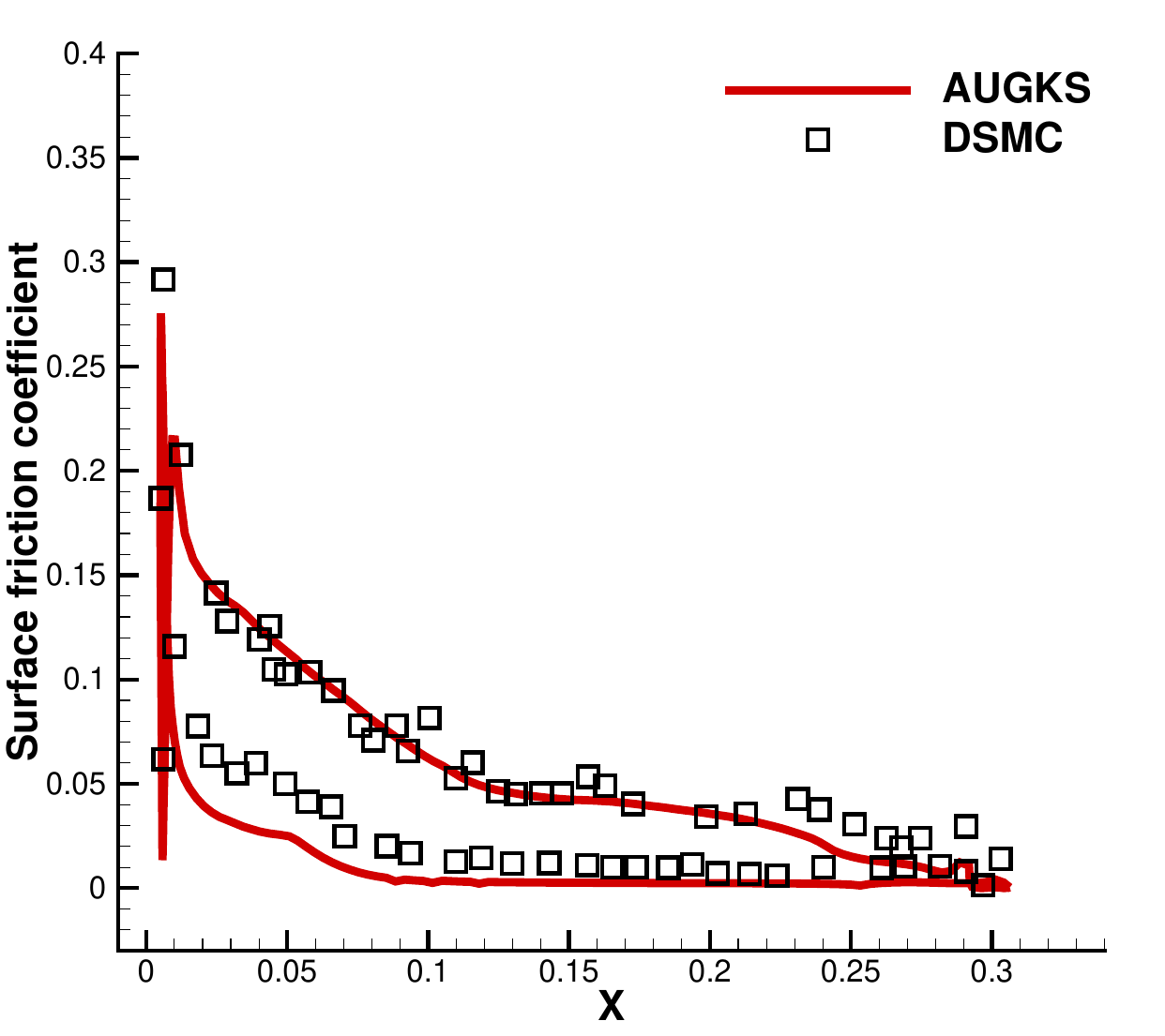}}
\subfloat[]{\includegraphics[width=0.33\textwidth]
	{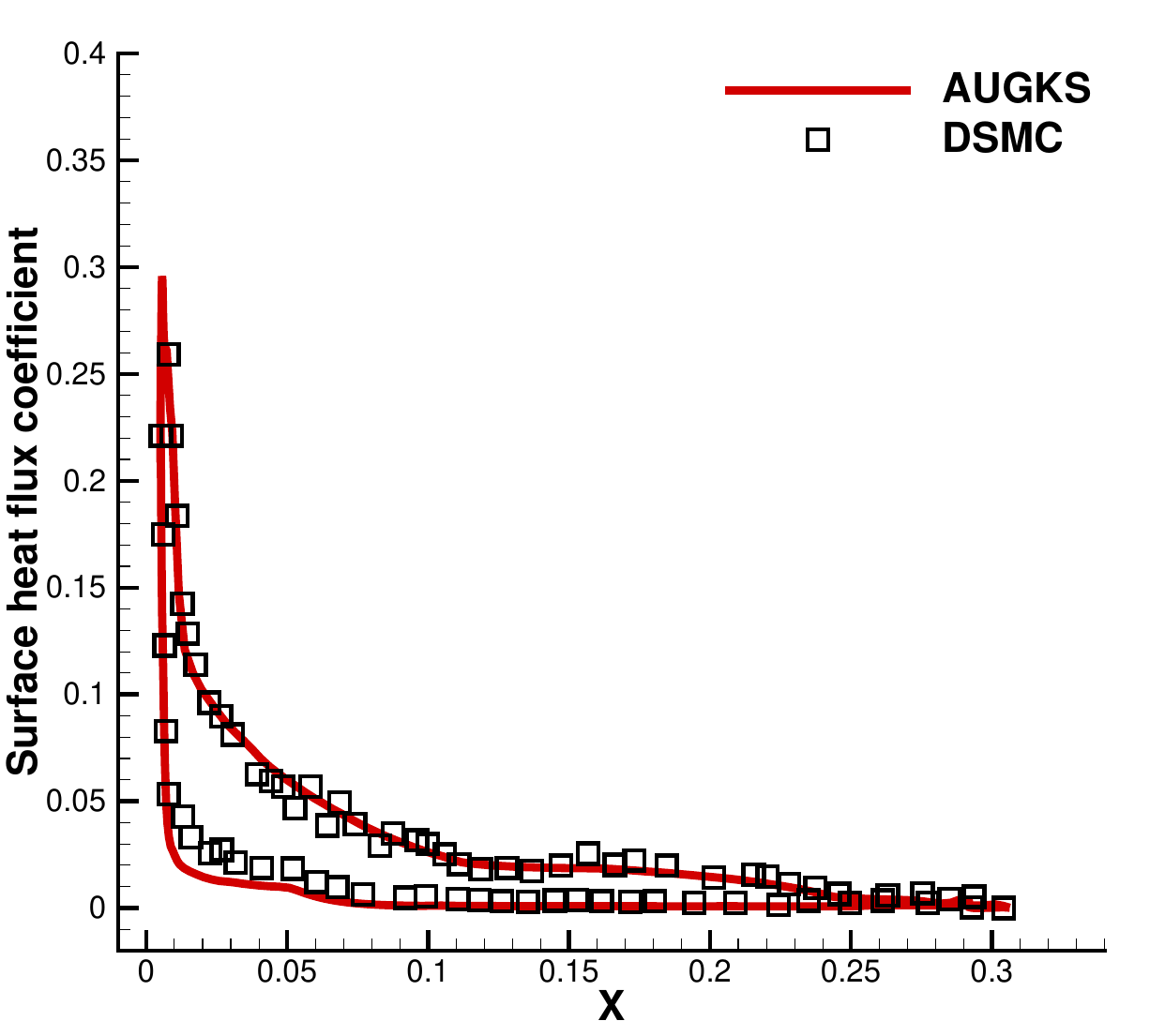}}\\
\caption{Surface quantities of hypersonic flow
	around a X38-like space vehicle at ${\rm Ma}_\infty = 8.0$ for ${\rm Kn}_\infty = 0.00275$ by the AUGKS method for argon gas compared with the DSMC method. (a) Pressure coefficient, (b) shear stress coefficient,
	and (c) heat flux coefficient.}
\label{fig:x38-surface-line}
\end{figure}

\begin{figure}[H]
	\centering
	\subfloat[]{\includegraphics[width=0.33\textwidth]
		{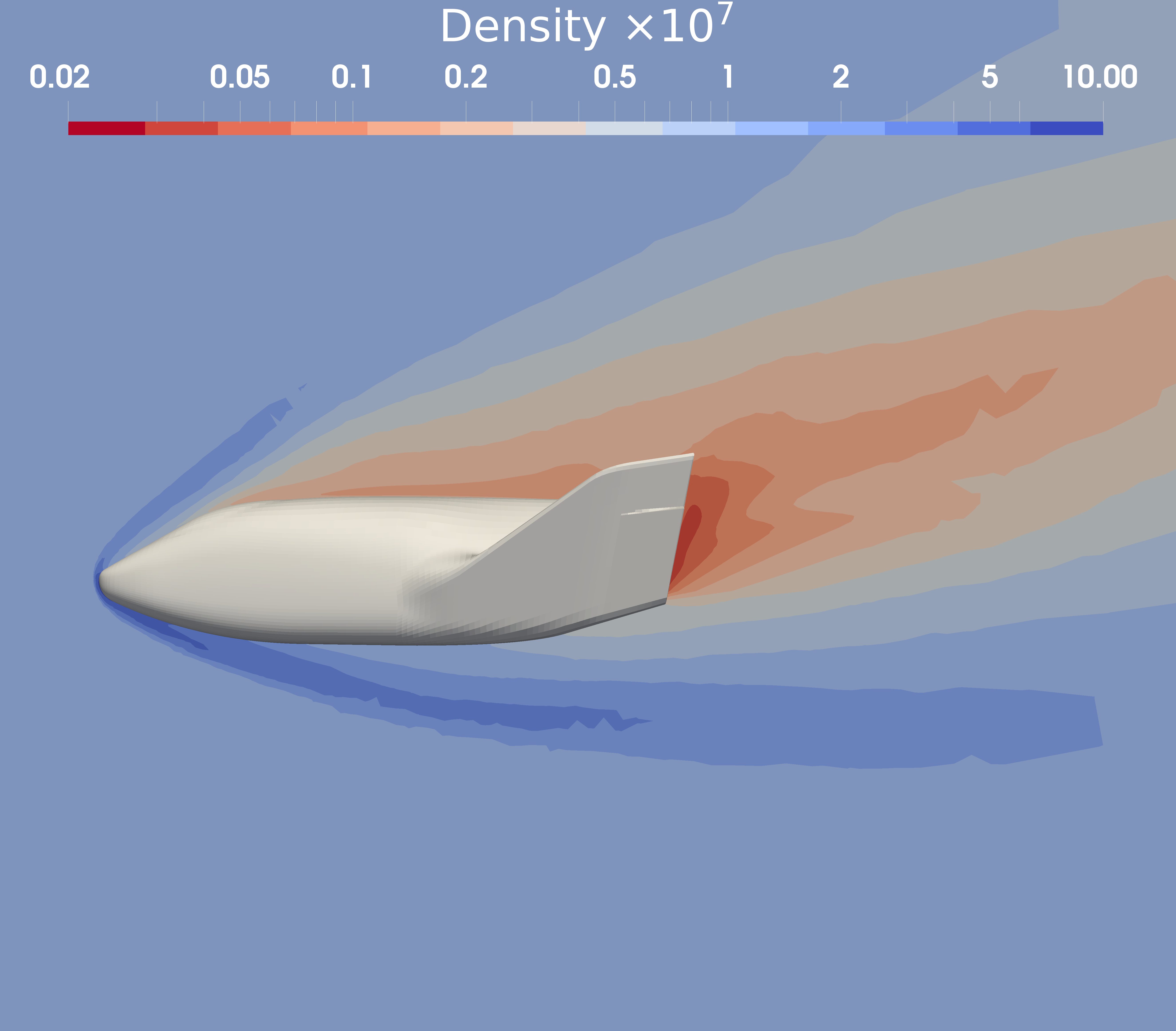}}
		~
	\subfloat[]{\includegraphics[width=0.33\textwidth]
		{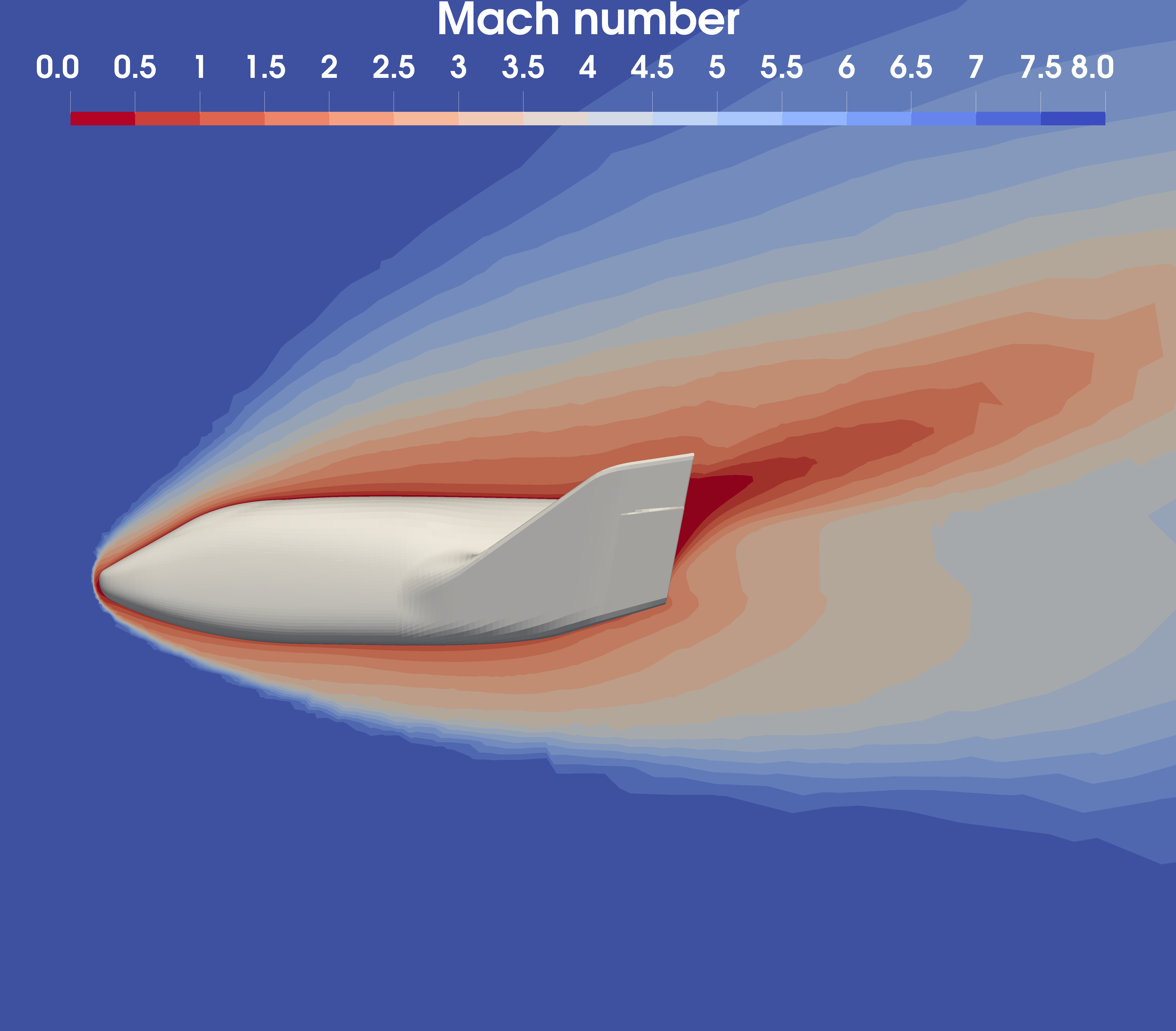}}
		~
	\subfloat[]{\includegraphics[width=0.33\textwidth]
		{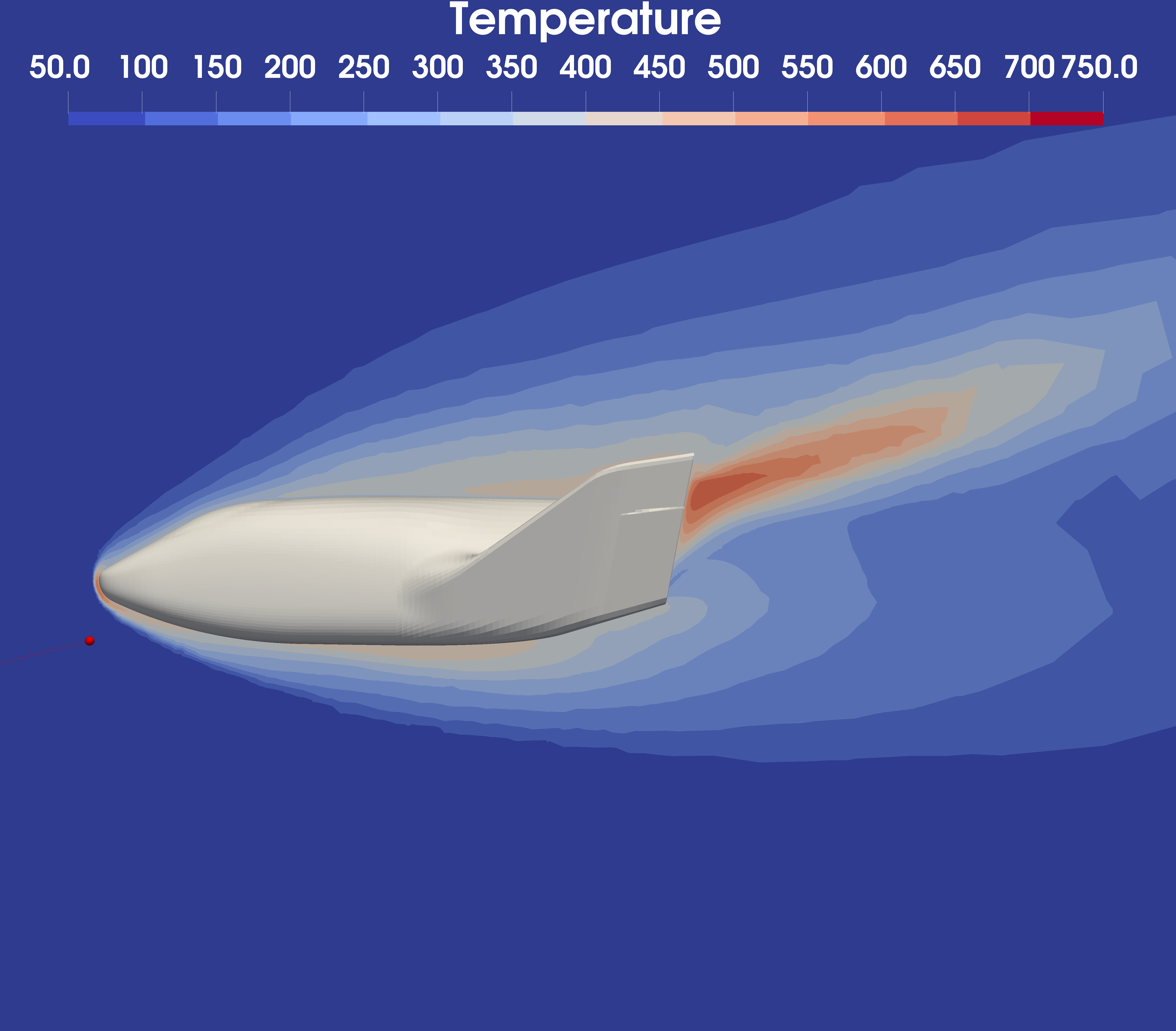}} \\
	\subfloat[]{\includegraphics[width=0.33\textwidth]
		{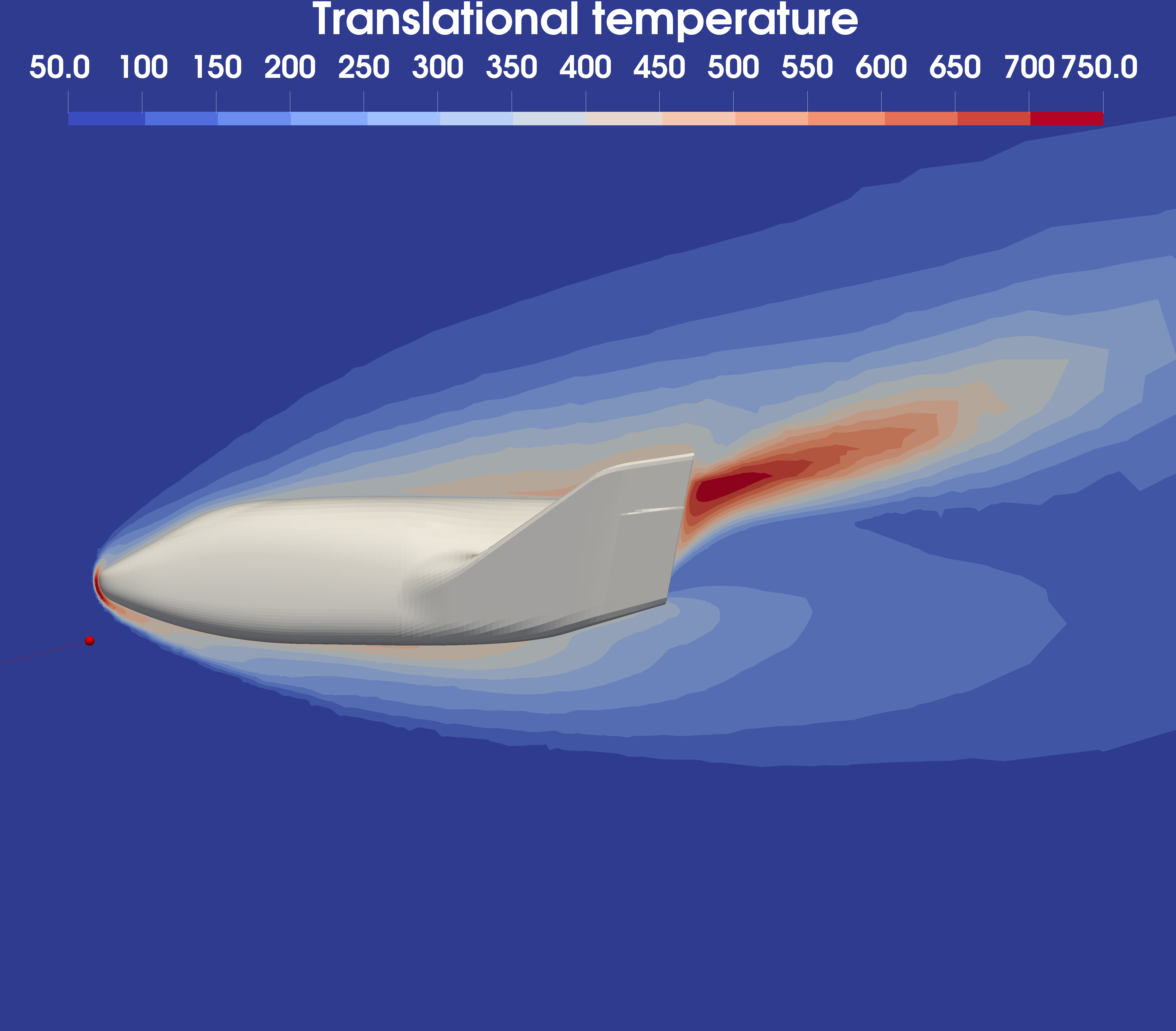}}
		~
	\subfloat[]{\includegraphics[width=0.33\textwidth]
		{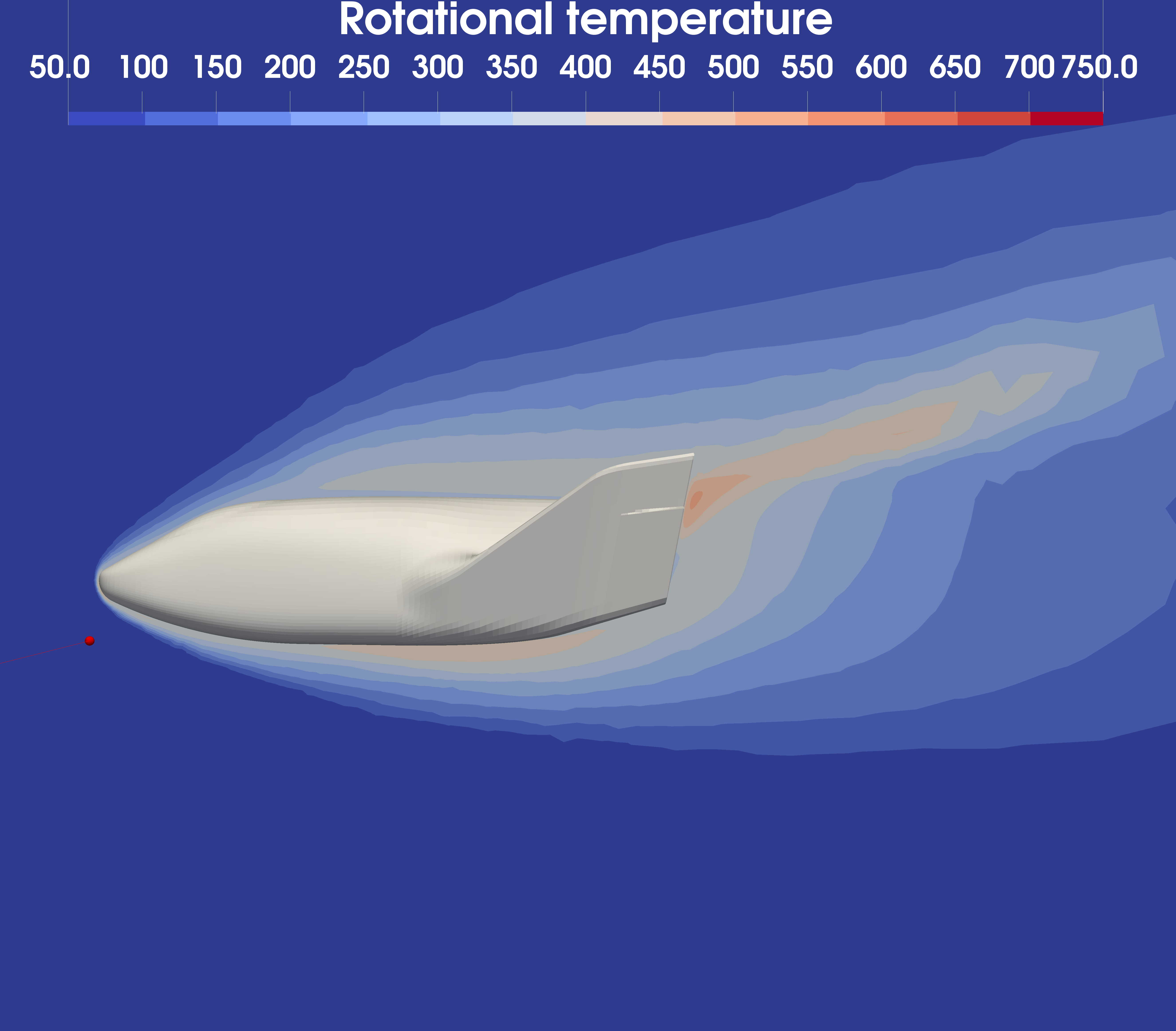}}
		~
	\subfloat[]{\includegraphics[width=0.33\textwidth]
		{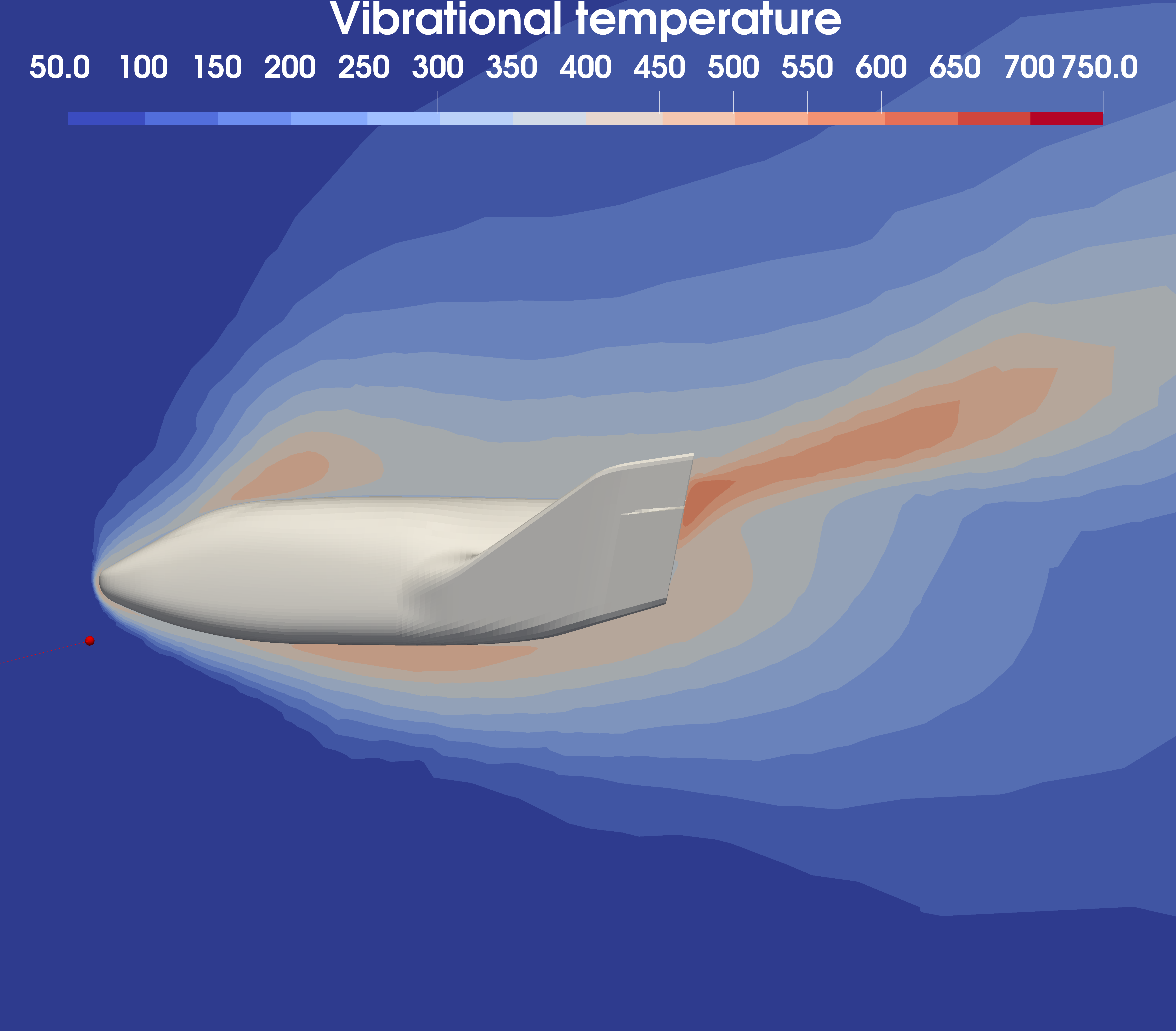}} \\
	\caption{Supersonic flow around a X38-like space vehicle at ${\rm Ma}_\infty = 8$ for ${\rm Kn}_\infty = 0.00275$. Distributions of (a) density, (b) Mach number, (c) average temperature, (d) translational temperature, (e) rotational temperature, and (f) vibrational temperatures.}
	\label{fig:x38-contour}
\end{figure}

\begin{figure}[H]
	\centering
	\includegraphics[width=0.4\textwidth]
		{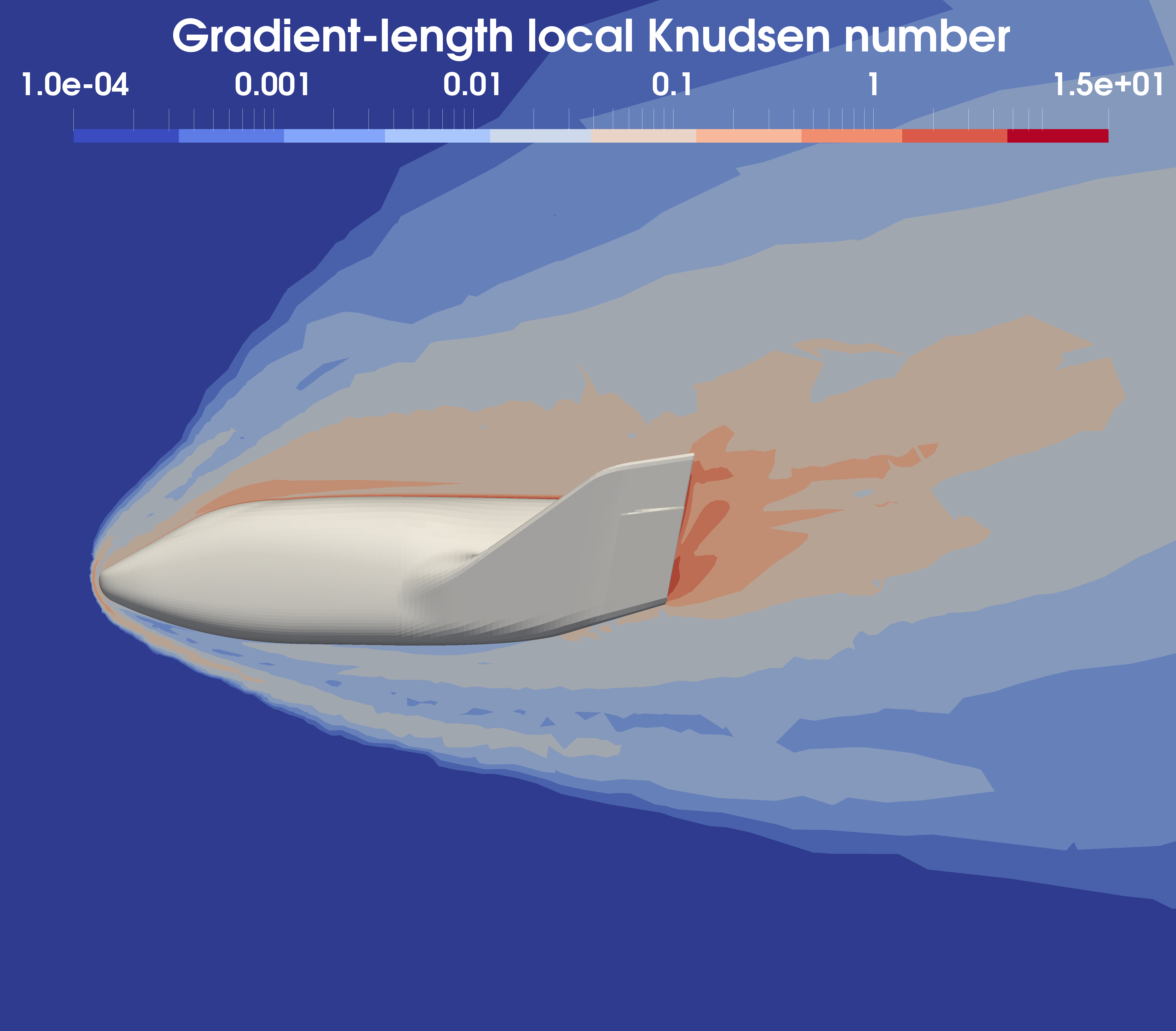}
	\caption{Hypersonic flow around a X38-like space vehicle at ${\rm Ma}_\infty = 8$ for ${\rm Kn}_\infty = 0.00275$. Distributions of gradient-length local Knudsen number ${\rm Kn}_{Gll}$ in the flow field.}
	\label{fig:x38-kn}
\end{figure}

\begin{figure}[H]
	\centering
\subfloat[]{\includegraphics[width=0.4\textwidth]
	{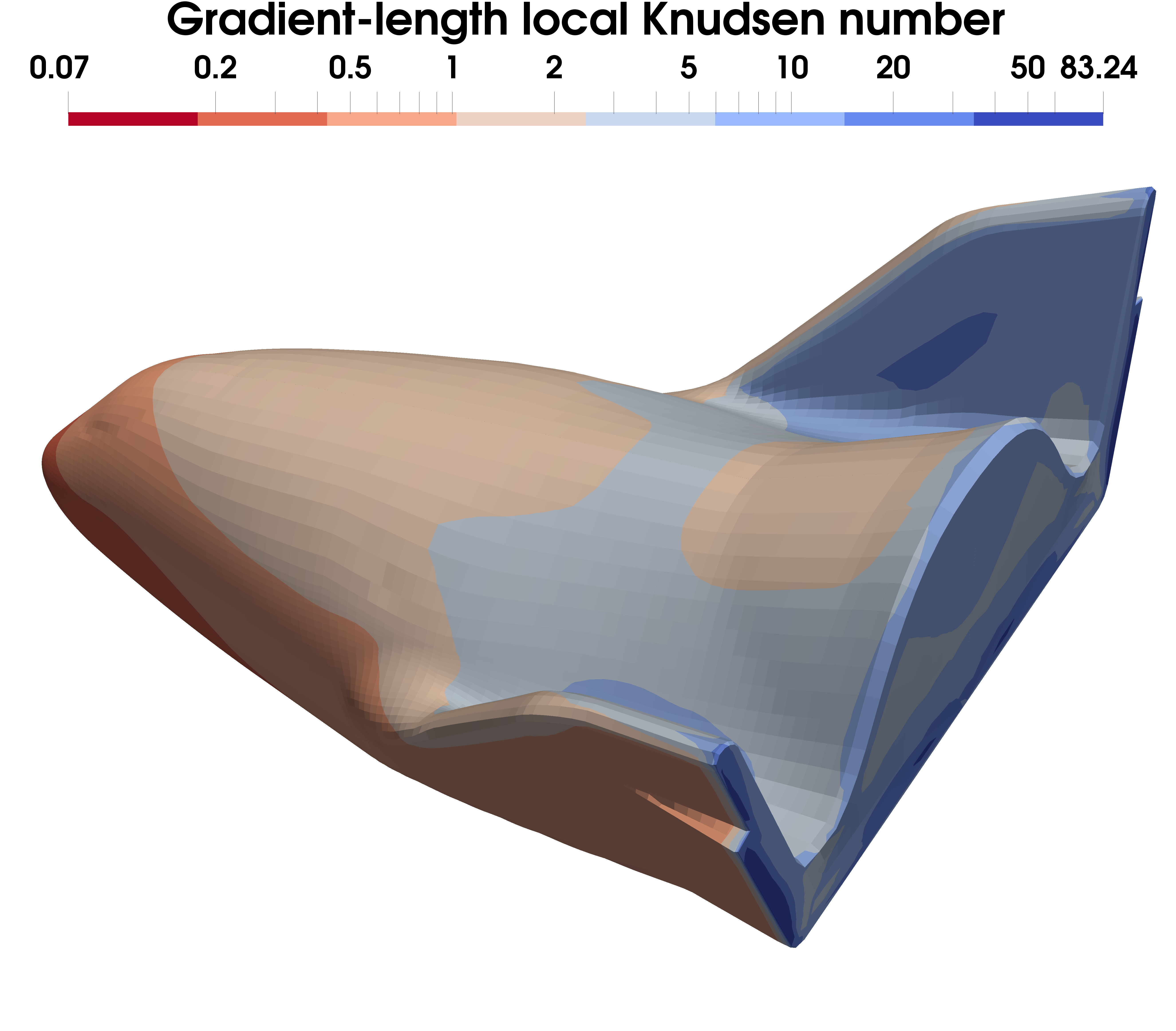}}
\subfloat[]{\includegraphics[width=0.4\textwidth]
	{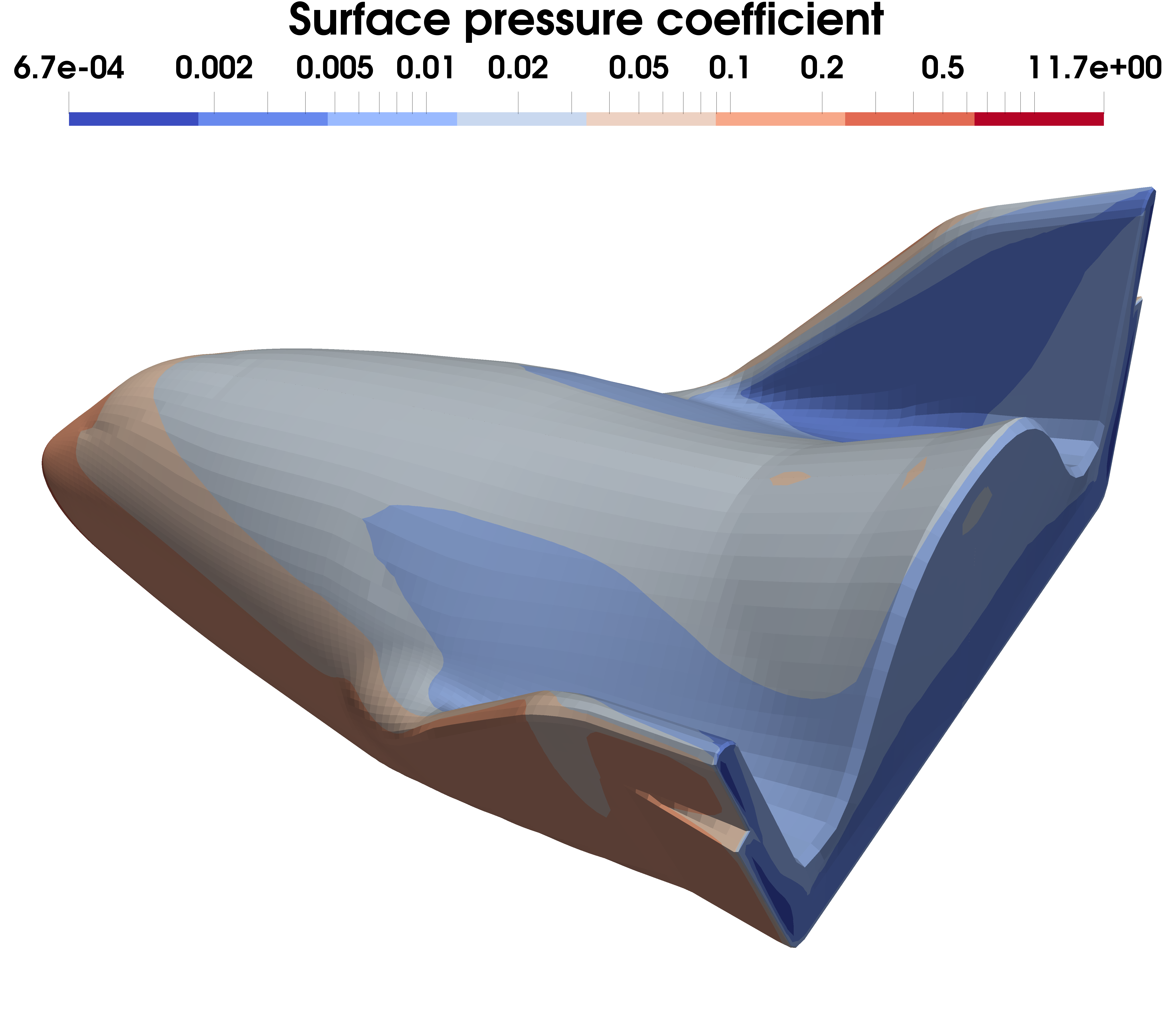}}\\
\subfloat[]{\includegraphics[width=0.4\textwidth]
	{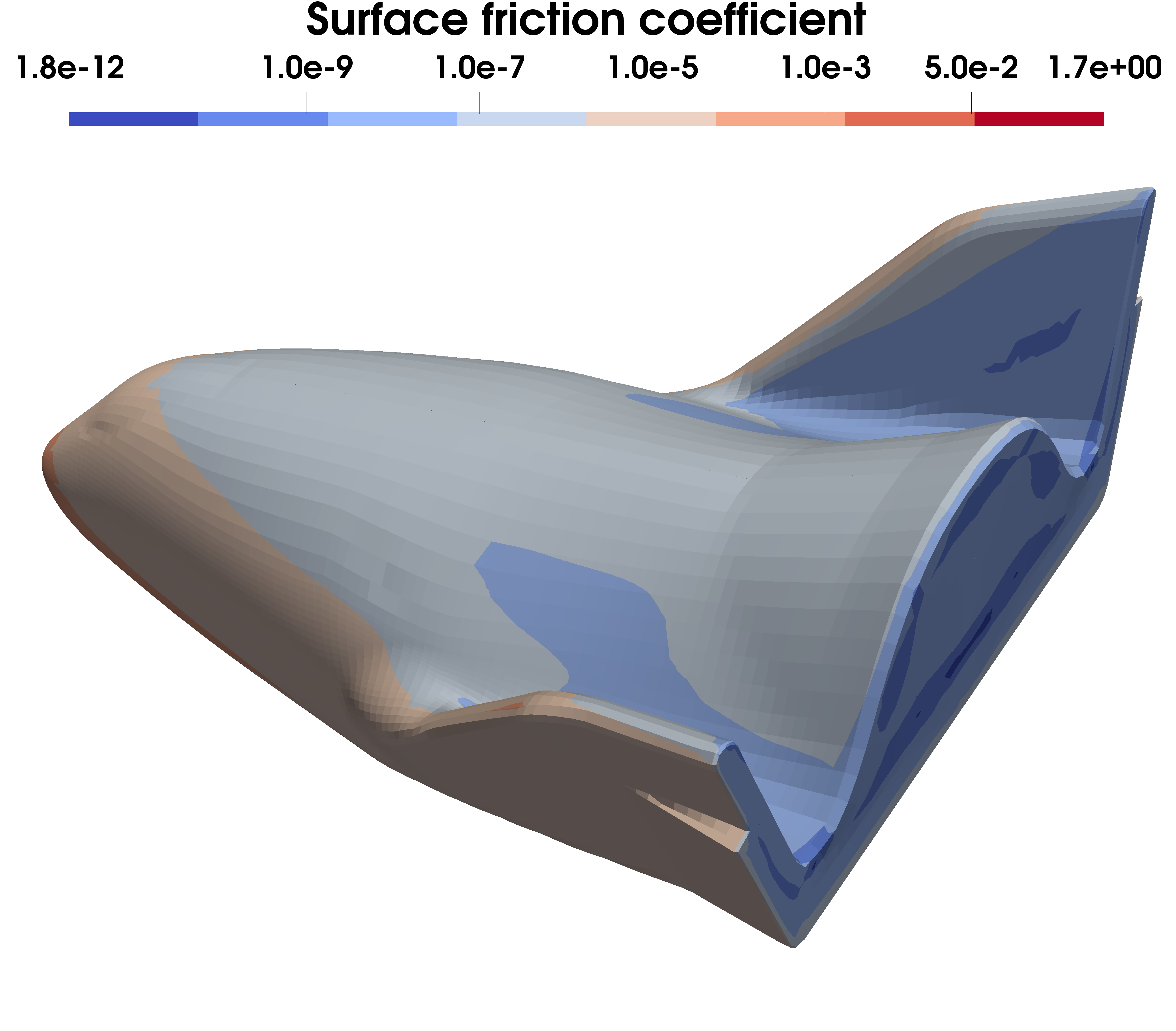}}
\subfloat[]{\includegraphics[width=0.4\textwidth]
	{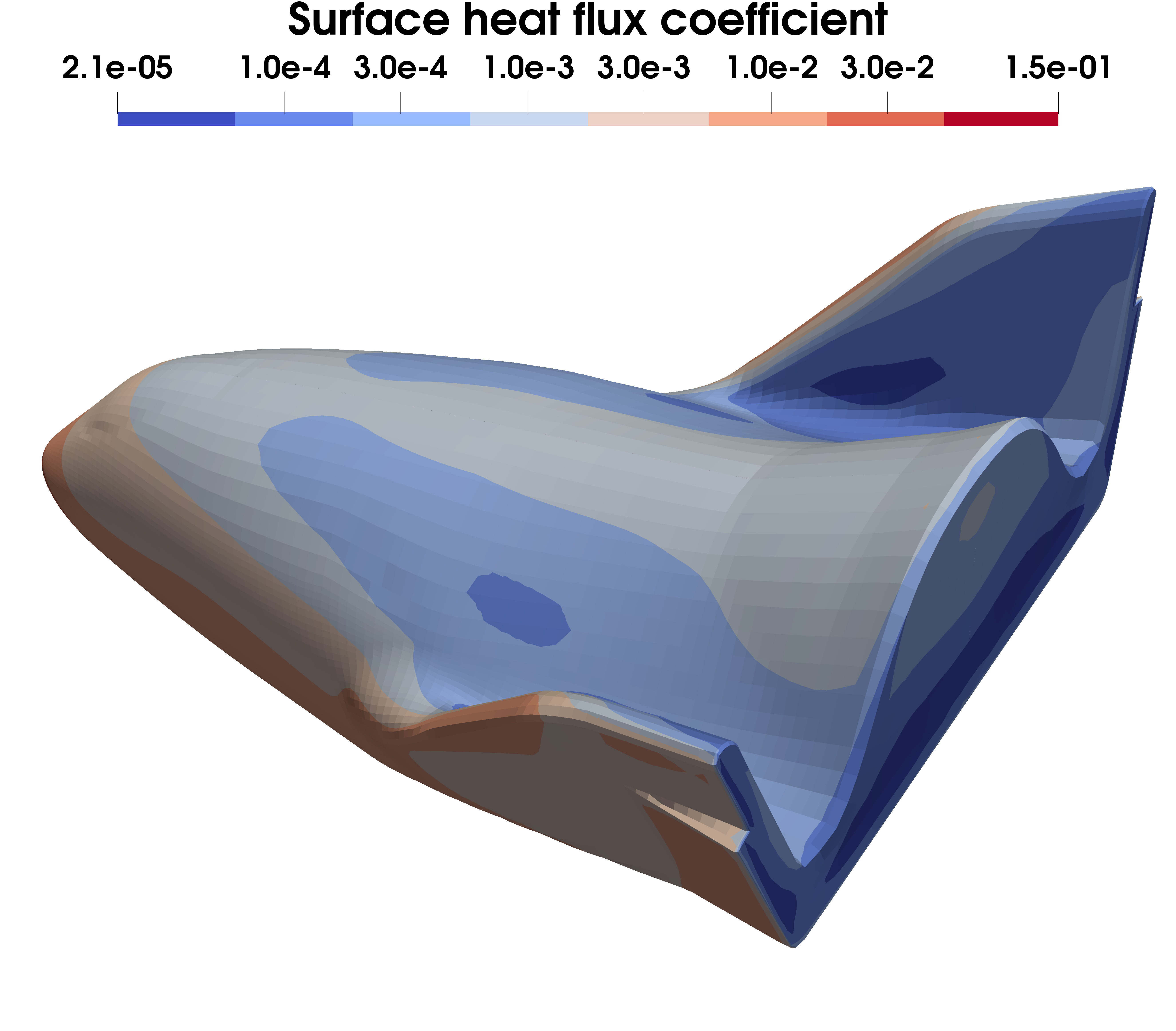}}
\caption{Surface quantities of hypersonic flow around a X38-like space vehicle at ${\rm Ma}_\infty = 8.0$ for ${\rm Kn}_\infty = 0.00275$ by the AUGKS-vib for nitrogen gas. (a) Gradient-length local Knudsen number, (b) surface pressure coefficient, (c) surface friction coefficient, and (d) heat flux coefficient.}
\label{fig:x38-surface}
\end{figure}

In the published available computational results of X38-like space vehicle, only the DSMC method for argon gas can be found \cite{li2021kinetic}. In this paper, to fully demonstrate the accuracy of the AUGKS-vib, it is necessary to first simulate argon gas. For the argon gas, its gas properties can be fixed by molecular mass $m = 6.63\times10^{-26}$ kg, rotational and vibrational degrees of freedom $K_r = 0$, $K_v = 0$, rotation and vibration collision number $Z_r\to\infty$, $Z_v\to\infty$, and reference dynamic viscosity $\mu_{ref} = 2.117\times10^{-5}$ ${\rm Nsm^{-2}}$, $\omega = 0.81$, $\alpha = 1.0$. The distributions of surface pressure, heat flux, and shear stress coefficients are shown in Fig.~\ref{fig:x38-surface-line}, which show good agreement with DSMC results. Then, after the correctness of the scheme is verified, the simulation of hypersonic flow is applied for nitrogen gas with the rotation and vibration collision number $Z_r = 23.5$, $Z_v = 95.0$. Fig.~\ref{fig:x38-contour} shows the contours of density, Mach number, average temperature, translational temperature, rotational temperature, and vibrational temperatures. Fig.~\ref{fig:x38-kn} shows the distribution of gradient-length local Knudsen number in the flow field. Fig.~\ref{fig:x38-surface} shows the distributions of gradient-length local Knudsen number, pressure coefficient, friction coefficient, and heat flux coefficients on the surface of the space vehicle.

Combining the characteristics of flow field and the surface quantities of the space vehicle, it can be observed that the gas in the windward is highly compressed and in the leeward region expands rapidly. At a flight altitude of 80 kilometers, the local Knudsen number ${\rm Kn}_{\rm Gll}$ on the surface of the space vehicle increases from 0.07 on the windward side to 84 on the leeward side. It can be concluded that although the incoming flow Knudsen number ${\rm Kn}_\infty$ is in the order of $10^{-3}$, the flow characteristics of complex geometry cannot be determined as slip flow regime. The flow state of a specific geometry should be determined according to its local non-equilibrium/equilibrium physics. Meanwhile, in the vicinity of the windward region of the fuselage surface, the local Knudsen number is between 0.1 and 1. At this position, the vibrational temperature increases, even exceeding the local rotational temperature. This is because although the larger vibrational collision number leads to a relatively small amount of energy allocated to the vibrational degree of freedom, when the vibrational degree of freedom $K_v$ is not fully excited, it remains a small quantity, resulting in a higher vibrational temperature. On the other hand, the rotational degree of freedom is fixed at 2, so the rotational temperature is proportional to the rotational energy.

\begin{figure}[H]
	\centering
	\includegraphics[width=0.4\textwidth]
		{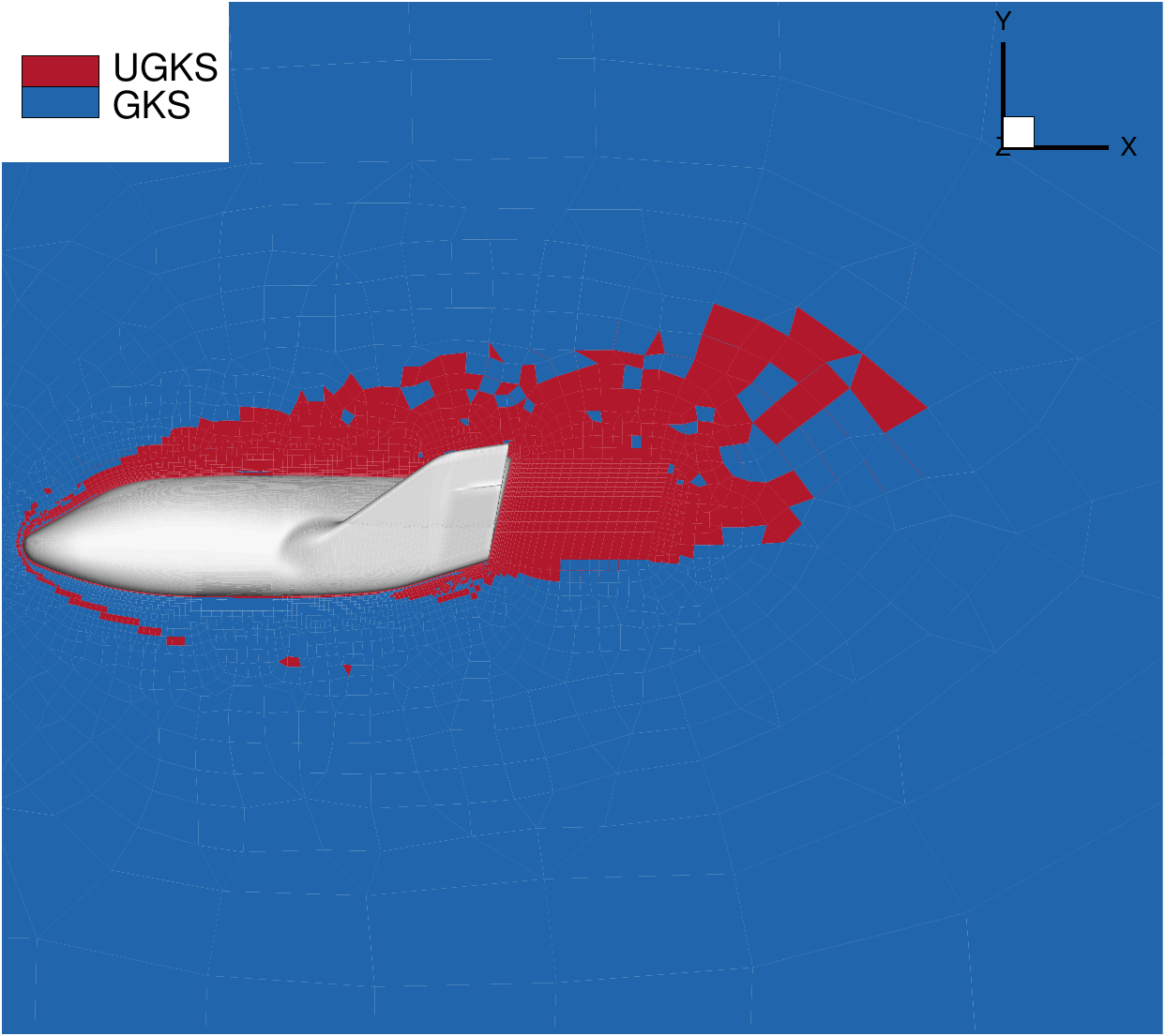}
	\caption{Hypersonic flow around a X38-like space vehicle at ${\rm Ma}_\infty = 8$ for ${\rm Kn}_\infty = 0.00275$. Distributions of velocity space adaptation with $C_t = 0.05$ where the discretized velocity space (UGKS) is used in 52.98\% of physical domain.}
	\label{fig:x38-isDisc}
\end{figure}

Fig.~\ref{fig:x38-isDisc} shows the distribution of velocity space adaptation where the 52.98\% of computational domain is covered by discrete distribution function solved by UGKS. The simulation takes 150,000 steps first-order GKS step and 160,000 steps AUGKS which is conducted on the SUGON computation platform which takes wall clock time 5.45 hours on 2560 cores of CPU 7285 32C 2.0GHz.

\subsection{Nozzle plume into the vacuum}
\begin{figure}[H]
	\centering
	\subfloat[]{\includegraphics[width=0.35\textwidth]
		{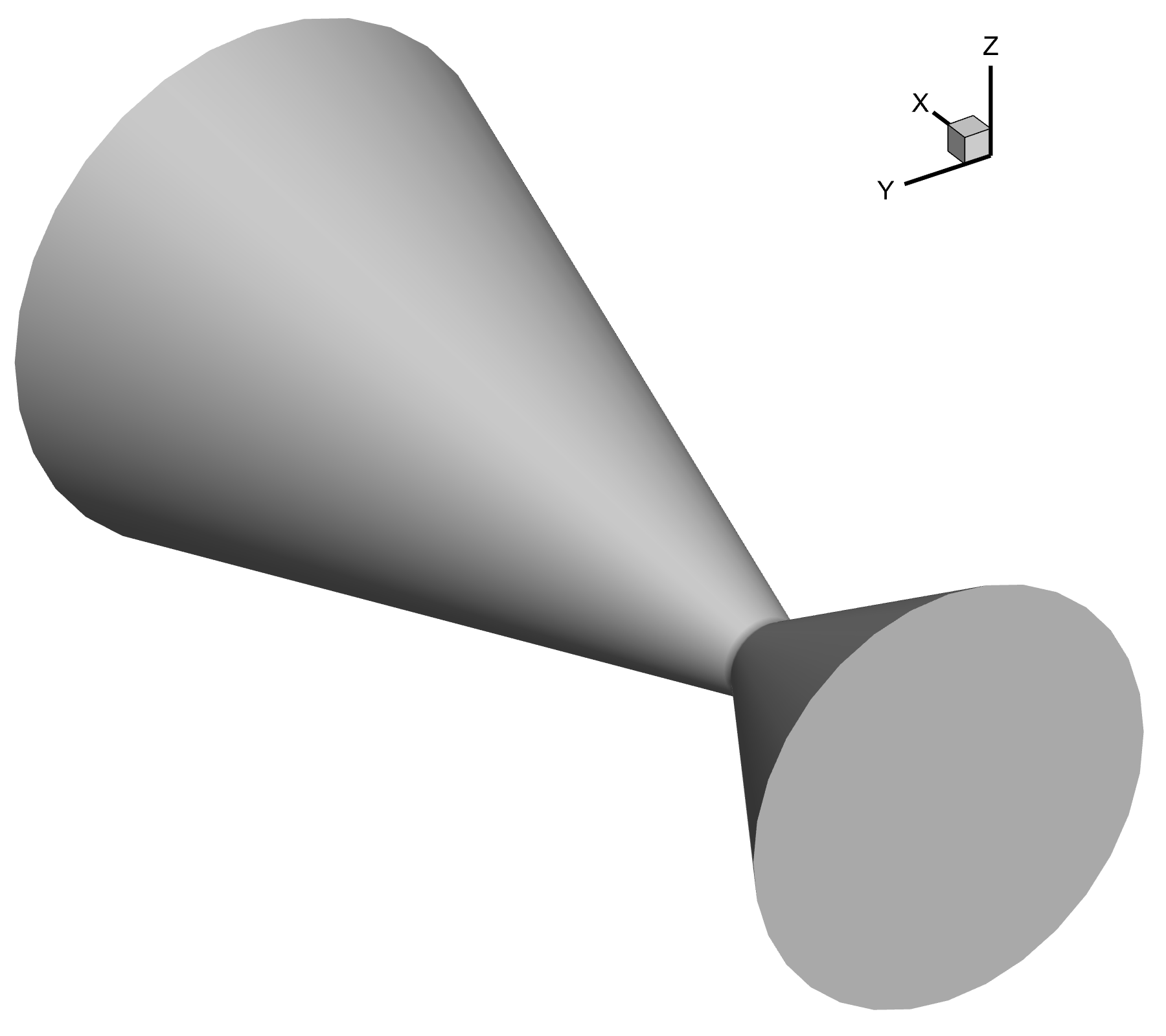}}
	\subfloat[]{\includegraphics[width=0.35\textwidth]
		{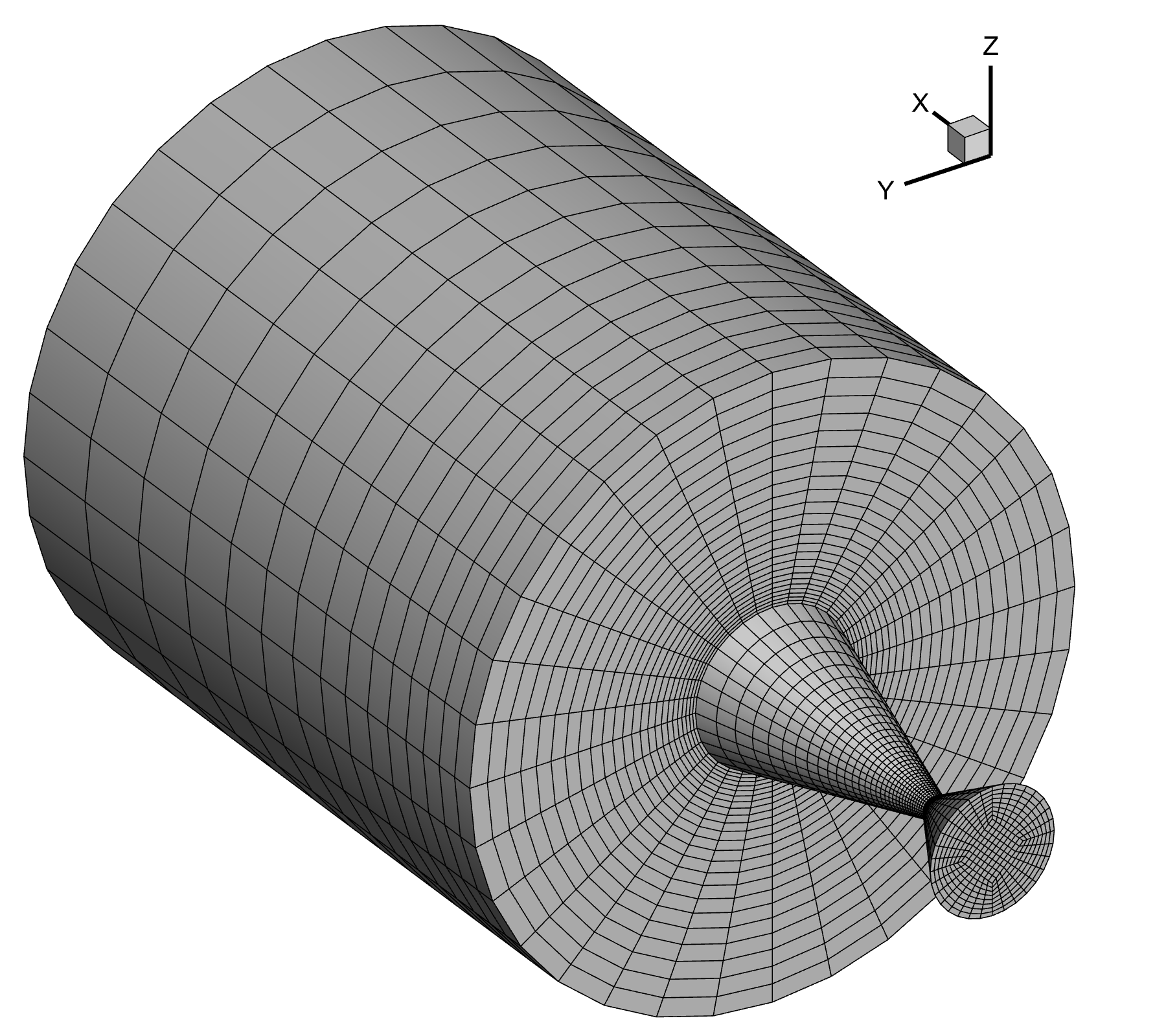}}
	\caption{Nozzle plume into the vacuum. (a) Shape, and (b) physical domain consists of 31,360 cells.}
	\label{fig:nozzleplume-mesh}
\end{figure}

The AUGKS-vib is applied to the ${\rm CO_2}$ expansions into a background vacuum. The unsteady and multiscale process of this plume flow is simulated. The geometric shape of the nozzle is based on the model used in Boyd et al.\cite{george1999simulation}, as shown in Fig.~\ref{fig:nozzleplume-mesh}. Inside the nozzle, the Knudsen number of the flow is small enough to approach the continuum flow regime. In the plume region, the Knudsen number gets to the free molecule regime. To capture the high Mach number jet in flow acceleration process through the nozzle, a wide range of discrete velocity space is needed. An adaptive spatial decomposition dynamically adjusts the AUGKS region in the whole unsteady process, and releases the limitation on the discrete velocity space.
The mesh adopted in this case is shown in Fig.~\ref{fig:nozzleplume-mesh}. The physical mesh of the entire calculation domain has 31,360 cells, and the discrete velocity space with the range $(-5\sqrt{R T_s}, 5\sqrt{R T_s})$ consists of 40$\times$20$\times$20 cells. The ambient pressure outside the nozzle is $P_{\infty}$ = 0.01 Pa, and the ambient temperature is $T_{\infty}$ = 300K. For the inlet boundary condition, the stagnation temperature and pressure are $T_s$ = 710 K, $p_s$ = 4866.18 Pa. The isothermal wall is applied for the nozzle wall with temperature $T_w$ = 500 K. The expansion gas ${\rm CO}_2$ is regarded as diatomic gas with molecular mass $m = 7.31\times10^{-26}$ kg, rotational degrees of freedom $K_r = 2.0$, vibrational characteristic temperature $\Theta_v = 1290$ K, and reference dynamic viscosity $\mu_{ref} = 1.38\times10^{-5}$ ${\rm Nsm^{-2}}$, $\omega = 0.67$, $\alpha = 1.0$.

\begin{figure}[H]
	\centering
	\subfloat[]{\includegraphics[width=0.35\textwidth]
		{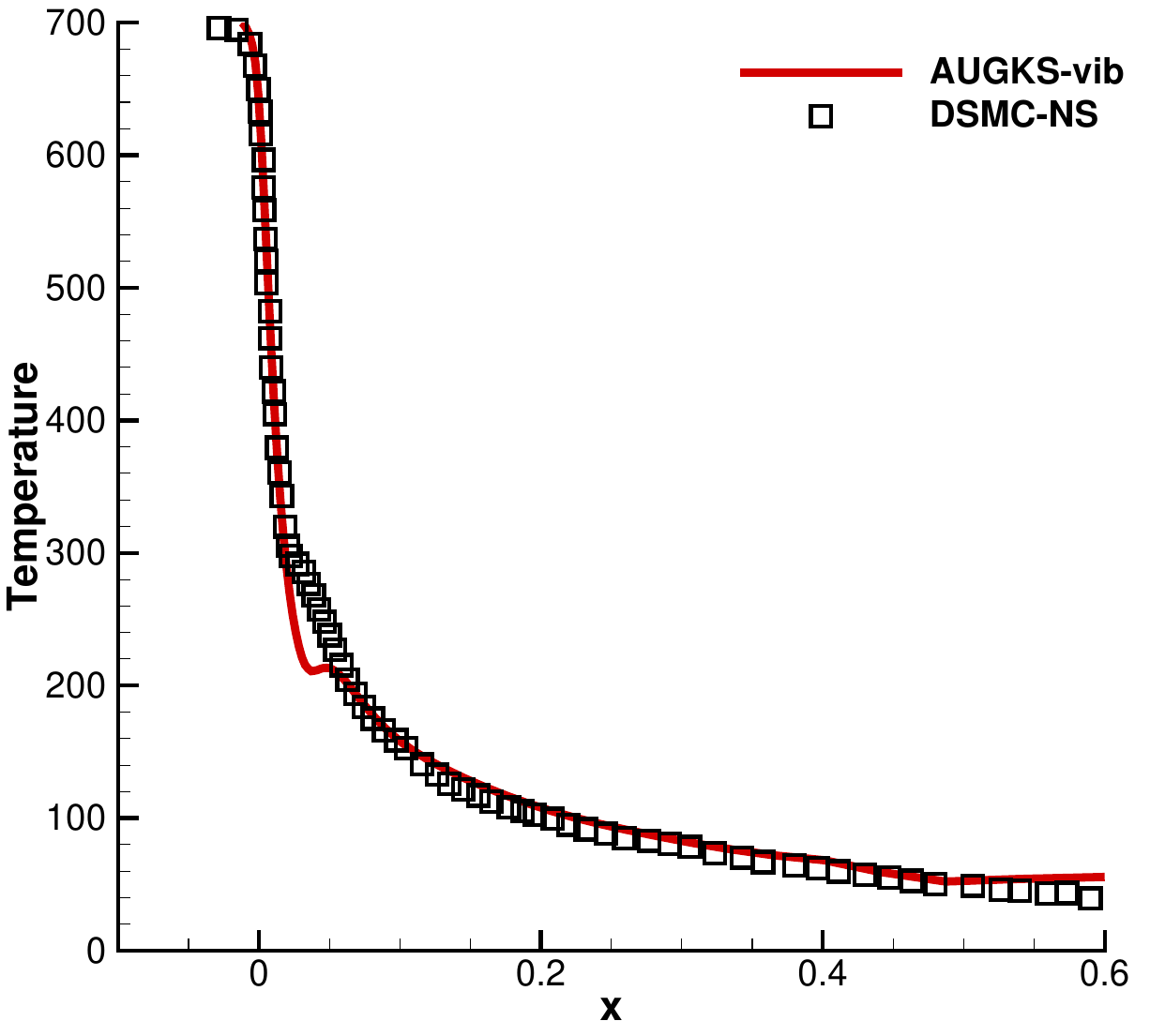}}
	\subfloat[]{\includegraphics[width=0.35\textwidth]
		{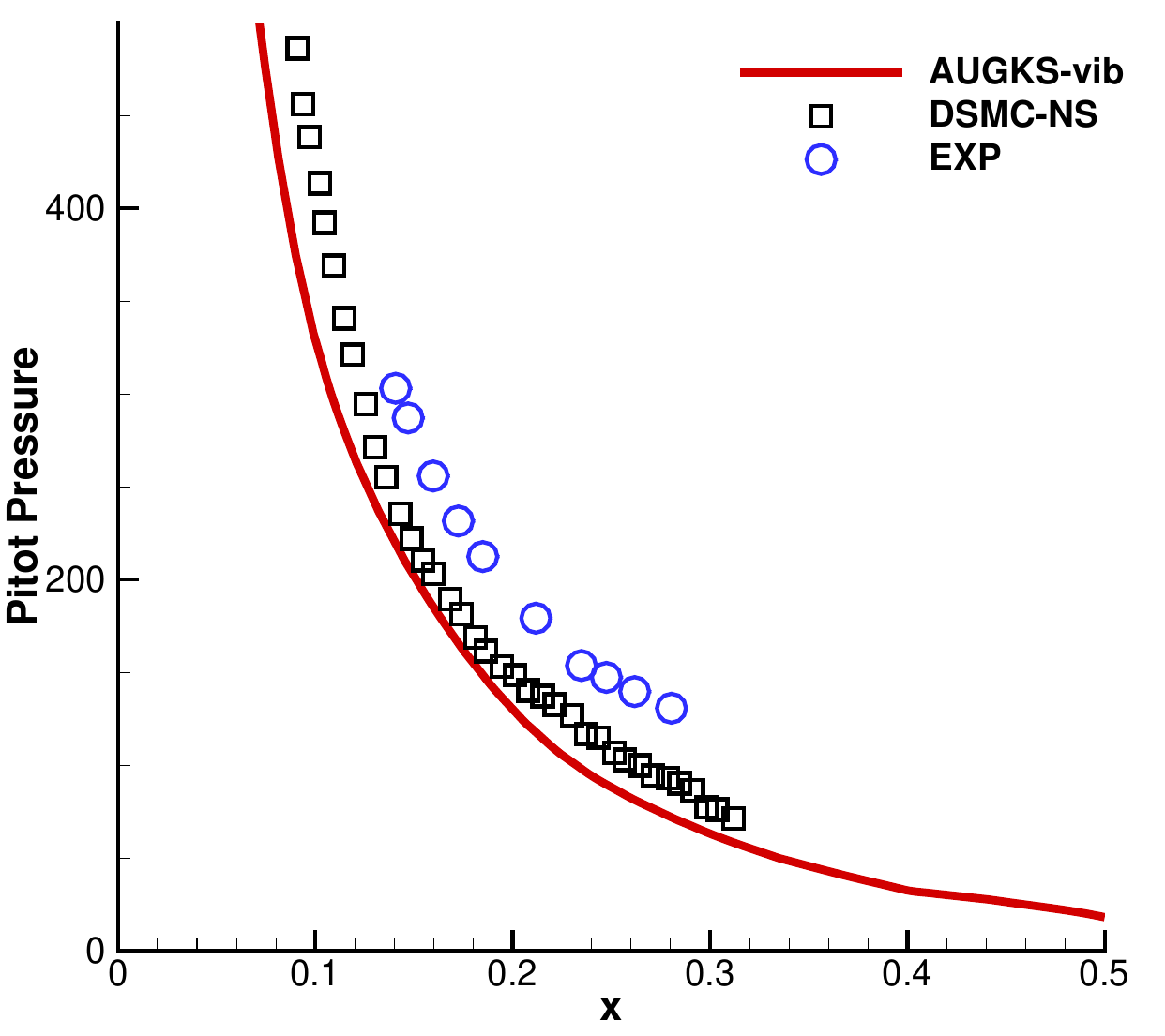}}
	\caption{Comparison of Pitot pressure and temperature along the central axis of the nozzle. (a) Temperature, and (b) Pitot pressure.}
	\label{fig:nozzleline}
\end{figure}

\begin{figure}[H]
	\centering
	\subfloat[]{\includegraphics[width=0.4\textwidth]
		{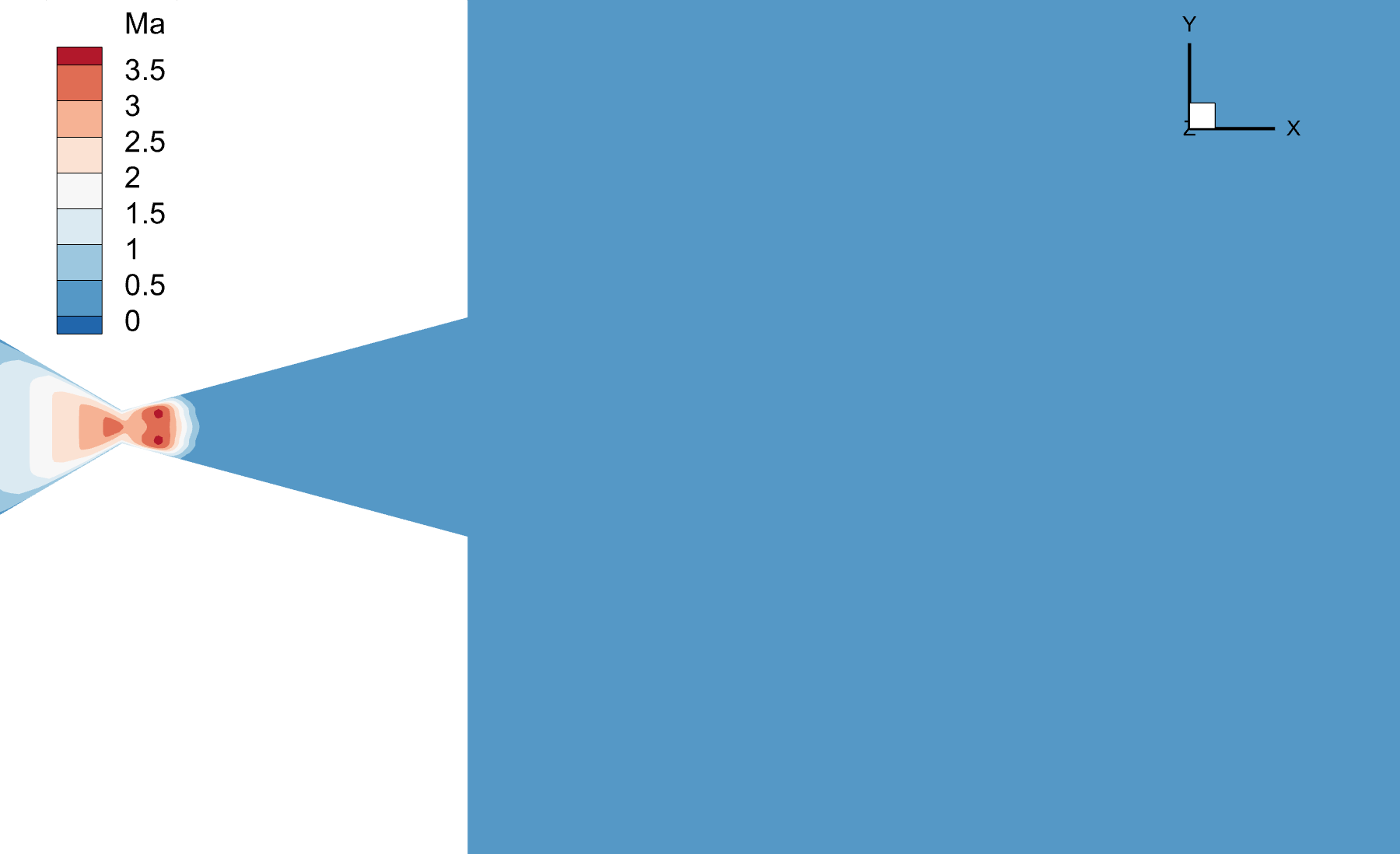}}
	\quad
	\subfloat[]{\includegraphics[width=0.4\textwidth]
		{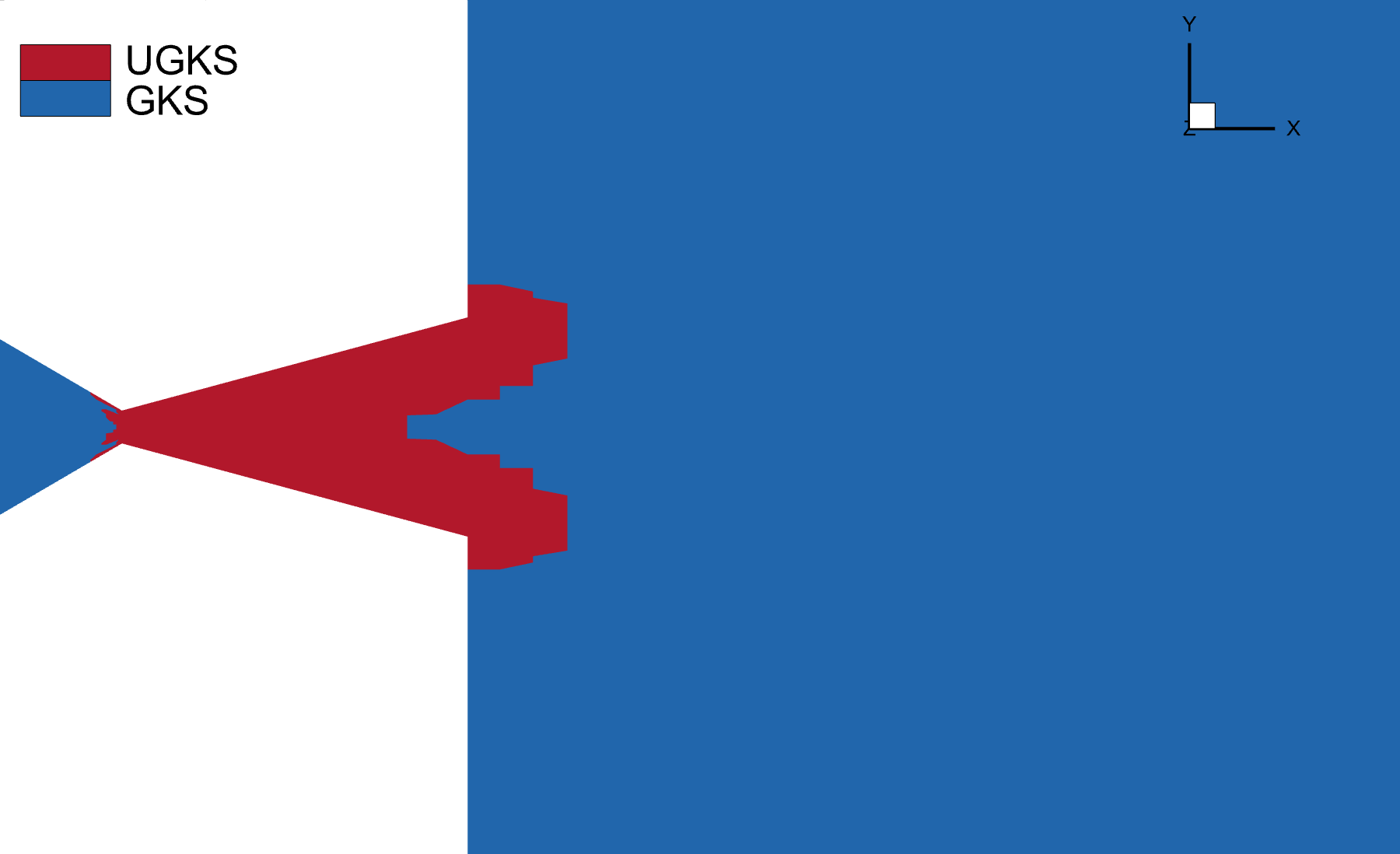}}
	\caption{Nozzle plume flow to a background vacuum $p_\infty = 0.01$ Pa at the initial stage. Distributions of (a) Mach number and (b) velocity space adaptation with $C_t= 0.05$, where the discretized velocity space (UGKS) is used in 22.10\% of physical domain.}
	\label{fig:nozzle-init}
\end{figure}

\begin{figure}[H]
	\centering
	\subfloat[]{\includegraphics[width=0.4\textwidth]
		{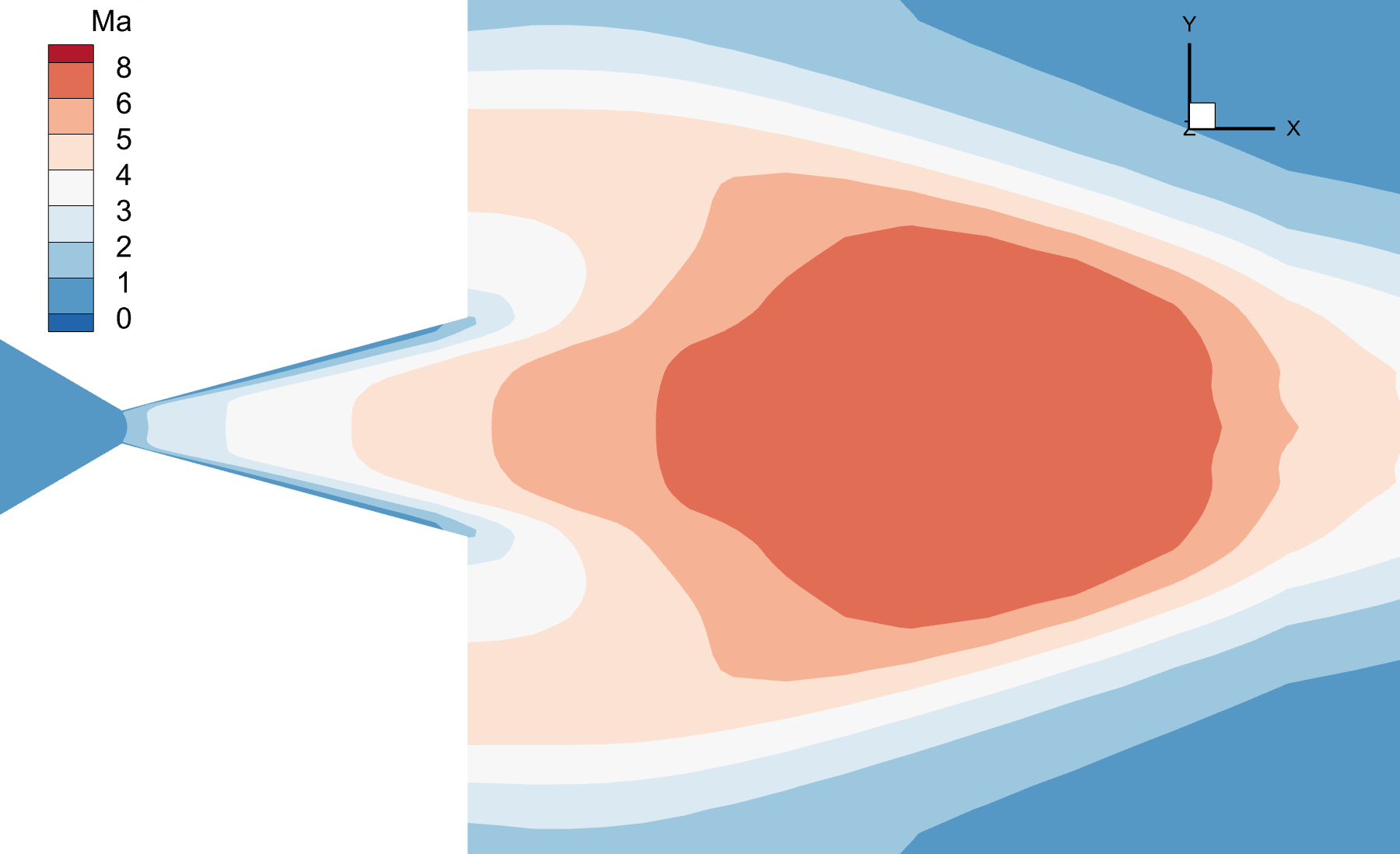}} 	
	\quad
	\subfloat[]{\includegraphics[width=0.4\textwidth]
		{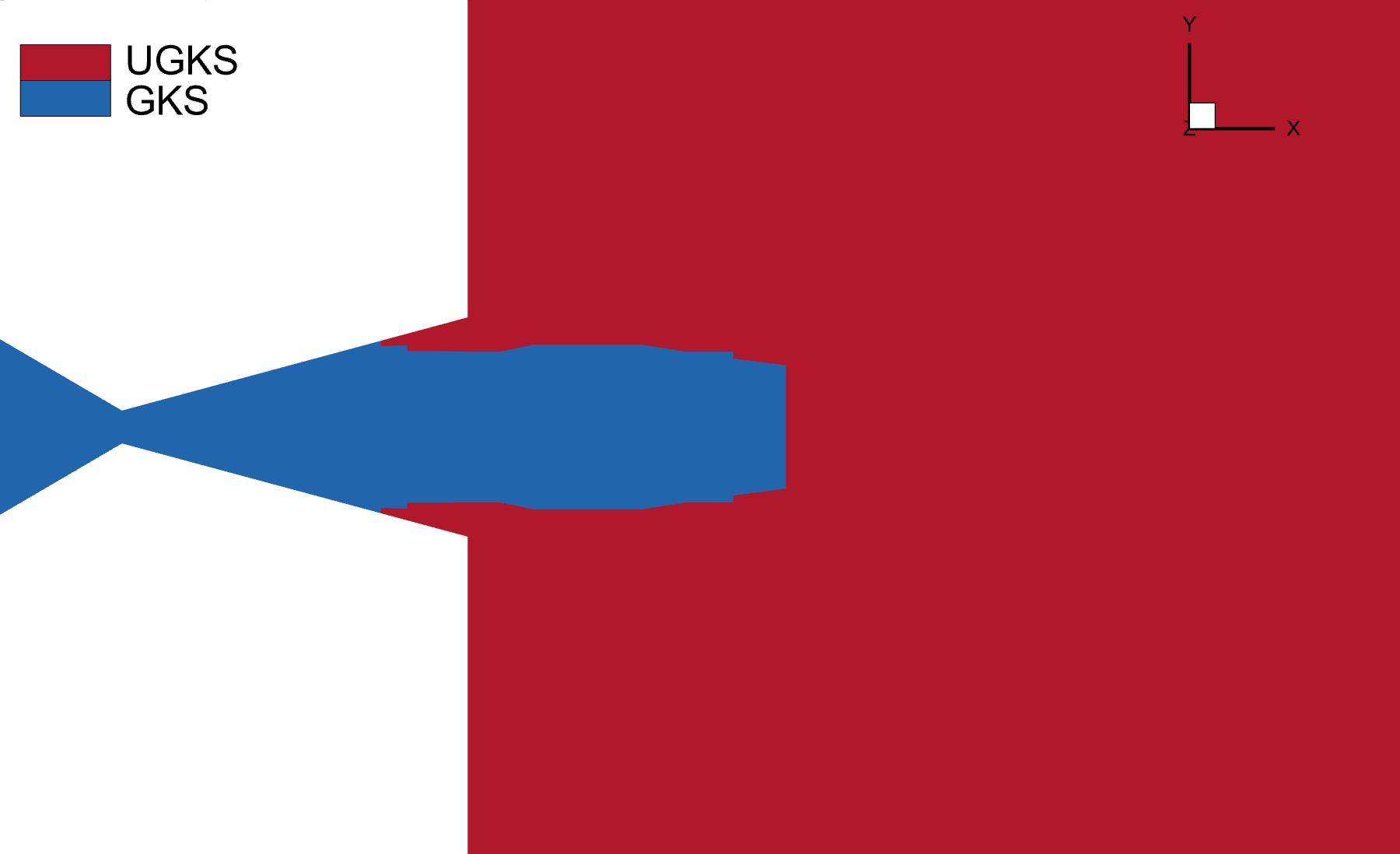}}
	\caption{Nozzle plume flow to a background vacuum $p_\infty = 0.01$ Pa at the developing stage. Distributions of (a) Mach number and (b) velocity space adaptation with $C_t= 0.05$ where the discretized velocity space (UGKS) is used in 46.28\% of physical domain.}
	\label{fig:nozzle-developing}
\end{figure}

\begin{figure}[H]
	\centering
	\subfloat[]{\includegraphics[width=0.4\textwidth]
		{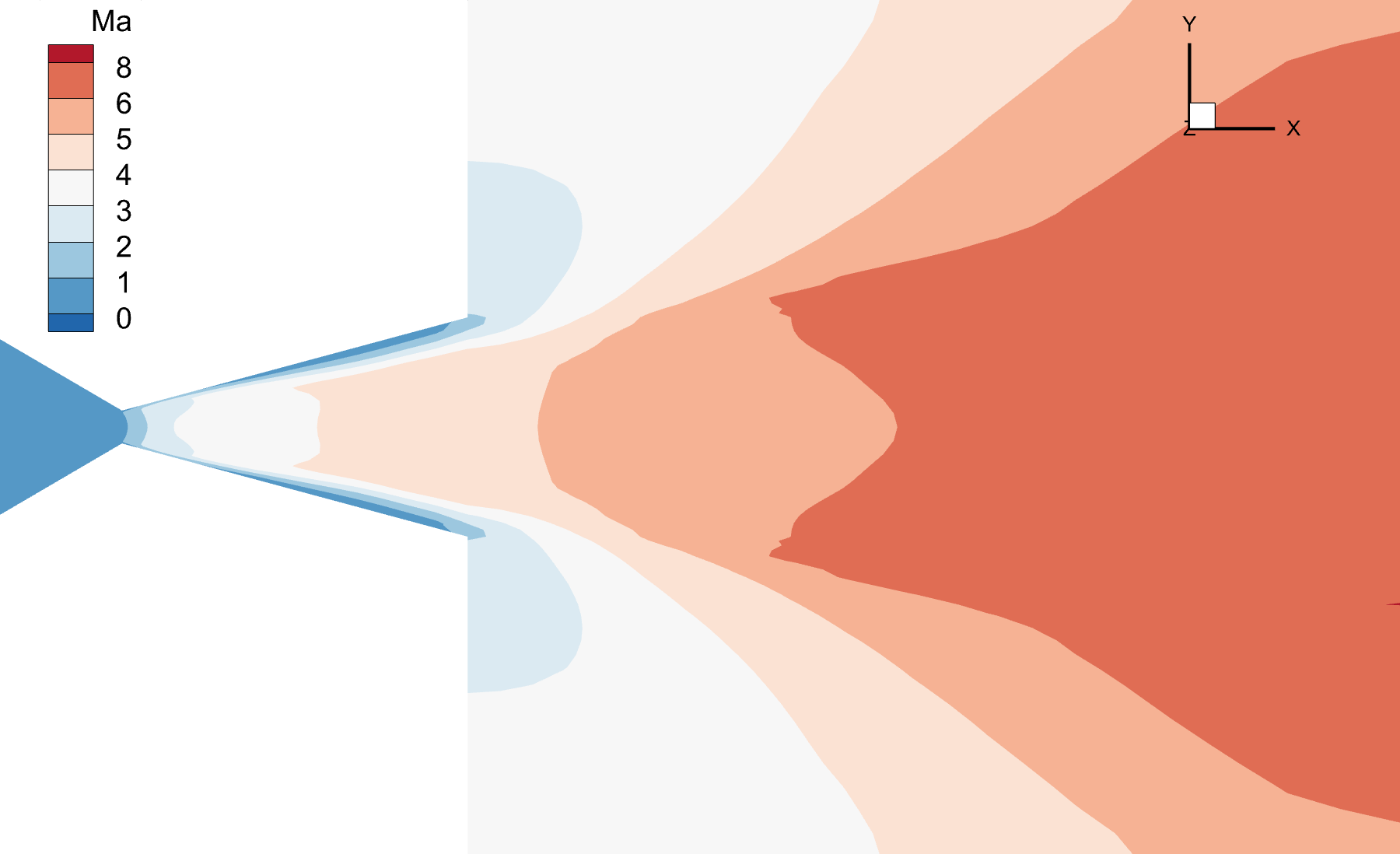}}
	\quad
	\subfloat[]{\includegraphics[width=0.4\textwidth]
		{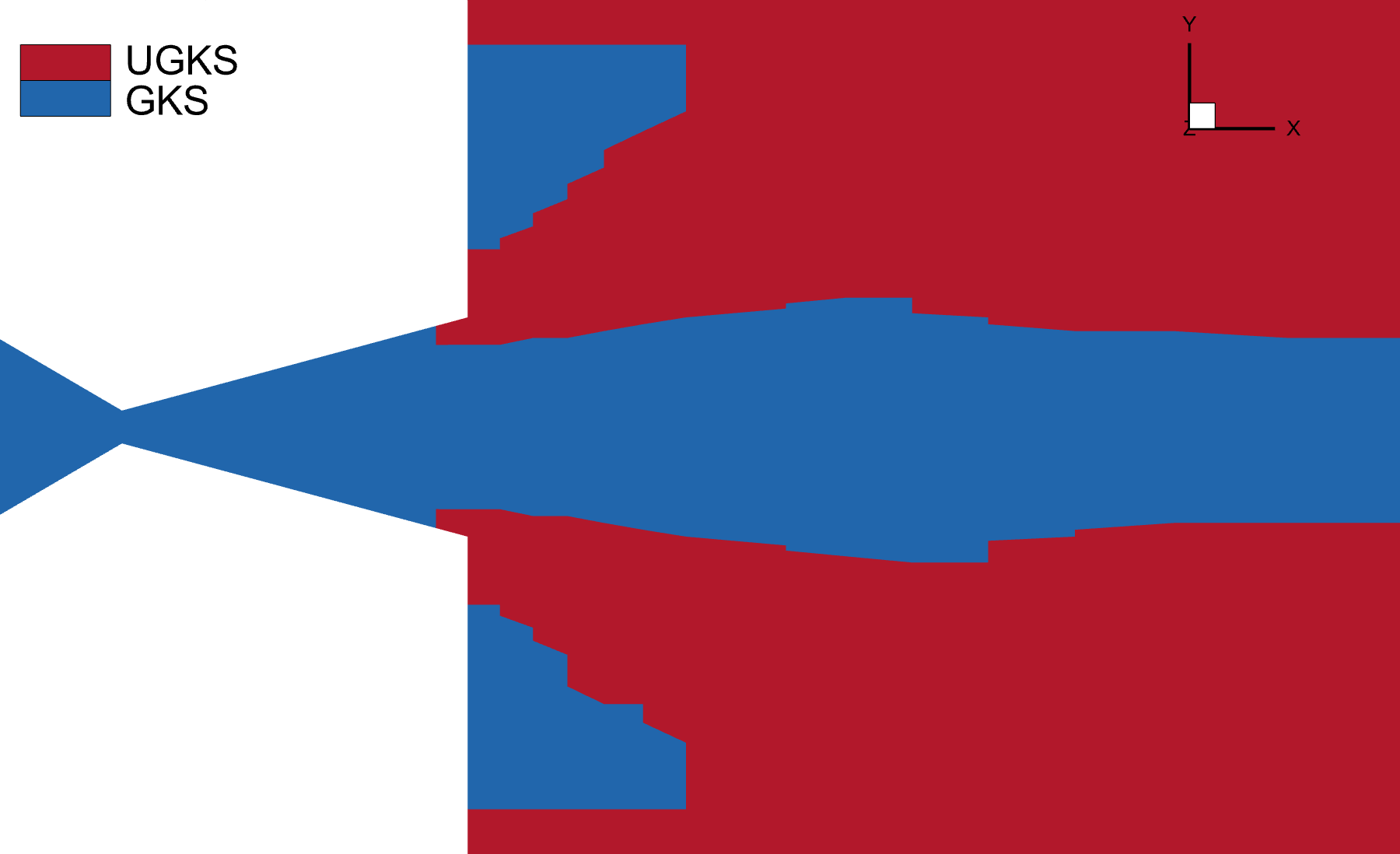}} \\
	\caption{Nozzle plume flow to a background vacuum $p_\infty = 0.01$ Pa at the steady stage. Distributions of (a) Mach number and (b) velocity space adaptation with $C_t= 0.05$ where the discretized velocity space (UGKS) is used in 36.60\% of physical domain.}
	\label{fig:nozzle-steady}
\end{figure}

The accuracy of the AUGKS method is verified by the comparison of Pitot pressure and temperature along the central axis of the nozzle shown in Fig.~\ref{fig:nozzleline}. Reasonable agreements with DSMC-NS data and experiment data have been observed. Figures~\ref{fig:nozzle-init}-\ref{fig:nozzle-steady} show the unsteady process of nozzle plume expansion. The AUGKS-vib provides clear pictures for velocity space adaptation. In the initial stage shown in Fig.~\ref{fig:nozzle-init}, particles with a large mean free path transport to the background vacuum, and the expansion gas forms a non-equilibrium region. The AUGKS-vib employs discretized distribution function with UGKS in this highly expanded region only, while the continuous distribution function using GKS is adopted in other regions. In this stage, 22.10\% of computational domain is covered by discretized velocity space. In the developing stage (see Fig.~\ref{fig:nozzle-developing}), a continuum flow regime appears near the nozzle exit, a transition regime forms around the high temperature expansion region, and a free molecular flow remains in the front of the plume. The discretized distribution functions (UGKS) is applied in the transition region which covers 46.28\% of the computational domain. Fig.~\ref{fig:nozzle-steady} shows the plume flow approaching a steady state, where the UGKS is used in the background flow region and covers 36.6\% of computational domain. The unsteady process shows the velocity space in the AUGKS-vib is dynamically adapted in computation, which coordinates with the local flow physics. During the computation, 13000 steps of AUGKS-vib simulation are conducted on the SUGON computation platform. 15 nodes (960 cores in total) of CPU 7285 32C 2.0GHz are used, and the total wall clock time is 9.69 h.

\section{Conclusion}\label{sec:conclusion}
In this paper, an adaptive unified gas-kinetic scheme (AUGKS) for diatomic gas with rotational and vibrational modes is constructed. Under the framework of UGKS, instead of using purely discretized particle velocity space in UGKS, the current adaptive scheme utilizes the GKS flux function based on a continuous particle velocity space in the continuum near-equilibrium region. 
The direct adaptation of gas distribution functions in AUGKS avoids the use of buffer zone with a mixture of different flow solvers. 
This compact property leads to an effective method for the unsteady multiscale flow simulation in complex geometries 
and with an automatic dynamic interface between UGKS and GKS.
Compared with the original UGKS, the AUGKS is more efficient and less memory demanding for multiscale flow computations. 
The AUGKS provides a useful tool for non-equilibrium flow study. The algorithm can be further developed with the acceleration techniques for convergence, such as implicit and multigrid, for the steady state flow simulation. 

\section*{Author's contributions}

All authors contributed equally to this work.

\section*{Acknowledgments}

This work was supported by National Key R$\&$D Program of China (Grant Nos. 2022YFA1004500), National Natural Science Foundation of China (12172316), Hong Kong research grant council (16208021,16301222).

\section*{Data Availability}

The data that support the findings of this study are available from the corresponding author upon reasonable request.

\appendix

\section{Moments and derivative of the Maxwellian distribution function for vibration model}\label{sec:appendix-1}
In the adaptive unified gas-kinetic scheme with the vibrational mode, the equilibrium flux $\vec{F}^{eq}_{ij}$ requires higher order moments of $ \vec{\xi}$ and $\varepsilon_v$. Here, we list the formula of the moments
\begin{equation*}
	\begin{aligned}
		&\int_{ -\infty }^{ \infty } \frac{\lambda_r}{\pi } \vec{\xi}^2
		e^{ - \lambda_r \vec{\xi}^2} {\rm{d}} {\vec{\xi}} = \frac{K_r}{2 \lambda_r} , \\
		&\int_{ -\infty }^{ \infty } \frac{\lambda_r}{\pi } \vec{\xi}^4
		e^{ - \lambda_r \vec{\xi}^2} {\rm{d}} {\vec{\xi}} = \frac{ K_r^2 + 2 K_r }{ 4 \lambda_r^2} ,\\
		&\int_0^{ \infty }           \frac{4 \lambda_v}{K_v(\lambda_v)} \varepsilon_v
		e^{ - \frac{ 4 \lambda_v }{ K_v(\lambda_v) }{ \varepsilon_v }} {\rm{d}} \varepsilon_v
		= \frac{ K_v(\lambda_v) }{4 \lambda_v}, \\
		&\int_0^{ \infty } \frac{4 \lambda_v}{K_v(\lambda_v)} \varepsilon_v^2
		e^{ - \frac{ 4 \lambda_v}{K_v(\lambda_v)}\varepsilon_v} {\rm{d}} \varepsilon_v
		= 2 \left( \frac{K_v(\lambda_v)}{ 4 \lambda_v} \right)^2.
	\end{aligned}
\end{equation*}
The distribution of the equilibrium state in space and time $( \vec{r}, t)$ can be expanded by the Taylor expansion
\begin{equation*}
	g(\vec{r}, t) = g_0 + \vec{r} \cdot \frac{\partial g}{\partial \vec{r}}
	                   + \frac{\partial g}{\partial t} t.
\end{equation*}
As an example, taking the $x-$ as the normal direction of the cell interface, the micro-slope $a$ can be defined by
\begin{equation*}
	a=\frac{1}{g}\bigg(\frac{\partial g}{\partial x}\bigg),
\end{equation*}
with the form
\begin{equation*}
	a =a_1
	+a_2 u
	+a_3 v
	+a_4 w
	+\frac12 a_5\vec{u}^2
	+\frac12 a_6\vec{\xi}^2
	+ a_7{\varepsilon_v}.
\end{equation*}
Applying the chain rule, the micro-slope $a$ can be determined by the derivative of macroscopic quantities evaluated at $({\vec r},t)$
\begin{equation*}
	\begin{aligned}
	&a_1 = \frac{1}{\rho }\frac{{\partial \rho }}{{\partial x}}
			 - {a_2}U - {a_3}V - {a_4}W
			 - \frac{1}{2}{a_5}\left( {{{\vec{U}}^2} + \frac{3}{{2{\lambda _t}}}} \right)
			 - \frac{1}{2}{a_6}\frac{{{K_r}}}{{2{\lambda _r}}}
			 - {a_7}\frac{K_v(\lambda_v)}{4\lambda_v}, \\
	&a_2 = \frac{\lambda_t}{\rho} R_1 - a_5 U,\\	
	&a_3 = \frac{\lambda_t}{\rho} R_2 - a_5 V,\\
	&a_4 = \frac{\lambda_t}{\rho} R_3 - a_5 W,\\
	&a_5 = \frac{4\lambda_t^2}{3 \rho} \left( B- U R_1 - V R_2 - W R_3 \right),\\
	&a_6 =  \frac{ 4 \lambda_r^2 } { K_r \rho }
	       \left( \frac{4}{ K_r } \frac{ \partial (\rho E_r)}{ \partial x }
		        - \frac{1}{ \lambda_r} \frac{ \partial \rho }{ \partial x} \right),\\
	&a_7 = \frac{ 4 e^{ 2 \Theta_v R \lambda_v } \lambda_v^2 }
	            { ( 4 \lambda_v R \Theta_v + K_v(\lambda_v)) \rho }
			\left( \frac{4}{K_v(\lambda_V)} \frac{ \partial (\rho E_v) }{ \partial x}
			     - \frac{1}{\lambda_v}      \frac{ \partial \rho }{ \partial x} \right),
	\end{aligned}
\end{equation*}
with the defined variables
	\begin{align*}
	\begin{aligned}
		B &= 2 \frac{\partial (\rho E - \rho E_{r} - \rho E_v )}{\partial x}
		   - (\vec{U}^2 + \frac{3}{2\lambda_t}) \frac{\partial \rho}{\partial x} ,\\
		R_1 &= 2 \frac{\partial \rho U}{\partial x}  - 2U \frac{\partial \rho}{\partial x} ,\\
		R_2 &= 2 \frac{\partial \rho V}{\partial x}  - 2V \frac{\partial \rho}{\partial x} ,\\
		R_3 &= 2 \frac{\partial \rho W}{\partial x}  - 2W \frac{\partial \rho}{\partial x}.\\
		\end{aligned}
	\end{align*}

	\section{Upstream and downstream condition of a shock structure with vibrational mode}
	\label{sec:app-shock}
	
	Since the vibrational degrees of freedom depend on the temperature, the specific heat ratio is not a constant in the computational domain. For normal shock structure, the Rankine--Hugoniet relation under the constant specific heat ratio $\gamma = 7/5$ is no longer valid. Instead, the relation between upstream and downstream states should be obtained by imposing conservation laws with a non-constant specific heat ratio
	\begin{equation}\label{eq:shock-T}
		\frac{\lambda_2}{\lambda_1}
		=
		\frac{\left( {\rm Ma}_2^2 \gamma_2 \right) / 2 + \gamma_2 / \left( \gamma_2 - 1 \right)}
			 {\left( {\rm Ma}_1^2 \gamma_1 \right) / 2 + \gamma_1 / \left( \gamma_1 - 1 \right)},
	\end{equation}
	\begin{equation} \label{eq:shock-u}
		\frac{u_2}{u_1}
		=
		\sqrt{ \frac{ \left[ 1/2 + {\rm Ma}_2^2 / \left(\gamma_1 - 1\right)\right] }
					{ \left[ 1/2 + {\rm Ma}_1^2 / \left(\gamma_2 - 1\right)\right] } },
	\end{equation}
	\begin{equation}\label{eq:shock-p}
		\frac{p_2}{p_1}
		=
		\frac{ 1 + \gamma_1 {\rm Ma}_1^2}{ 1 + \gamma_2 {\rm Ma}_2^2 },
	\end{equation}
	\begin{equation}\label{eq:shock-Ma}
		\frac{ \left( 1 + \gamma_1 {\rm Ma}_1^2 \right)^2 }
			 { \gamma_1 {\rm Ma}_1^2 \left[ \gamma_1 / \left( \gamma_1 - 1 \right)
								   + \left( \gamma_1 {\rm Ma}_1^2 \right) / 2 \right] }
		=
		\frac{ \left( 1 + \gamma_2 {\rm Ma}_2^2 \right)^2 }
			 { \gamma_2 {\rm Ma}_2^2 \left[ \gamma_2 / \left( \gamma_2 - 1\right)
								   + \left( \gamma_2 {\rm Ma}_2^2 \right) / 2 \right] },
	\end{equation}
	where the subscripts “1” and “2” denote the state at upstream and downstream, respectively. The relation between specific heat ratio and the internal degrees of freedom is
	\begin{equation} \label{eq:gamma}
		\gamma = \frac{ 7 + K_v }{ 5 + K_v}.
	\end{equation}
	Substituting Eq.~\eqref{eq:Kv} into Eq.~\eqref{eq:gamma}, the expression for specific heat ratio with respect to temperature $\lambda$ can be obtained
	\begin{equation}\label{eq:gamma-T}
		\gamma = \frac{ 7 \left( e^{ 2 R \lambda \Theta_{v} } - 1 \right) + 4 R \lambda \Theta_v }
					  { 5 \left( e^{ 2 R \lambda \Theta_{v} } - 1 \right) + 4 R \lambda \Theta_v }.
	\end{equation}
	Due to the complexity of Eq.~\eqref{eq:gamma-T}, explicit determination of the downstream is difficult,
	therefore, implicit iteration of Eqs~\eqref{eq:shock-Ma}, \eqref{eq:shock-T} and \eqref{eq:gamma-T} is carried out to get the downstream temperature and Mach number. Then, the velocity and pressure in the downstream are determined by Eqs~\eqref{eq:shock-u} and \eqref{eq:shock-p}.
	




\bibliographystyle{elsarticle-num}
\bibliography{augks.bib}







\end{document}